\documentclass[twocolumn, twocolappendix]{aastex63}

\usepackage{graphicx}
\usepackage{multirow,booktabs}
\usepackage{amsmath}
\usepackage{amssymb}
\usepackage{xcolor}

\graphicspath{{Figures/}}

\makeatletter
\newcommand\footnoteref[1]{\protected@xdef\@thefnmark{\ref{#1}}\@footnotemark}
\makeatother

\shorttitle{Ophiuchus 163131 Radiative Transfer Modeling}
\shortauthors{Wolff et al.}

\begin{document}

\title{The Anatomy of an Unusual Edge-on Protoplanetary Disk I. Dust Settling in a Cold Disk}

\correspondingauthor{Schuyler G. Wolff}
\email{sgwolff@arizona.edu}

\author{Schuyler G. Wolff}
\affil{Leiden Observatory, Leiden University \\ 2300 RA Leiden, The Netherlands}
\affil{University of Arizona \\ Tucson, AZ}

\author{Gaspard Duch\^{e}ne}
\affil{University of California, Berkeley}
\affil{Universit\'e Grenoble-Alpes, CNRS \\ Institut de Plan\'etologie et d'Astrophyisque (IPAG) \\ F-38000 Grenoble, France}

\author{Karl R. Stapelfeldt}
\affil{Jet Propulsion Laboratory \\ California Institute of Technology \\
Mail Stop 321-100, 4800 Oak Grove Drive
Pasadena, CA 91109  USA}

\author{Francois M\'{e}nard}
\affil{Universit\'e Grenoble-Alpes, CNRS \\ Institut de Plan\'etologie et d'Astrophyisque (IPAG) \\ F-38000 Grenoble, France}

\author{Christian Flores}
\affil{Institute for Astronomy, University of Hawaii at Manoa \\ 640 N. Aohoku Place, Hilo, HI 96720, USA}

\author{Deborah Padgett}
\affil{Jet Propulsion Laboratory \\ California Institute of Technology}

\author{Christophe Pinte}
\affil{UMI-FCA, CNRS/INSU, France (UMI 3386)}
\affil{Dept. de Astronom\'{\i}a \\ Universidad de Chile \\ Santiago, Chile}

\author{Marion Villenave}
\affil{Universit\'e Grenoble-Alpes, CNRS \\ Institut de Plan\'etologie et d'Astrophyisque (IPAG) \\ F-38000 Grenoble, France}

\author{Gerrit van der Plas}
\affil{Universit\'e Grenoble-Alpes, CNRS \\ Institut de Plan\'etologie et d'Astrophyisque (IPAG) \\ F-38000 Grenoble, France}

\author{Marshall D. Perrin}
\affil{Space Telescope Science Institute \\ Baltimore, MD 21218, USA}



\begin{abstract}
As the earliest stage of planet formation, massive, optically thick, and gas rich protoplanetary disks provide key insights into the physics of star and planet formation. When viewed edge-on, high resolution images offer a unique opportunity to study both the radial and vertical structures of these disks and relate this to vertical settling, radial drift, grain growth, and changes in the midplane temperatures. In this work, we present multi-epoch HST and Keck scattered light images, and an ALMA 1.3 mm continuum map  for the remarkably flat edge-on protoplanetary disk SSTC2DJ163131.2-242627, a young solar-type star in $\rho$ Ophiuchus. We model the 0.8 $\mu$m and 1.3 mm images in separate MCMC runs to investigate the geometry and dust properties of the disk using the MCFOST radiative transfer code. In scattered light, we are sensitive to the smaller dust grains in the surface layers of the disk, while the sub-millimeter dust continuum observations probe larger grains closer to the disk midplane. An MCMC run combining both datasets using a covariance-based log-likelihood estimation was marginally successful, implying insufficient complexity in our disk model. The disk is well characterized by a flared disk model with an exponentially tapered outer edge viewed nearly edge-on, though some degree of dust settling is required to reproduce the vertically thin profile and lack of apparent flaring. A colder than expected disk midplane, evidence for dust settling, and residual radial substructures all point to a more complex radial density profile to be probed with future, higher resolution observations.\\
\end{abstract}


\section{Introduction}
\label{Sec:intro}

The evolution of young stellar objects with gas rich envelopes to dust-dominated, optically thin disks (i.e. from Class I - III systems) is not a simple homologous reduction in disk mass for a fixed disk geometry. As the disk evolves, large changes in the disk structure are expected to occur \citep[e.g.][]{1996ApJ...462..439K,1999MNRAS.304..425A}. 
Once accretion onto the central star is disrupted, the gas is expected to dissipate quickly.
Disk evolution through magnetospheric accretion and photoevaporation will move the inner edge of the disk to larger radii, creating disks with large inner disk holes. 
Additionally, one of the key theoretically predicted stages of planet formation is dust settling, whereby large grains migrate preferentially to the disk midplane, causing dust disks to become geometrically thin \citep{1977Ap&SS..51..153W,2004ApJ...608.1050G}.
Both processes leave clear observational signatures in scattered light images and the spectral energy distribution (SED) of the disks \citep{2004A&A...421.1075D, 2006ApJ...638..314D}.
Recent millimeter observations of the HH 30 \citep{2012A&A...543A..81M} and HL Tau \citep{2016ApJ...816...25P} disks show significant dust settling of the larger grains to the disk midplane with very flat surface density distributions.

When viewed close to edge-on (typically within $15 \degr$ with a strong dust mass dependence), observations of dust rich protoplanetary disks provide a unique opportunity to study the vertical and radial structures in these systems. The central star is occulted at wavelengths where a coronagraph would typically be required and issues of inner working angle or PSF subtraction artifacts are avoided. In scattered light, the shape of the flared surface of the disk can be directly related to the mid-plane gas temperatures \citep{2007prpl.conf..523W}. 
Observations at longer wavelengths probe structures closer to the disk midplane. When combined with radiative transfer modeling, disk observations at optical through millimeter wavelengths have been used to constrain disk geometries (inclination, radius, scale height), dust masses, and grain properties \citep[for examples see][]{2003ApJ...588..373W, 2009A&A...505.1167S, 2012A&A...543A..81M}.

This work is part of a larger  Hubble Space Telescope (HST) observation program designed to double the sample of edge-on protoplanetary disks resolved in scattered light. 
Targets were selected from young, nearby sources based on a characteristic double-peaked spectral shape from literature compiled spectral energy distribution. The results are summarized in \citet{2014IAUS..299...99S}.
Followup ALMA continuum observations were also obtained for select sources to probe the larger dust grains in the disk midplane and are presented in \citet{2020arXiv200806518V}.
In this work, we focus on characterizing the dust in the individual source, 2MASS J16313124-2426281.
In a companion paper, Flores et al.~(in prep; hereafter Paper II) focus on the gas content of the disk, using CO maps to examine its underlying temperature structure and use optical and IR spectra to determine that the central source is roughly a solar mass star.

2MASS J16313124-2426281 (hereafter Oph163131) was first identified by \citet{2008ApJS..179..249D} during a survey of embedded sources in the Ophiuchus star forming region (SFR). However, the authors determined it was unlikely to be a low-luminosity embedded source because it was not located in a region of high volume density within Ophiuchus. \citet{2009ApJS..181..321E} later confirmed this was a young stellar object in Ophiuchus because of the IR excess greater than $3 \sigma$ at $8 \mu m$. The extremely weak optical and near-infrared fluxes relative to Ophiuchus YSOs of similar type make it a candidate edge-on disk, since these are typically underluminous by 3--5\,mag compared to directly visible young stars in the same star-forming region \citep{1997IAUS..182..355S}.

\citet{2017ApJ...834..141O} establish the distance to the Ophiuchus SFR 
from trigonometric parallaxes using the VLBA as part of the Gould's Belt Distances Survey. 
Oph163131 sits closest to the Eastern streamer of Ophiuchus (L1689) for which the authors provide a distance of $147.3 \pm 3.4 \, pc$.
More recently, GAIA has provided very precise distance determinations. Unfortunately, parallax data for Oph163131 is not yet available, but the distances to co-located Ophiuchus cloud members suggest a distance of $\sim 144 \, pc$, which agrees well with the Gould Belt Survey results \citep{2018ApJ...869L..33O}. 
We adopt a distance of 147 pc for this source.  

While Ophiuchus is thought to be one of the youngest nearby SFRs \citep[0.1 - 1 Myr;][]{1999ApJ...525..440L}, the constituents outside of the dense core are older with a median age of 2.1 Myrs \citep{2005AJ....130.1733W}. 
\citet{2011AJ....142..140E} report a disk fraction in Ophiuchus of $27 \pm 5$\% based on Spitzer MIR excesses. 
Even at such a young age, the disks in Ophiuchus show signs of dust evolution. 
\citet{2009ApJ...703.1964F} find a higher rate of disk evolution (as characterized by IR excesses and the equivalent width of the 10 $\mu m$ silicate emission feature) in the off-core region of Ophiuchus than other SFRs of similar ages \citep[e.g. Chamaeleon I;][]{2004ApJ...602..816L}.

\begin{figure*}[hpt!]
\begin{center}
\includegraphics[width=3.3in]{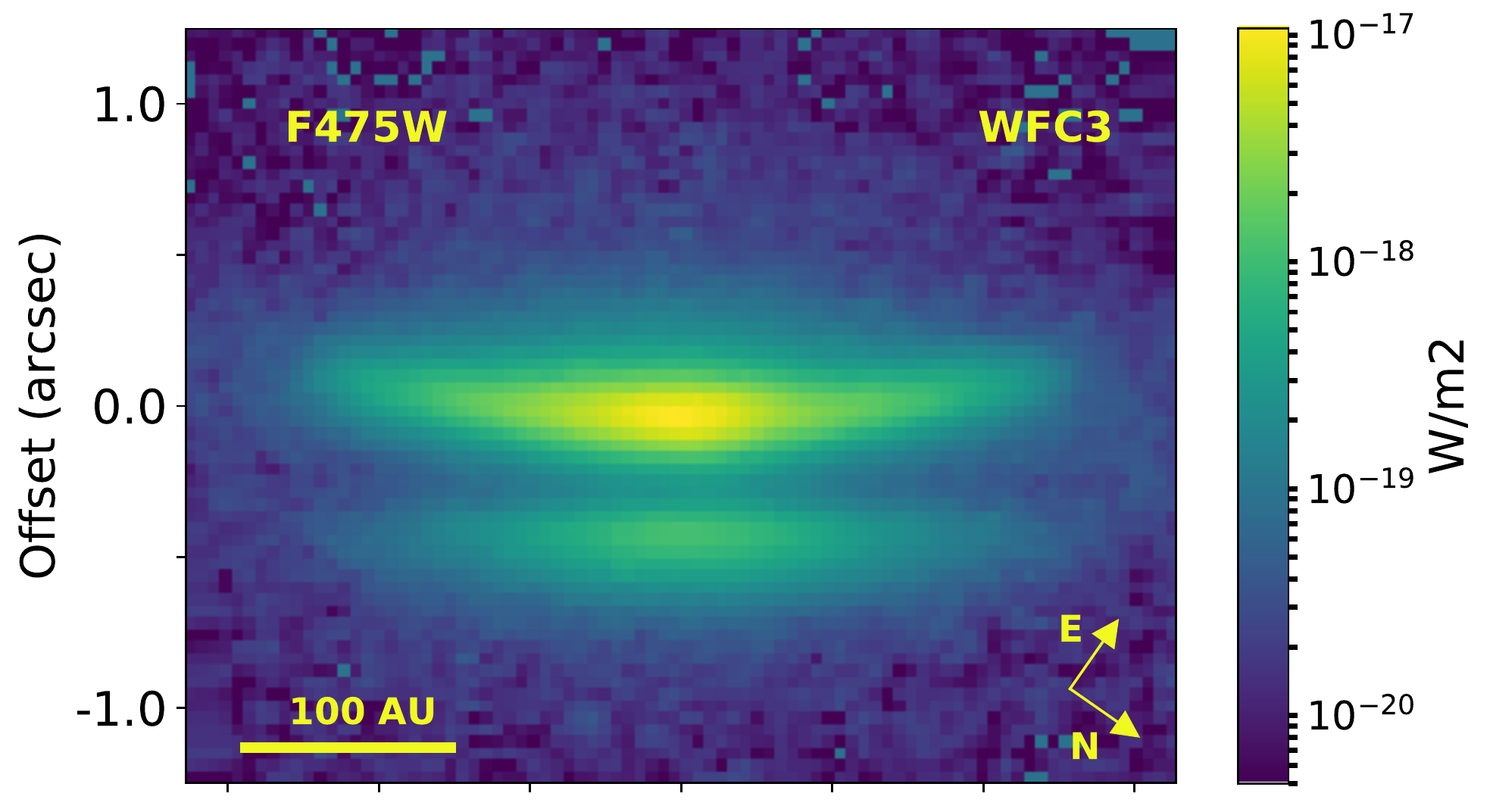}
\includegraphics[width=3.3in]{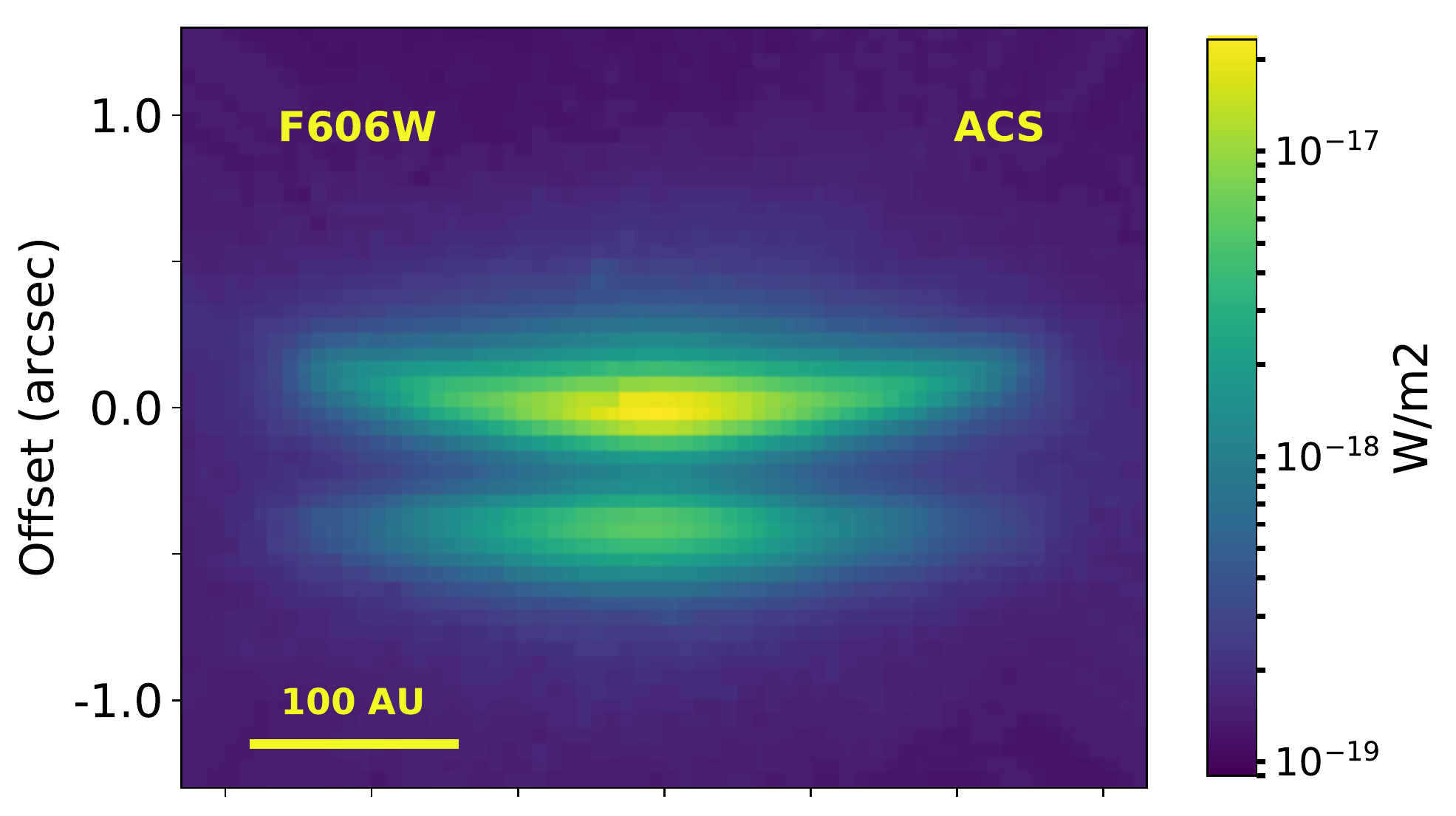}
\includegraphics[width=3.3in]{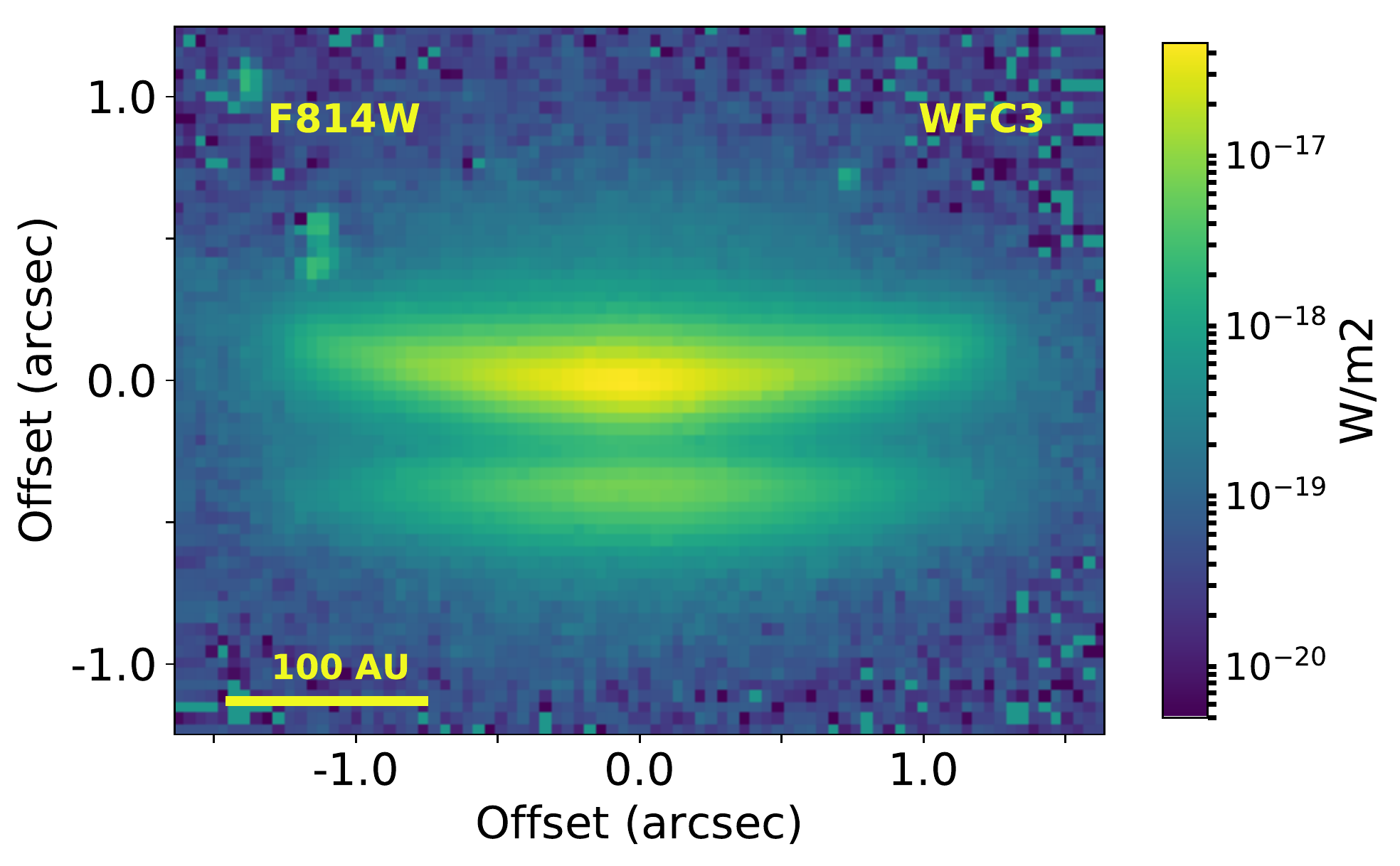}
\includegraphics[width=3.3in]{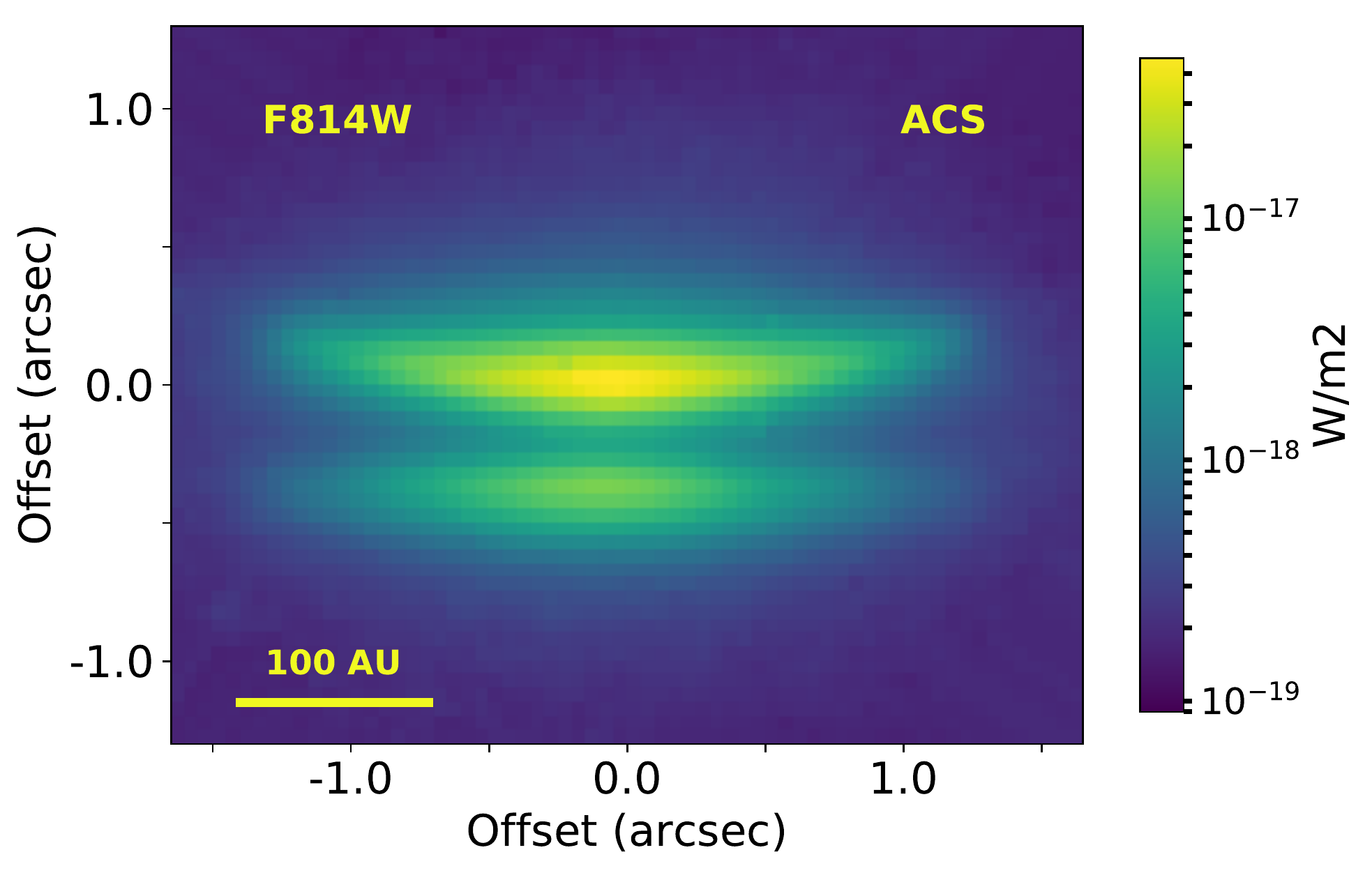}
\caption{HST scattered light observations obtained for Oph163131. For a complete description of the observations, see Table 1. Filters and instrument are shown, and scale bars are provided assuming a distance of 147 pc. \label{fig:scatteredobs}}
\end{center}
\end{figure*}

In this work, we investigate the geometry and dust properties of Oph163131 and relate the flattened appearance to the temperature and evolutionary stage of the disk.
Section \ref{Sec:obs} presents a comprehensive set of observations for Oph163131 including scattered light images obtained with the Hubble Space Telescope (HST) with the F475W, F606W, and F814W filters and Keck at 2.2 $\mu$m, ALMA continuum observations, and a spectral energy distribution compiled from the literature. 
In Section \ref{Sec:model} we describe the disk model and our scheme for the radiative transfer modeling of both the HST scattered light and ALMA continuum maps. The results are presented in Section \ref{Sec:results}. Implications for the structure of the Oph163131 including dust settling are discussed in Section \ref{Sec:discussion}. Finally, a summary of our results is provided in Section \ref{Sec:summary}.

\section{Observations and Analysis}
\label{Sec:obs}

\subsection{HST Scattered Light Images}
\label{Sec:hstdata}

Scattered light images of the Oph163131 disk were obtained using HST ACS in both the F814W and F606W broadband filters on March 4th, 2012 in Cycle 19 as part of the program GO 12514 (PI: Stapelfeldt). The total exposure times were 1320.0s for F606W and 800.0s for F814W, with each filter's exposure split into two integrations for cosmic ray rejection. Follow-up observations were conducted in HST Cycle 22 with WFC3 on October 3rd, 2014 as part of program GO 13643 (PI:  Duch\^{e}ne) in the F475W and F814W filters with exposure times of 2190.0s and 300.0s, respectively. 
Table \ref{table:obs} provides a summary of HST observations, while Figure \ref{fig:scatteredobs} shows the resolved disk images for each instrument/filter combination.
The WFC3 data shown are the calibrated data products produced by the default HST pipeline available via the Mikulski Archive for Space Telescopes (MAST) \citep{wfc3handbook}. Due to the unfortunate coincidence of a bad column on the ACS CCD with our disk observations, the MAST data products were re-drizzled using the DrizzlePac python package to combine the individual exposures with optional subsampling \citep{astrodrizzle}. 
The ACS datasets have a pixel scale of 0.05 arcseconds per pixel while the WFC3 datasets have a pixel scale of 0.039 arcseconds per pixel.

\begin{figure}[hbpt!]
\begin{center}
\includegraphics[width=3.3in]{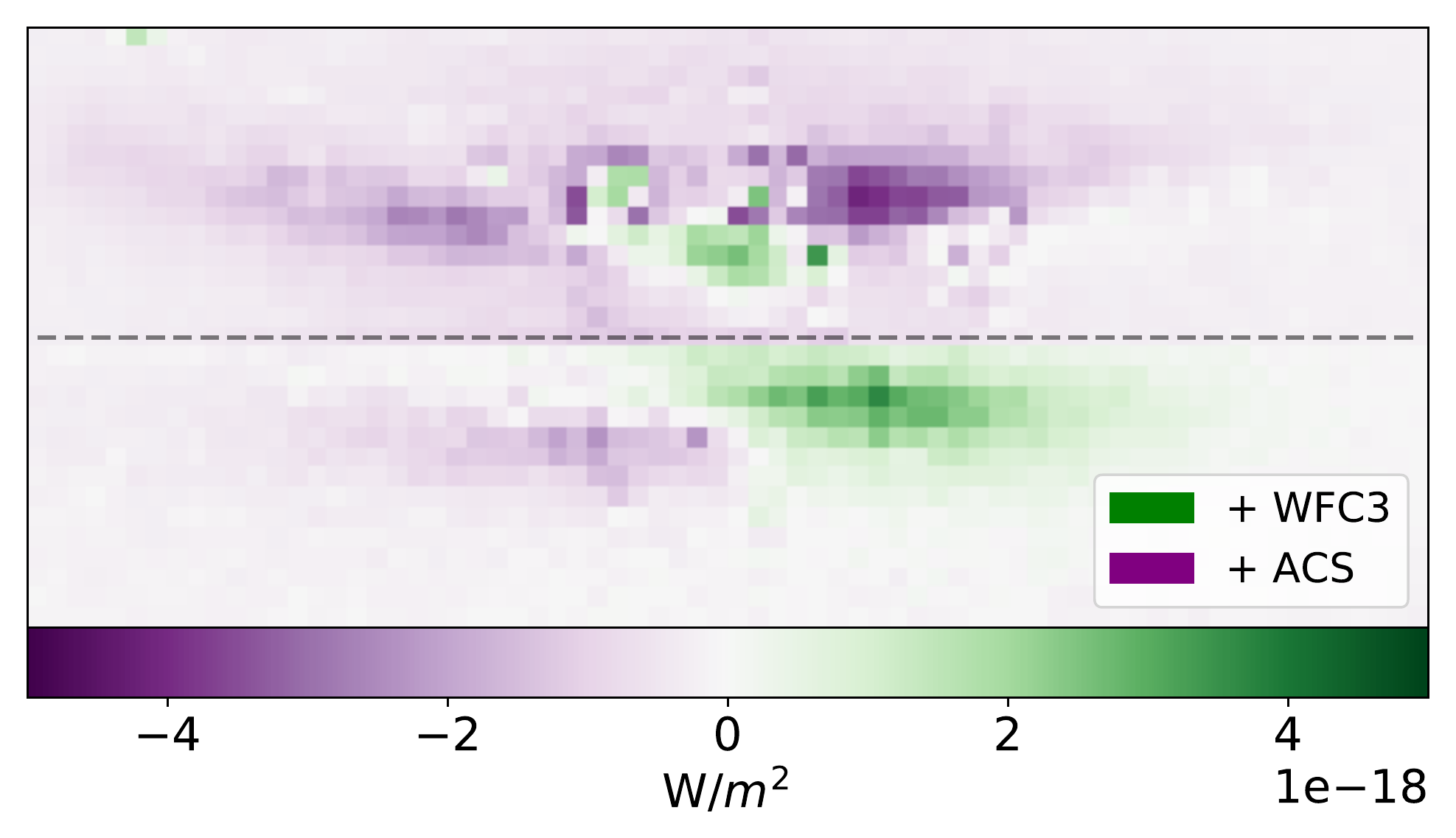}
\caption{Difference image between the two F814W data epochs. The WFC3 image has been recentered to the ACS image using an FFT-based sub-pixel algorithm. The top and bottom nebulae have been scaled separately prior to subtraction of the two epochs to highlight different features. The top/bottom panels have been scaled to the peak value in the top/bottom nebulae. The bottom nebula shows a vertical shift between the two epochs resulting from the different top/bottom flux ratios between the two datasets (See Figure \ref{fig:darklane}). The ACS data is more sensitive to low surface brightness disk emission. Both panels show a left/right asymmetry indicating variability.
\label{fig:difference}}
\end{center}
\end{figure}

Each image shows the characteristic double nebula structure typical of an optically thick protoplanetary disk viewed nearly edge-on.
However, the disk has a clear left/right (SW/NE) asymmetry that shifts between the two epochs spaced 19 months apart. 
The left/right flux ratio between the two nebulae is $\sim 1.2$ across all the scattered light observations (Table \ref{table:obs}).
Figure \ref{fig:difference} shows an image difference between the two F814W epochs that has been scaled to highlight the variations. 
The bottom nebula surface brightness peaks $\sim 1''$ to the northeast of the center in the WFC3 images and to the southwest in the ACS images. This implies some structure in the inner regions of the disk is producing variable illumination at the disk's outer surface, as has been observed in HH\,30 \citep{2004ApJ...602..860W}. 
The top/bottom integrated flux ratio is $3.5 \pm 0.9$ for both ACS images and approximately twice this value for the WFC3 images, though the values are consistent within $1 \sigma$ uncertainties (see Table \ref{table:obs}). The higher SNR ACS data is more sensitive to low surface brightness disk emission. 

The disk is atypically flat for a source of such a young age when compared across our HST sample \citep{2014IAUS..299...99S}. To quantify this, we introduce an aspect ratio metric defined as the maximum radial extent divided by the maximum vertical extent of the disk dark lane. An example of this measurement is provided in \citet[][Appendix D]{2020arXiv200806518V}. Oph163131 has an aspect ratio of $\sim 4.5$ while most other edge-on disks have an aspect ratio under 3, though HK Tau B and LkHa 263C are notable exceptions with aspect ratios of 4.4 and 5, respectively. Like Oph163131, both HK Tau B and LkHa 263 C are uncharacteristically flat \citep{2011ApJ...727...90M, 2002ApJ...571L..51J}. This has interesting implications for the temperature structure and evolution of the disk that will be discussed further below. Also notable is the lack of any jet emission visible in the F606W filter. For a disk of this age, we would expect the central star to be actively accreting with stellar jets observed perpendicular to the disk in H$\alpha$ or forbidden line emission \citep{2005A&A...434.1005A}. 
In Paper II, we find very little in the emission line spectrum of the source that supports ongoing accretion and no other conclusive signatures of accretion have been published for this source. 
It is possible that some mechanism has cut off accretion onto the central star (e.g., an embedded proto-planet or disk gap) but there is no clear evidence for an inner clearing of disk material in the spectral energy distribution (see Section \ref{sec:sed}). 

We measure the position angle (PA) of the disk by minimizing the difference between the left and right sides of the nebula as a function of the angle. The PA is $49 \pm 1$ degrees in all bands measured relative to North. 
The disk is red with both the peak and integrated flux of the disk increasing with longer wavelengths. Table \ref{table:obs} lists these values for each HST scattered light image. 
We construct a noise map combining both background and the photon noise for each HST scattered light image. 
The integrated flux was computed from the weighted sum of the pixels above a $5 \sigma$ uncertainty. The two F814W measurements agree within the uncertainties. Using the ACS/F814W flux density, this gives AB magnitude colors of [F606W]-[F814W]=1.5 mag, 
[F475W]-[F814W]=2.6 mag, and [F475W]-[F606W]=1.1 mag.
For a conservative estimate of the outer disk radius, we determine the outermost point at which the disk flux is above the $3 \sigma$ noise level and average this value for the four instrument/filter combinations to arrive at a value of $191 \pm 8$ au, or $\sim 1.3''$ at a distance of 147 pc. 
The vertical extent of the disk is more difficult to quantify and we discuss this in more detail with the Keck data below.

\begin{deluxetable*}{ccccccc}  
\tablecaption{Summary of HST Observations \label{table:obs}}
\tablehead{ 
      \colhead{Date of} &  \colhead{Instrument/} & \colhead{Exposure} & \colhead{Peak Surface Brightness} & \colhead{Integrated Flux} & \colhead{Top/Bottom} & \colhead{Left/Right} \\  [-2mm]      
      \colhead{Observation} & \colhead{Filter} & \colhead{ Time (s)} &  \colhead{($mJy/sq \arcsec$)} &  \colhead{($mJy$)}  & \colhead{Flux Ratio} & \colhead{Flux Ratio} }
\tablewidth{7.2in}
\startdata
2014-10-03 & WFC3/F475W & 2190.0 & 1.1 $\pm$ 0.1 & 0.2 $\pm$ 0.1 & 6.7 $\pm$ 0.2 & 1.1 $\pm$ 0.1 \\ 
2014-10-03 & WFC3/F814W & 300.0 & 8.2 $\pm$ 0.5 & 1.7 $\pm$ 0.4 & 5.5 $\pm$ 0.2 & 1.2 $\pm$ 0.2 \\ 
2012-03-04 & ACS/F606W & 1320.0 & 3.1 $\pm$ 0.2 & 0.6 $\pm$ 0.1 & 3.7 $\pm$ 0.2 & 1.1 $\pm$ 0.2 \\ 
2012-03-04 & ACS/F814W & 800.0 & 8.4 $\pm$ 0.5 & 2.1 $\pm$ 0.2 & 3.5 $\pm$ 0.2 & 1.3 $\pm$ 0.2 
\enddata
\end{deluxetable*}

\subsection{Keck Image}
\label{sec:keck}

\begin{figure}[hbpt!]
\begin{center}
\includegraphics[width=3.3in]{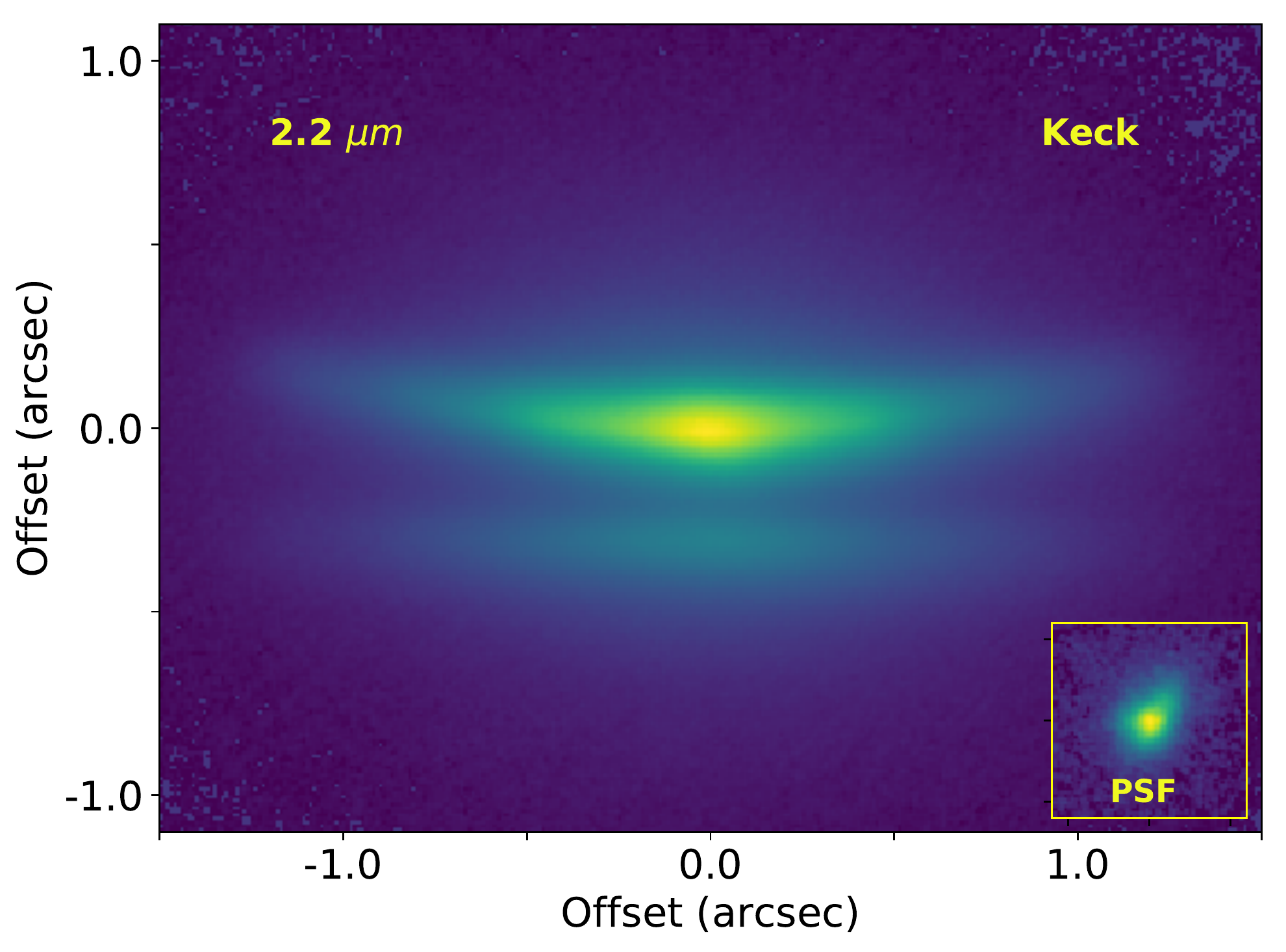}
\caption{Keck K' band scattered light image obtained as part of the edge-on disks campaign. These data exhibit a more pronounced halo of light outside of the disk and a narrower dark lane than the lower wavelength HST scattered light images. The Keck PSF is shown in the lower right corner. The asymmetric PSF is likely a result of using a faint extended source as an AO guide star. \label{fig:keckobs}}
\end{center}
\end{figure}

\begin{figure}[hbpt!]
\begin{center}
\includegraphics[width=3.3in]{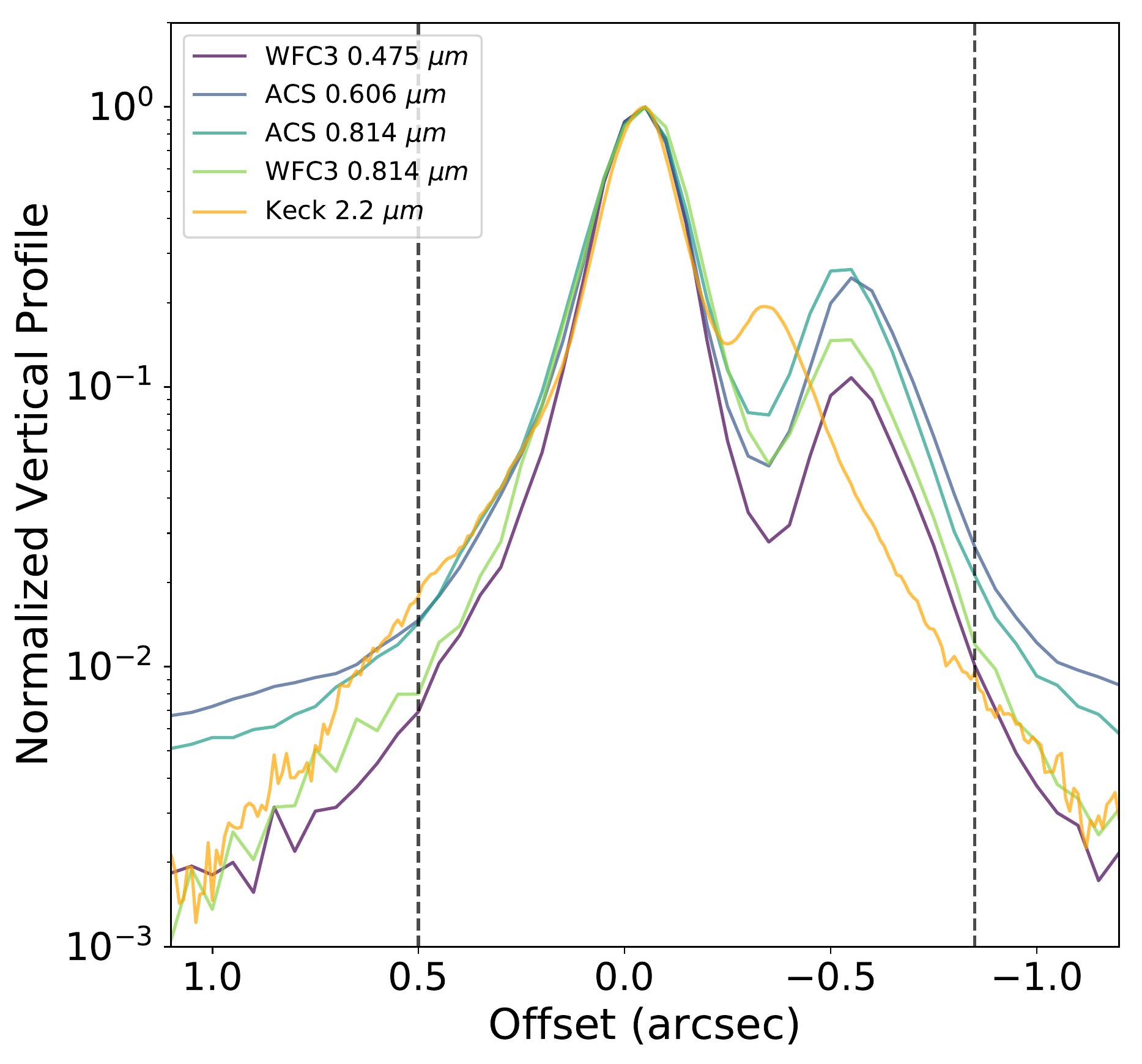}
\caption{Vertical profile for the scattered light images of Oph163131. While the HST data shows no strong change in the darklane width, the Keck data show a much narrower profile. Both the depth of the darklane and  the top/bottom flux ratio for the two disk nebulae increase with decreasing wavelength. The vertical dashed lines indicate the extent of the mask used when modeling the disk. \label{fig:darklane}}
\end{center}
\end{figure}

On April 25, 2019, we obtained a $K'$ ($2.124 \, \mu m$) image of Oph163131 with the NIRC2 instrument, with its Narrow camera (0\farcs00994/pix), installed behind the facility laser-guided adaptive optics system on the Keck\,II telescope (PI: Duch\^{e}ne). Observations were obtained at the highest airmass accessible from Mauna Kea, $z\approx1.4$, under 0\farcs5--0\farcs6 seeing conditions. The target itself was bright enough to use as the on-axis tip-tilt star. However, given its faint optical brightness ($R \approx 17$), the adaptive optics correction was imperfect and we measured an FWHM of $\approx $0\farcs09 from the images of point sources in the field of view. Indeed, the PSF appears elongated along the North-South direction (where the FWHM approaches 0\farcs13), possibly as a consequence of a double-lobed object being used for the adaptive optics tip-tilt correction.

We obtained a total of 15 exposures, each with an integration time of 60\,s, using a 3-position on-chip dither to obtain a simultaneous measurement of the sky background. Processing of the frames included sky subtraction, flat fielding, image registration based on the centroid location of one of the nearby stars, and median combination. The resulting final image is shown in Figure\,\ref{fig:keckobs} along with a PSF reconstructed using a nearby bright point source. We note that flux calibration for moderate-Strehl Keck adaptive optics images is impracticable and does not impact our model comparisons.

The Keck data show the same double nebulae structure as the HST scattered light images, with a diffuse, roughly circular halo (likely induced by the AO PSF) and a narrower dark lane. We estimate the disk PA using the same method described for the HST dataset and find a consistent value of $\sim$ 50 degrees. The left/right asymmetry observed in the HST observations does not appear in the Keck data with a flux ratio of 1.00. Either the feature that is causing the asymmetry is less visible at longer wavelengths, or was not active at the time of the Keck observations. We find an integrated flux ratio of 3.3 between the top/bottom disk nebulae. 

The vertical extent of the disk in scattered light is confused with a diffuse halo seen most clearly in the Keck 2.2 $\micron$ and WFC3/F814W images. These Keck observations show photons out to at least an arcsecond beyond the disk mid-plane in the vertical direction. We expect this to be a low-Strehl PSF feature rather than a physical halo of material associated either with the Ophiuchus cloud or the protostellar envelope. A true halo from a nascent envelope or disk wind would be comprised of smaller dust grains and consequently, would appear more prominent in the optical than the NIR.

Figure \ref{fig:darklane} shows the vertical profiles for all of the scattered light images. Each dataset has been normalized to the peak of the emission in the upper nebula.
The width of the dark lane does not change significantly in the HST data, but is markedly narrower in the Keck 2.2 $\mu m$ image where we are probing a population of larger grains with a $\tau=1$ surface closer to the disk mid-plane. In the HST dataset, the depth of the darklane and the top/bottom nebulae flux ratio both decrease with increasing wavelength (with the exception of the low-SNR WFC3 F814W image). However, at 2.2 $\mu m$ while the depth of the darklane continues to decrease, the flux ratio exhibits a slight increase.
Previous multi-wavelength scattered light observations of edge-on disks generally exhibit a narrowing darklane and a decrease in the top/bottom nebulae flux ratio at longer wavelengths \citep[e.g.][]{2004ApJ...602..860W,2010ApJ...712..112D}. However, an increase in the flux ratio between the top/bottom nebulae into the NIR was observed for HK Tau B \citep{2011ApJ...727...90M} and interpreted as evidence for a preferentially forward scattering grain population.

\subsection{ALMA Continuum Map}

\begin{figure*}[hbpt!]
\begin{center}
\includegraphics[width=2.97in]{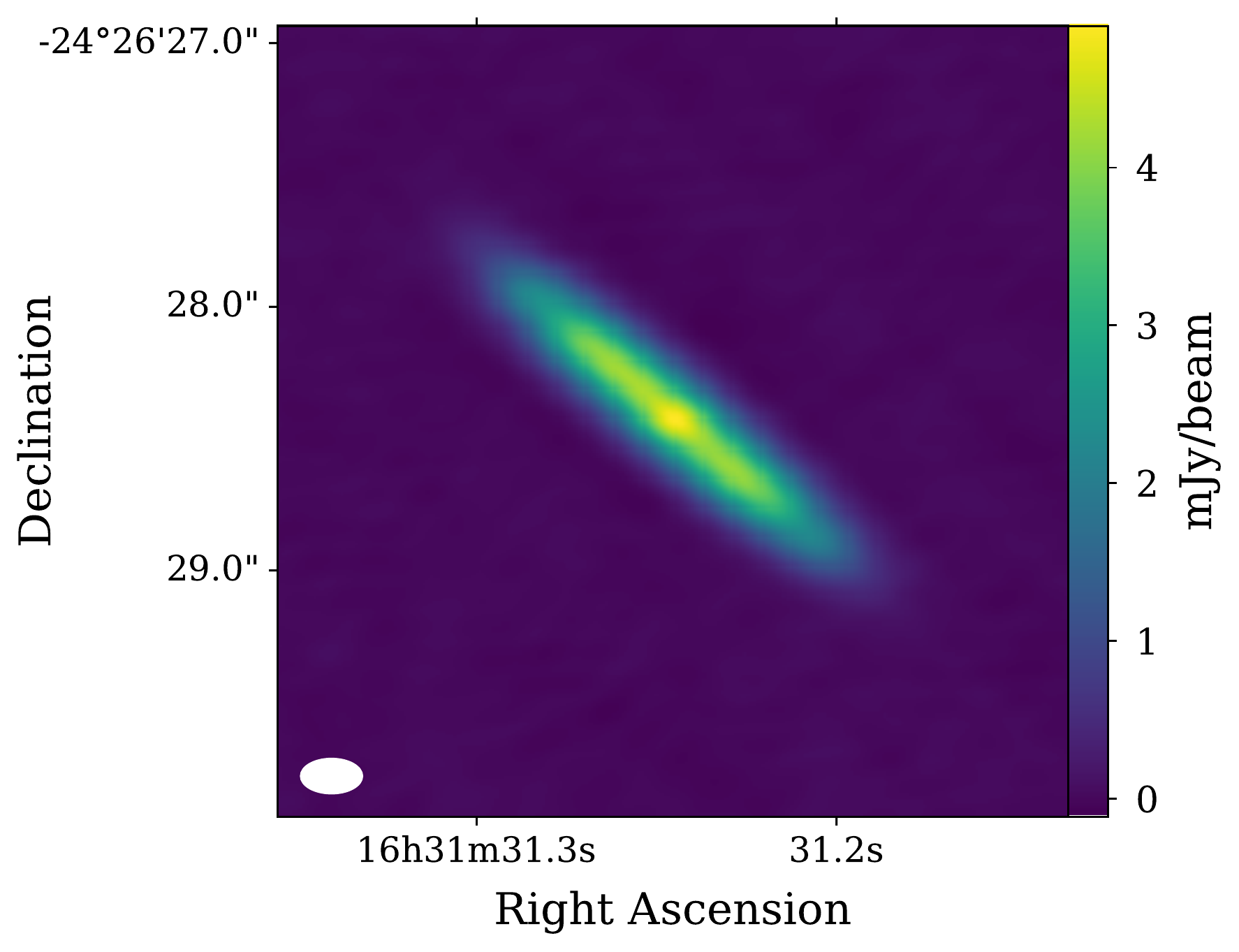}
\includegraphics[width=3.96in]{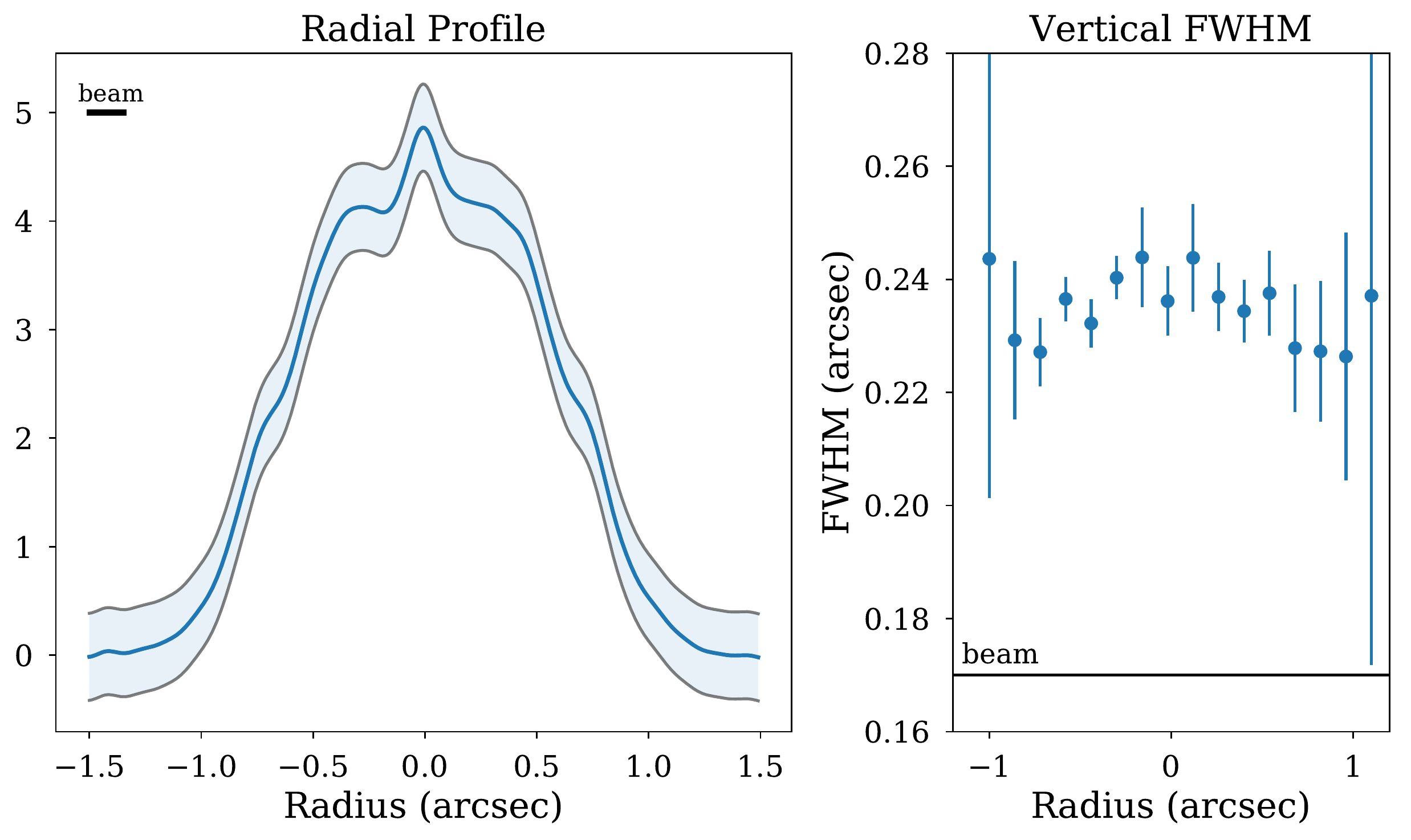}
\caption{The ALMA Band 6 Continuum data for Oph163131. The left panel shows the flux in mJy/beam. The middle panel shows the radial profile of the disk with a shaded region representing the $\pm \, 3 \sigma$ uncertainties. The right panel shows the the apparent vertical extent of the disk derived from a gaussian fit to cuts along the disk minor axis as a function of radius. The projected vertical span of the disk is constant with radius and is marginally resolved. The beam size is $\approx$ 0.15'' projected at the orientation of the disk major axis and $\approx$ 0.17'' for the disk minor axis.   \label{fig:alma}}
\end{center}
\end{figure*}

\begin{figure}[hbpt!]
\begin{center}
\includegraphics[width=3.3in]{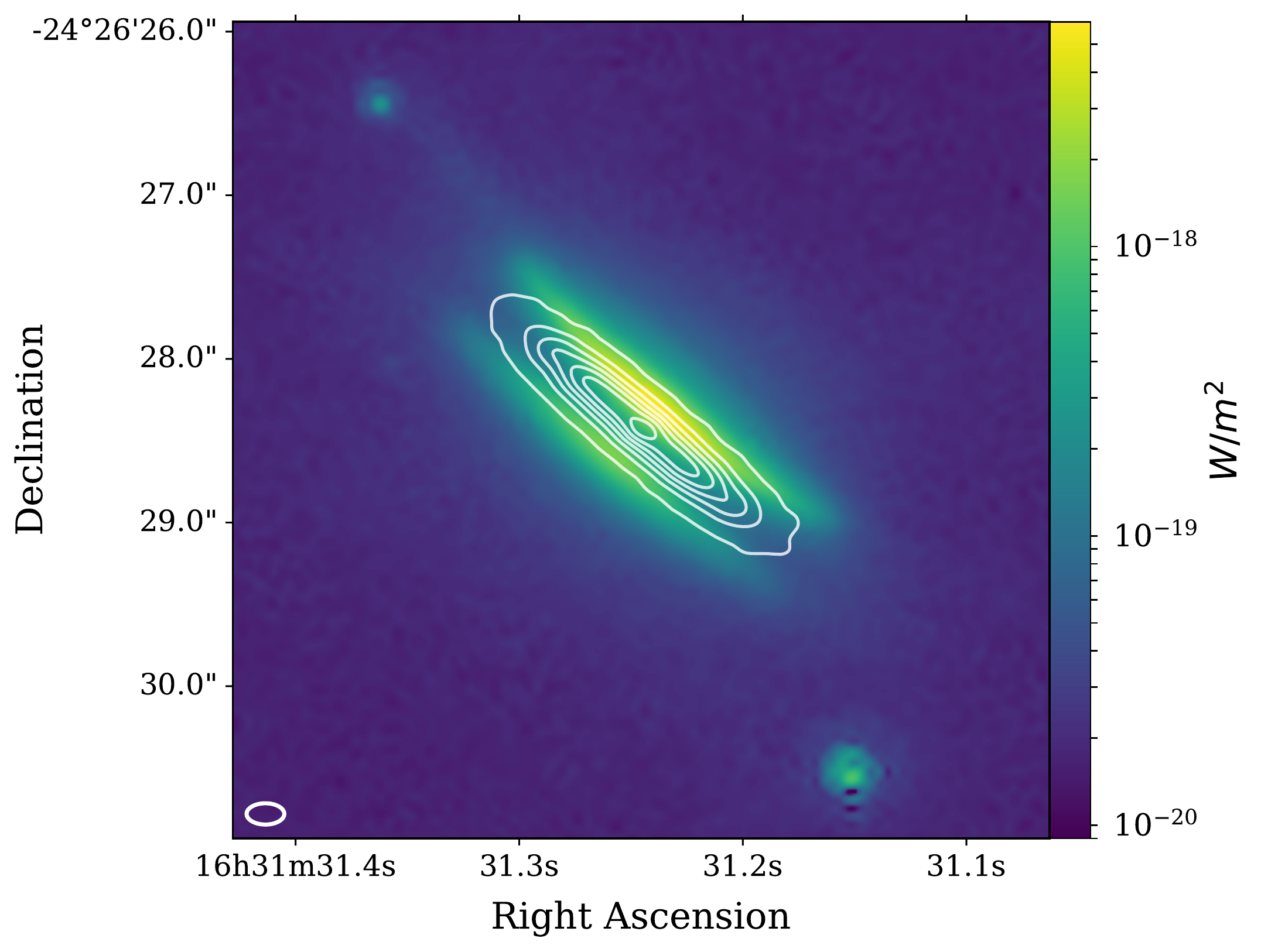}
\caption{A composite image of the ALMA continuum data, shown in contours, and the F814W ACS scattered light image. The small dust grains appear to extend further than the large dust grains, though this may be a result of the different sensitivities in the two datasets. Contours extend from 0.1 to 4.6 mJy/beam in increments of 0.75 mJy/beam. 
\label{fig:composite}}
\end{center}
\end{figure}

We obtained ALMA band 6 continuum data (1.3\,mm) as part of a larger proposal to map the 2-1 transitions of the three main CO isotopologues in two large edge-on protoplanetary disks in nearby, young star forming regions (Project 2016.1.00771.S, PI: Duch\^{e}ne). These disks were chosen to complement the existing high resolution HST data in scattered light. A full description of the ALMA continuum data reduction processes and results is presented in \citet{2020arXiv200806518V}, where we analyze a sample of 12 edge-on disks, including Oph163131. 
In this work, we model the continuum emission to probe the degree of vertical stratification and radial migration of the large dust grains in the system. 

The ALMA continuum map (beam size of 0\farcs23 x 0\farcs13) is shown in the left panel of Figure \ref{fig:alma}. 
Given the profile of the disk (sharply peaked followed by a plateau and a gradual taper; see Fig. \ref{fig:alma}), we were concerned a gaussian profile would not provide an accurate estimate of the flux in the image. Instead, we perform aperture photometry in the CASA software package with a 3$\sigma$ cut (with an rms of 0.134 mJy). 
Using this method, we find a flux density of 44.8 $\pm$ 4.5 mJy where we have assumed a conservative flux error of 10\%. Note that this is consistent with the integrated flux value of $\sim 47$ mJy provided by the CASA imfit package assuming gaussianeity.
We also find a deconvolved disk major axis FWHM of 1.22 $\pm$ 0.01 arcsec, a minor axis FWHM of 0.11 $\pm$ 0.01 arcsec, and a position angle of 49.0 $\pm$ 0.1 degrees. 

ALMA observations for this source have previously been published.
\citet{2017ApJ...851...83C} observed Oph163131 in band 7 as part of a high sensitivity survey of the Ophiuchus SFR and clearly
show the elongated source with size 1\farcs3 x 0\farcs16 for the major and minor disk axis with a 0\farcs2 beam. They reported an integrated flux of 124.8 $\pm$ 2.4 mJy which is included in the SED shown in Appendix \ref{appendix:sed}. 
Likewise, \citet{2019MNRAS.482..698C} provided a comprehensive ALMA survey of $\sim$ 300 sources in Ophiuchus in band 6. 
The flux and PA values are consistent with our results within uncertainties, though they reported a disk that is more radially compact and vertically extended. This is likely related to the larger beam (0\farcs28 x 0\farcs19) and lower sensitivity (10x smaller integration time) of this dataset.

We measured the semi-major and semi-minor axis profiles, which we interpret as tracing the radial and vertical distributions, respectively.
The middle panel of Figure \ref{fig:alma} shows the radial profile of the disk. 
The radial profile shows a symmetric disk with a peak at the location of the central star, a plateau of flux at 80\% of the peak value extending out to $\sim$ 0\farcs5 in radius, and an apparent outer radius of $\sim$ 1.2'' (180 au). 
This outer radius is smaller than inferred from the HST scattered light images ($\sim$ 190 au) signifying some amount of radial drift for the larger dust particles. However, of all the disks in our sample, Oph163131 has the smallest difference in the outer radius measured in the HST and ALMA datasets \citep{2020arXiv200806518V}. We note that in both cases, this outer radius does not represent a sharp cutoff of material but rather the point at which the surface brightness drops below the noise level of each image. Thus this discrepancy in the outer radius could be due to the different datasets' sensitivities, rather than a physical difference resulting from e.g., radial drift. 

The shape of the radial profile suggests the presence of unresolved rings, though higher resolution observations are needed to confirm.
For comparison, Figure \ref{fig:composite} shows a composite image of the ALMA continuum map (contours) and the 0.8 $\mu m$ scattered light image. 
In the right panel of Figure \ref{fig:alma}, we show the FWHM derived from a gaussian fit to vertical cuts of the data obtained at different radii. 
The vertical FWHM of the disk does not vary with radius with $\langle FWHM \rangle = 0.237''$. For comparison, the beam size is $\approx$ 0.17'' projected at the orientation of the disk minor axis.

\subsection{CARMA Detection}

Oph163131 was observed in the 2.6 mm continuum with the Combined Array for Research in Millimeter Astronomy (CARMA) under project cf0024  (PI: Stapelfeldt).  The observations took place with 23 antennas on 13 May 2013.  The synthesized beam was 2.5"x1.2". The total observing time including calibrators was 2.4 hours. The data show a putative detection of the unresolved disk. We find a flux density of $7.2 \pm 4.2$ mJy using the CASA imfit software package and interpret this as a $3\sigma$ upper limit of 13 mJy. This photometry is included in the spectral energy distribution (see Section \ref{sed:sedres} and Appendix \ref{appendix:sed}).

\subsection{Spectral Energy Distribution}
\label{sec:sed}

A well sampled spectral energy distribution (SED) has been compiled from the literature. A full description of the relevant observations is provided in Appendix \ref{appendix:sed}. 
The SED exhibits the double peaked structure typical of these optically thick disks when viewed nearly edge-on. 
While this dataset is not included in the radiative transfer modeling efforts, a comparison to the literature SED is given in Section \ref{sed:sedres}.

\section{Radiative Transfer Modeling}
\label{Sec:model}

In this work, we aim to fit for the disk geometry and dust properties simultaneously across two independent datasets; the 0.8 $\mu m$ HST/ACS scattered light image and the 1.3 mm ALMA continuum map. The combination of heterogeneous datasets in this manner is not often done because of the large variations in noise properties and degeneracies between the disk properties. For a more in depth discussion of the problem, see \citet{2017ApJ...851...56W}. In that work, we combined a scattered light image with a spectral energy distribution using a covariance-based likelihood estimation coupled with an MCMC. Here we extend that framework to include resolved interferometric millimeter continuum emission. In the next sections, we describe the disk model and the different modeling stages. 

\subsection{Stellar Properties}

The stellar properties of sources embedded in optically thick protoplanetary disks are, by nature, difficult to identify.
There is no way to directly determine the luminosity from observations because of the
uncertain amount of dilution of starlight by disk scattering. 
In Paper II, we report mid-resolution optical and IR spectroscopy for Oph163131 and compare to template spectral libraries to arrive at a spectral type of K4-5 for the central star. 
We then construct a catalogue of K3.5\,--\,K5.5 
spectral type stars in the Ophiuchus SFR from the literature \citep{2005AJ....130.1733W,2011AJ....142..140E}, and derive average values of 4500 K and 0.96 $L_{\odot}$ for the temperature and luminosity, with a corresponding radius of and 1.7 $R_{\odot}$. Finally, a fit to the PMS tracks on the HR diagram from \citet[][See Fig. 9]{2003ApJ...593.1093L} gives a stellar mass of $1.0 \pm 0.1 \, M_{\odot}$. 
This is consistent with the dynamical mass of $\rm 1.2\pm0.2 \, M_{\odot}$ measured by Paper II based on ALMA $^{12}$CO observations.
This source is under-luminous for its location and spectral type, as expected for a star seen only indirectly via scattered light from its edge-on disk.

\subsection{MCFOST Disk Model}

The MCFOST radiative transfer code  \citep{2006A&A...459..797P, 2009A&A...498..967P} was used to construct synthetic SEDs (unused when determining goodness of fit), 1.3 mm ALMA continuum maps and 0.8 $\mu m$ HST scattered light images to explore disk properties. 
Of the scattered light observations, we elected to fit only the F814W HST/ACS image. 
We aim to maintain the scattered light and mm continuum fitting on reasonably even footing, and the inclusion of multiple scattered light images would create systematic biases towards disk surface features.  
The HST scattered light observations all display similar features (e.g., no pronounced chromatic change in the darklane thickness, no strong jet signatures). 
Thus we chose to represent the HST dataset with the F814W HST/ACS image because it has the highest signal-to-noise ratio.  
We chose not to include the Keck image in the modeling because of the image artifacts characteristic of the poor AO correction from guiding on a faint, multi-lobed object.
We compare the best fit model to these observations in the discussion section.

We chose a tapered-edge axisymmetric disk model in which the surface density $\Sigma$ is described by a power law in radius ($R$) with an exponential taper outside of the critical radius $R_{c}$:\footnote{Note that previous works have used the notation $\alpha = - \gamma$. We avoid this here to limit confusion with the $\alpha$ viscosity parameter.} 
\begin{equation}
\Sigma = \Sigma_{c} \bigg(\frac{R}{R_{c}}\bigg)^{-\gamma} \exp \bigg[\bigg(- \frac{R}{R_{c}}\bigg)^{2-\gamma}\bigg]
\label{Eq:taper}
\end{equation} 

The disk gas scale height is also defined as a power law in radius by $H(R) = H_{0} (R/R_{0})^{\beta}$ where $\beta$ is the flaring exponent describing the curvature of the disk surface with a reference radius $R_{0}$ = 100 au.

Several model parameters were held fixed to minimize the degrees of freedom and to save computation time. Values for these parameters were either measured directly from the HST images or taken from the literature. The disk is assumed to be at a distance of 147 pc with an inner radius fixed at 1 au and an outer radius of 191 au. 
We expect a protoplanetary disk of this evolutionary stage to be optically thick in the inner few au at these wavelengths and we choose not to populate the inner au with dust in order to improve the speed of the radiative transfer calculations.
For comparison to data sets sensitive to thermal emission in regions close to the central star (e.g. the spectral energy distribution), a more conservative estimate of the inner radius is given by the sublimation radius $R_{\mathrm{sub}} = R_{\mathrm{star}} (T_{\mathrm{star}}/T_{\mathrm{sub}})^{2.1} \sim 0.07$\,au where $T_{\mathrm{sub}} = 1600$\,K \citep{2006ApJS..167..256R}. 

For a disk of this age, Oph163131 is uncommonly flat. By coupling the scattered light and mm continuum observations, we are able to probe the degree of dust settling within the disk.  
We use the \citet{2009A&A...496..597F} prescription for settling in which dust is diffused away from the midplane via MHD turbulence using a vertically varying diffusion coefficient to account for the larger velocity fluctuations in the outer layers of the disk. In MCFOST, this is parameterized using the $\alpha$ viscosity parameter \citep{1973A&A....24..337S}. 
The standard Gaussian vertical profile is then multiplied by an exponential term that depends both on the turbulent properties of the disk and the grain size distribution. In this way, the models are able to produce different scale heights for different grain sizes. We expect that the smallest dust grains should be well coupled with the gas and the larger grains (probed by the ALMA data) should have markedly smaller dust scale heights with $h_{1 \mu m} \simeq H \gg h_{1mm}$. 
For a more detailed description of how this is implemented in MCFOST see \citet{2016ApJ...816...25P}.

The dust population is described by a single species of amorphous olivine particles \citep{1995A&A...300..503D} with a particle size distribution described by a power law extending from $a$ = 0.03 - 1000 $\mu$m in size: $\frac{dN(a)}{da} \propto a^{-p}$ with 70 grain size bins spaced evenly in log(a). 
We assume non-porous grains. 

The inclination (with the edge-on orientation defined as 90\degr), gas scale height ($H_{0}$), dust mass ($M$), surface density exponent ($\gamma$), disk vertical flaring exponent ($\beta$), dust grain size exponent ($p$), the viscous settling parameter ($\alpha$), and the critical radius ($R_{C}$) were left as free parameters. 

\subsection{Model Likelihood Estimation and the MCMC Framework}
\label{Sec:covar}

In order to quickly and efficiently explore the parameter space we employ a parallel-tempered Monte Carlo Markov Chain approach using the {\tt emcee} package \citep{2013PASP..125..306F} and the Affine Invariant sampler \citep{goodmanandweare}. 
We anticipate complex, multi-modal posterior distributions and thus employ a parallel-tempered MCMC sampler where the posterior is modified by a temperature 
to artifically broaden the distribution and avoid being grounded in local minima.
We couple this with the {\tt mcfost-python} package developed by our team to interface with MCFOST, parse datasets from different instruments, and compute several goodness-of-fit metrics to be used in conjunction with the MCMC. The code is publicly available on Github.

Radiative transfer modeling of optically thick disks exhibits highly degenerate parameter relationships. Exacerbating this problem, is that we have two datasets with wildly different noise properties and model sensitivities. 
For a more complete understanding of the likelihood of each dataset given our model, we conduct three independent MCMC runs. We first perform individual fits to (a) the HST scattered light 0.8 $\mu m$ image and (b) the 1.3 mm ALMA continuum image. In both cases we use a $\chi^{2}$-based log-likelihood estimation. In the final run, we combine these observables using a covariance-based log-likelihood to inform the MCMC. 
By combining the information provided by both observables in a single MCMC, we aim to find a consistent solution for the disk geometry and grain properties across both datasets including settling and radial drift.

For the individual fits to the two datasets, we rely on a $\chi^{2}$-based log-likelihood estimation given by Equation \ref{eq:chisqr}.
\begin{equation}
\ln[P(\mathrm{D}| \Theta)] = -\frac{1}{2} \bigg[\chi^{2} + \sum_{i=1}^{N}  \ln{(\sigma_{i}^{2})} +  N \ln{(2 \pi)} \bigg]
\label{eq:chisqr}
\end{equation}

Here, D and $\Theta$ represent the data and model, respectively, $N$ is the number of data points, and $\sigma$ is our uncertainty. To determine the $\chi^{2}$ value for each model, we first generate a synthetic image with MCFOST at the specified wavelength, convolve this model with an instrumental PSF, normalize the model to the sum of the observed disk signal, recenter the model via cross correlation with the science frame and compute the sum of the $\chi^{2}$ values for each unmasked pixel.
For the 0.8 $\mu$m instrumental PSF we use a simulated Tiny Tim PSF \citep{1995ASPC...77..349K} and for the 1.3 mm PSF we use a 2D Gaussian corresponding to the ALMA beam size. 
In order to improve the computational efficiency, we mask each image to include only those pixels that are $3 \sigma$ above the background noise level. 
By marginalizing over the absolute flux scaling, we aim to fit the shape and structure of the disk without being limited by the relatively uncertain properties of the central star (e.g., luminosity or distance).

For both initial runs we use a parallel-tempered sampler with 2 temperatures and 50 walkers. The number of iterations varied between runs with 2000 required for the F814W scattered light run and only 1000 needed for the ALMA continuum run. A burn in stage of 100 iterations was used in both cases. 
The priors are described in Table \ref{table:params}. We use uniform priors for all parameters except for the inclination (uniform in $\cos{i}$), and the dust mass and viscosity parameter where log-uniform priors are used. Parameter values were chosen to be physically reasonable given the observations and analytical description (e.g., no critical radii outside of our masked image). 
These MCMC results are presented in Sections \ref{Sec:scatres} and \ref{Sec:almares} for the 0.8 $\mu$m and 1.3 mm optimized runs respectively. These results will also be used to inform the covariance-based method used in the combined fit as described below.

\begin{deluxetable}{l|c}[htbp!]  
\tablecolumns{2}
\tablecaption{MCMC Parameter Priors \label{table:params}
}
\tablehead{   
  \colhead{Parameter} &
    \colhead{Range} 
}
\tablewidth{2.7in}
\startdata
Inclination ($i$ in degrees) & 65.0 - 90.0  \\  
Scale Height ($H_0$ in au)  & 5.0 - 25.0  \\
Dust Mass* (M in $M_{\odot}$)   & $10^{-6}$ - $10^{-3}$ \\
Surface Density Exponent ($\gamma$) & 0.0 - 2.0  \\
Flaring Exponent ($\beta$) & 1.0 - 2.0  \\
Grain size exponent ($p$) & 2.5 - 4.0 \\
Viscous Settling Param.* ($\alpha$) & $10^{-5}$ - $10^{-1}$  \\
Critical Radius ($R_{C}$ in au) & 25.0 - 191.0 \\
\enddata
\tablecomments{Parameters marked with an * use log-uniform priors. All other parameters use uniform priors.}
\end{deluxetable}

The covariance-based method was first applied to an astrophysical context by \citet{2015ApJ...812..128C} for 1D spectral fitting. In \citet{2017ApJ...851...56W} we detail how to expand this to 2D disk observations and describe the benefits of including a global covariance in the modeling. 
In the third and final MCMC run, we combine the 0.8 $\mu$m and 1.3 mm images using this framework to leverage the information inherent in both datasets without preferentially biasing the fit towards one image or the other.

The matrix formalism for a Gaussian likelihood distribution describing the probability of the data given the model can be written as shown in Equation \ref{eq:likelihood_matrix}.
\begin{multline}
\ln[\mathrm{P}(\mathrm{D}| \Theta)] = -\frac{1}{2} \bigg[ E^{T} C^{-1} E + \ln(\det C) \\ + N \ln(2 \pi) \bigg]
\label{eq:likelihood_matrix}
\end{multline}
where $E$ represents the residuals (i.e., difference between pixel values in the model and observed images), 
and $C$ is the covariance matrix defined below. 
In all cases, the residuals are computed by first convolving the MCFOST model with an instrumental PSF,  normalizing the model to the sum of the observed disk signal, recentering the model via cross correlation with the science frame, and subtracting from the observations. These masked residuals are then flatted into a 1D array. For this combined MCMC, the ALMA data was downsampled to match the resolution of the HST image.

The covariance matrix $C$ (of size $N_{pix} \times N_{pix}$) combines both the noise in each individual pixel ($\sigma^{2}_{i,j}$) and a global covariances ($K^{G}_{i,j}$) between adjacent pixels (represented by $K^{G}$): $C_{i,j} = \delta_{i,j} \sigma^{2}_{i,j} + K^{G}_{i,j}$. The global covariance can encode correlated noise in datasets (for example any correlations between neighboring pixels from an instrumental PSF or beam) as well as any global limitations of the model to fit the data. To illustrate the power of this technique, consider the ability of these datasets to constrain the dust mass and flaring exponent. One would expect the ALMA continuum image to place a stronger constraint on the dust mass than the scattered light image. However, millimeter observations probing larger dust grains that have settled to the disk midplane are not sensitive to e.g. the flaring exponent and upper layers of the disk structure. 
Conversely, the scattered light image probes the disk flaring and vertical structure directly, but has a poor constraint on the mass since much of it is hiding in the optically thick interior and in the larger dust grains.
Here the familiar relationship between scattered light brightness and disk mass breaks down, but the structure of the nebula still provides a way to estimate the disk mass.

\begin{figure*}
\includegraphics[width=2.3in]{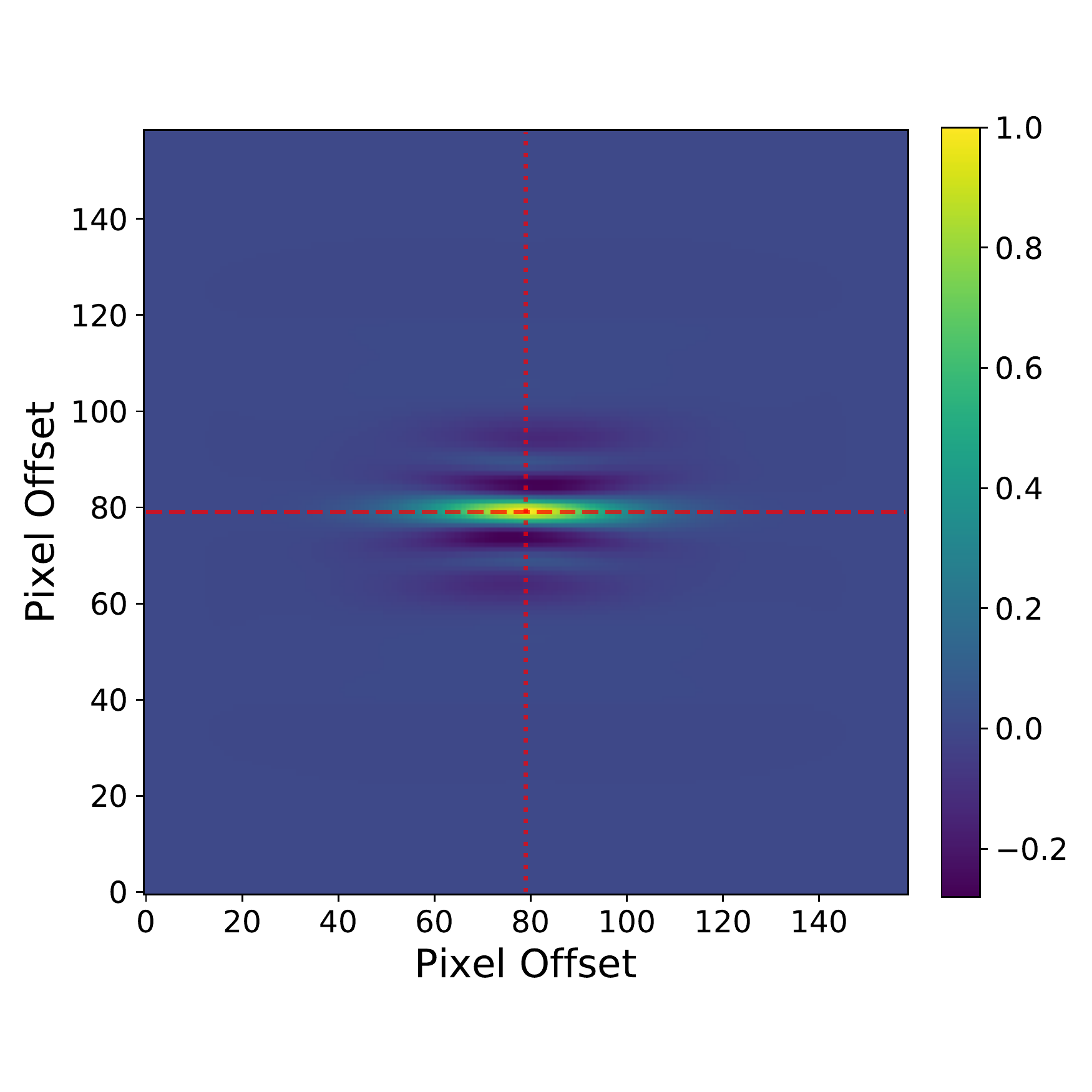}
\includegraphics[width=2.3in]{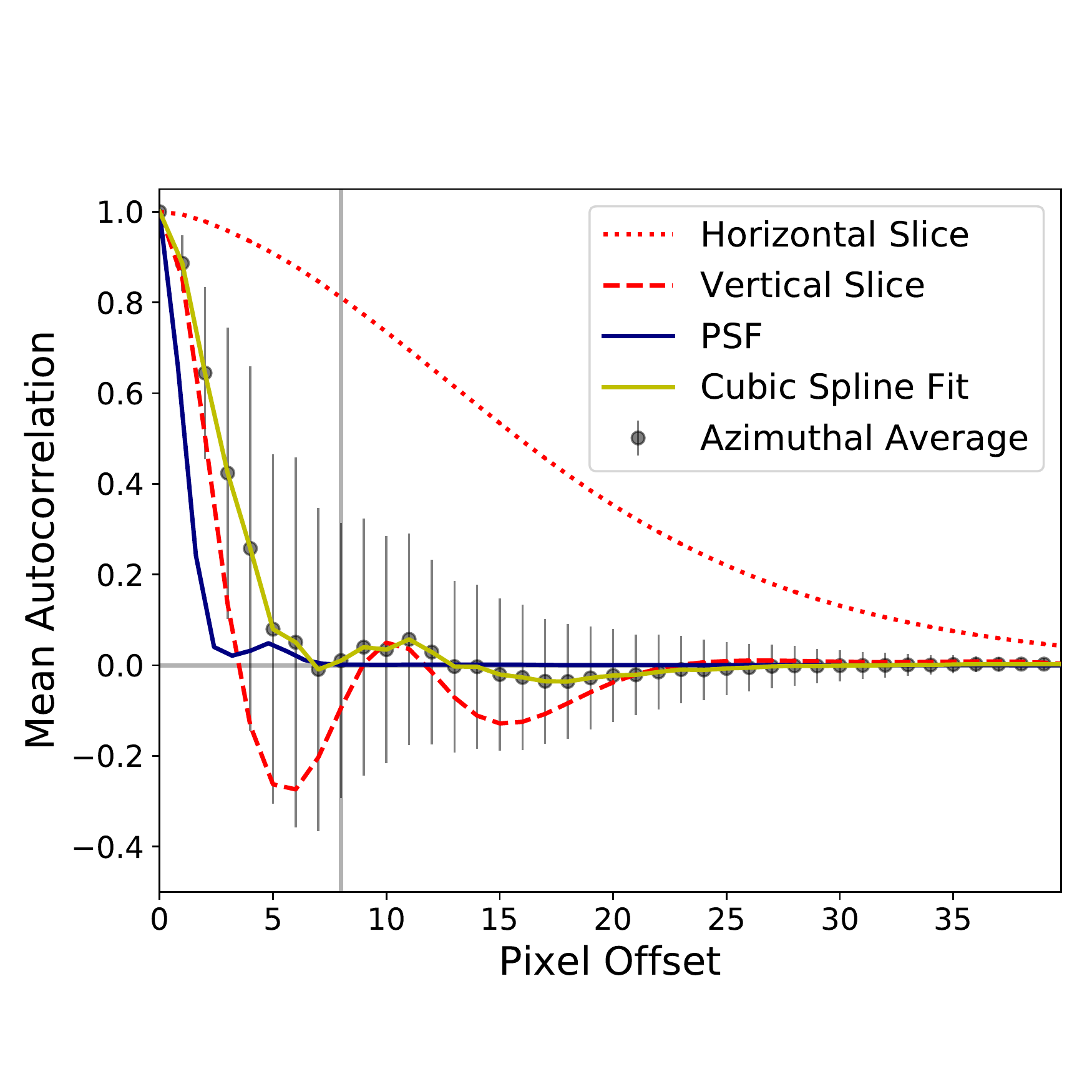}
\includegraphics[width=2.3in]{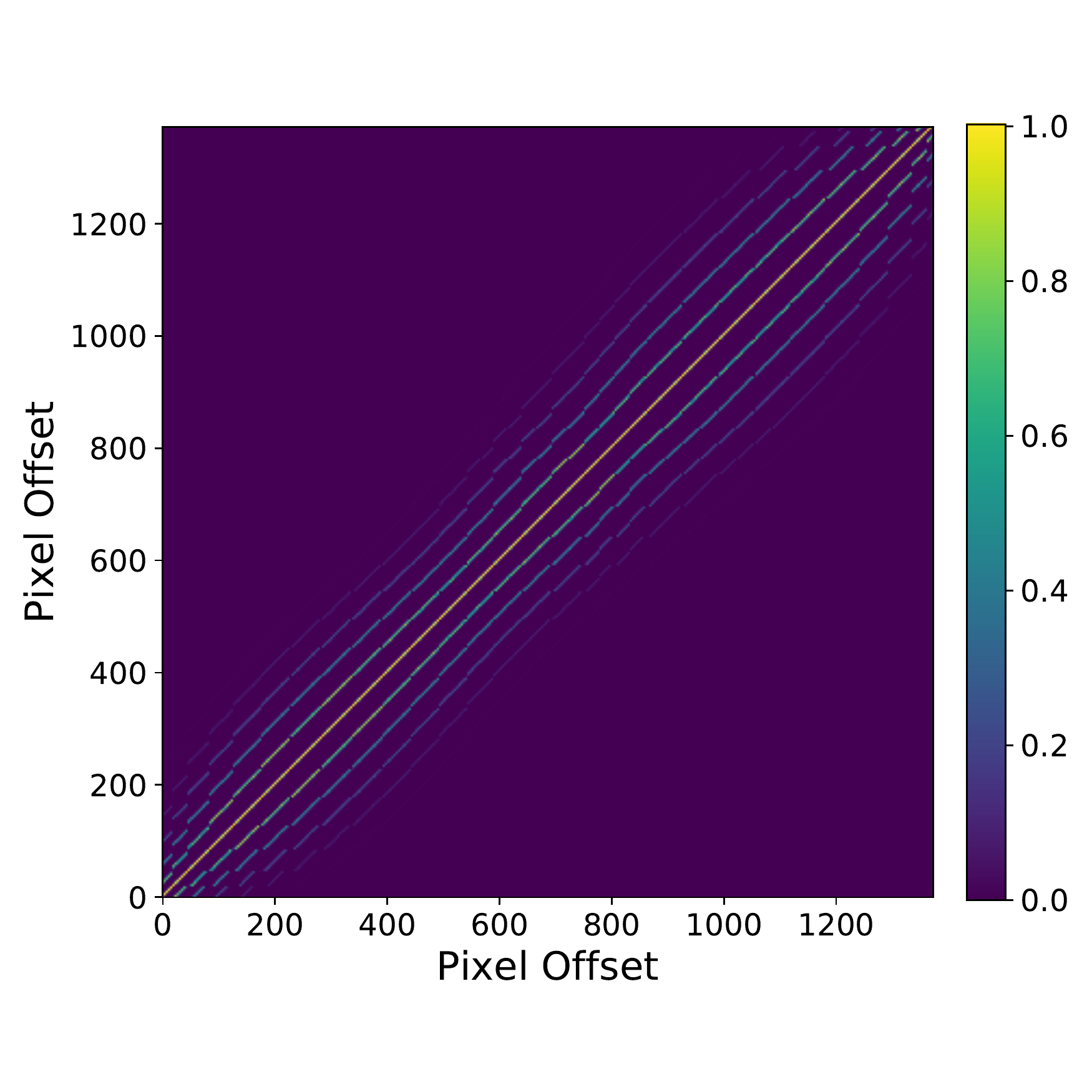}
\caption{Here we illustrate how the covariance matrix is generated from the initial MCMC fit to the F814W image using a $\chi^{2}$ - based log likelihood estimation. \textbf{Left:} Average autocorrelation map of the F814W observations - MCFOST model residuals for 1000 models drawn from the MCMC sampler. \textbf{Middle:} Vertical and horizontal slices through the autocorrelation map are shown in red, the PSF is given in blue, and the computed azimuthally averaged profile is given in grey. A cubic spline fit to the azimuthal average is shown in yellow. This was then used to describe the covariances between neighboring pixels when the covariance matrix is generated. \textbf{Right:} The resultant covariance matrix for the F814W dataset. \label{fig:covarhst}}
\end{figure*}

\begin{figure*}
\includegraphics[width=2.3in]{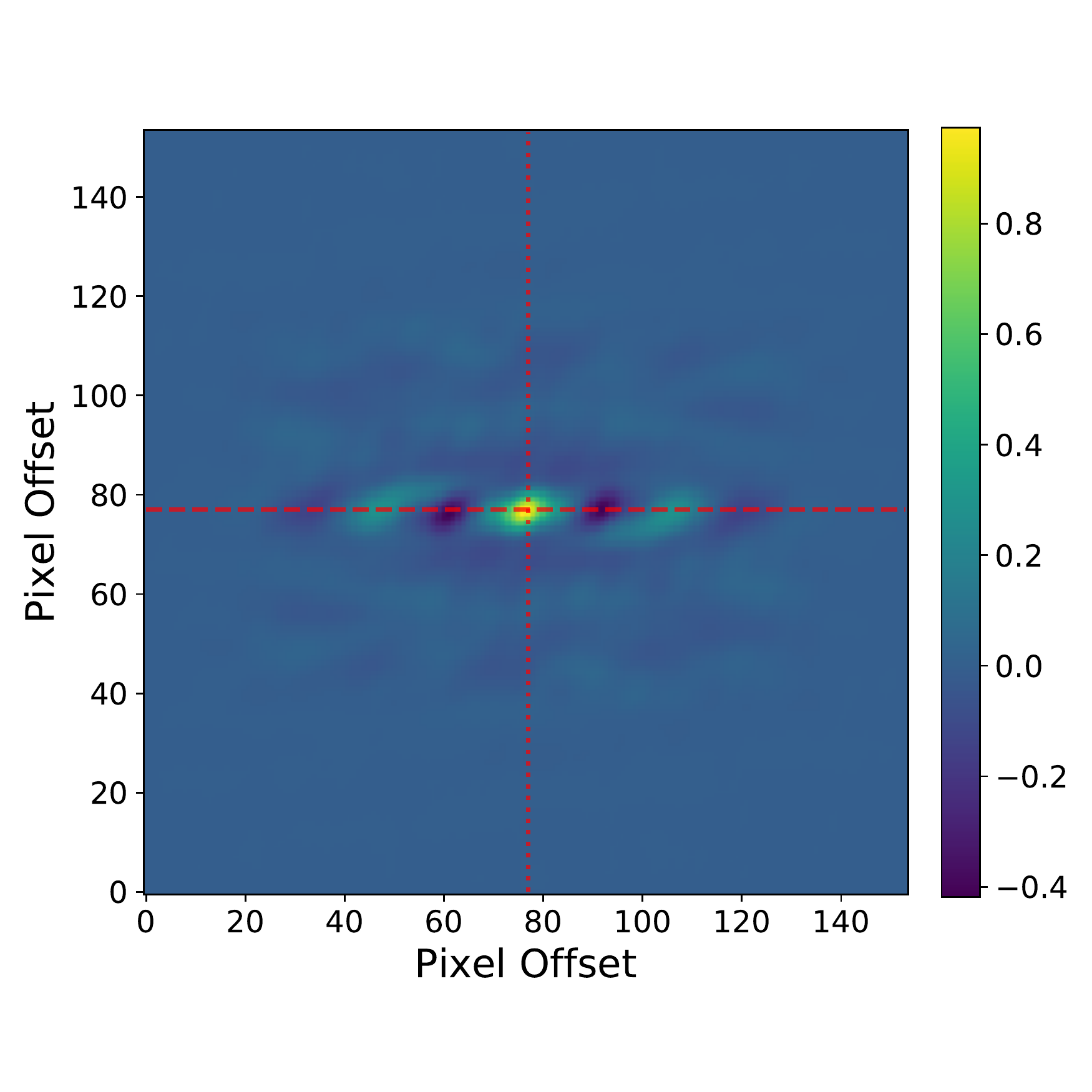}
\includegraphics[width=2.3in]{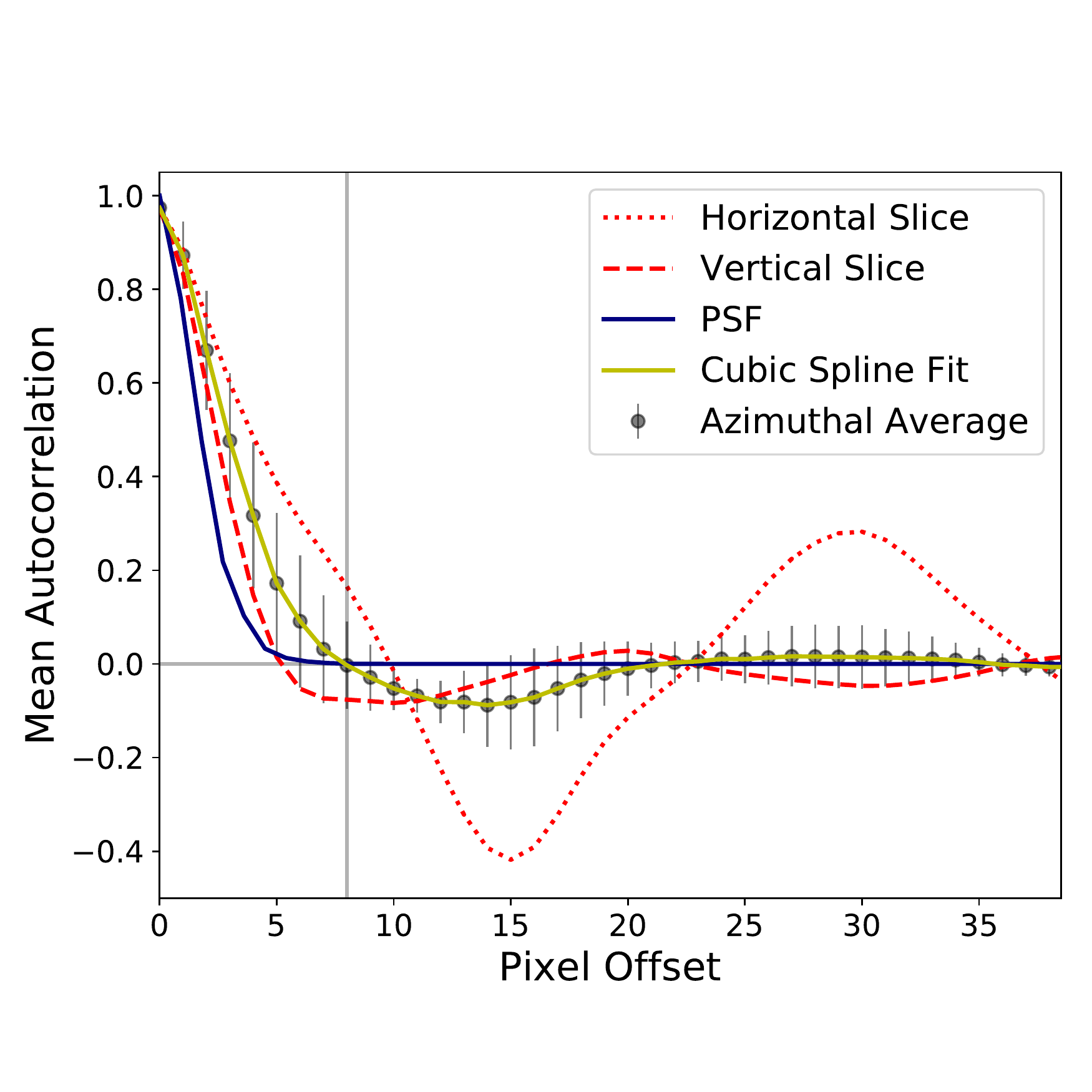}
\includegraphics[width=2.3in]{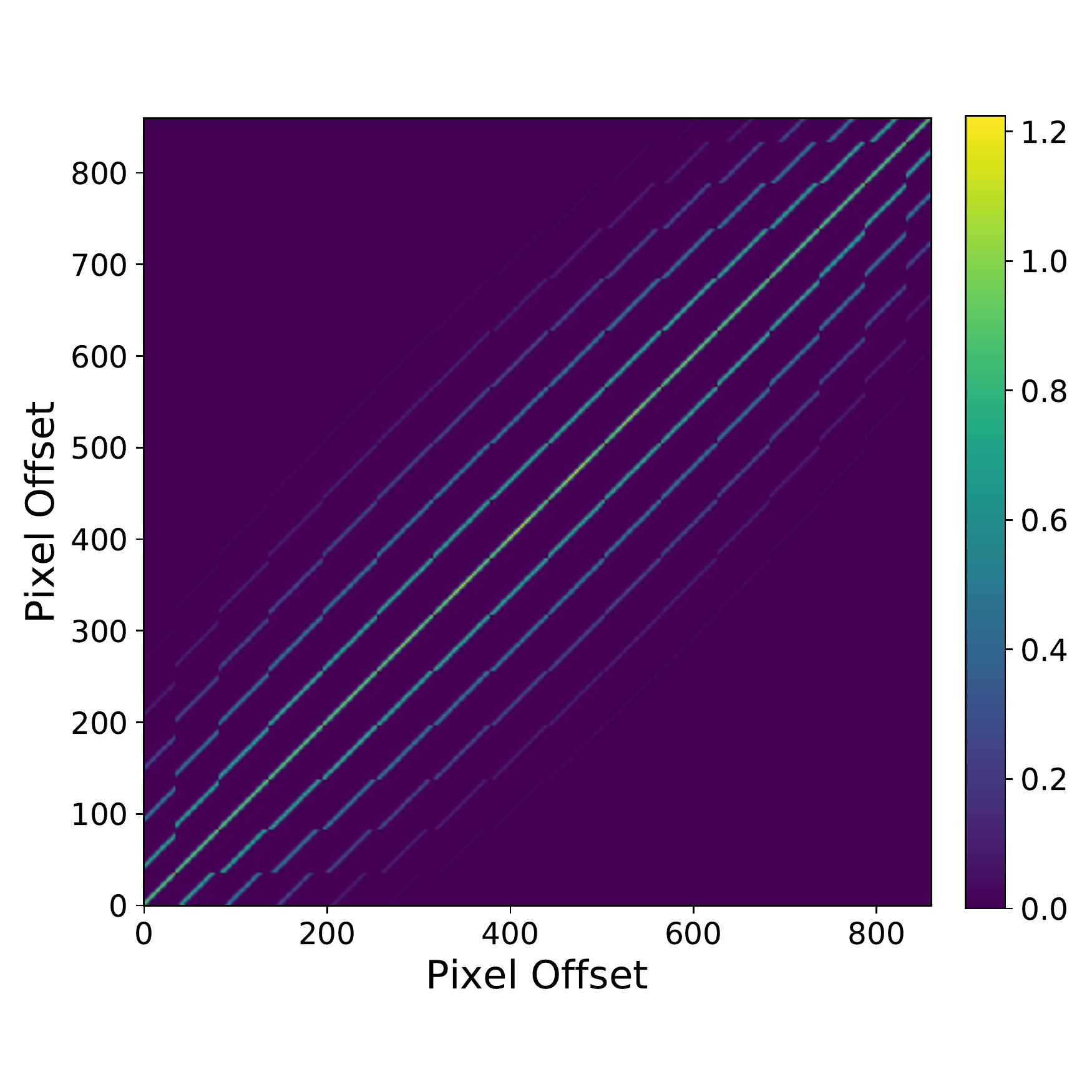}
\caption{Here we demonstrate how the covariance matrix is generated for the ALMA dataset in the same way as Figure \ref{fig:covarhst}. Radial substructure in the autocorrelation map hints at a more complex radial structure in the ALMA data than was allowed for in the disk dust models. \label{fig:almaautocorr}}
\end{figure*}

In order to quantify the limitations of our model at each wavelength, we return to the results of the first two individual MCMC runs. These provide a wavelength-dependent estimation of the global limitation of the chosen disk model to fit the observations. We quantify this via an autocorrelation function derived from the residuals. 
We generate an autocorrelation map by selecting 1000 models drawn from the MCMC chain, computing the observation - model residuals, and autocorrelating those residuals. To convert this map into a 1D autocorrelation as a function of offset, we take the azimuthal average of slices through the map center. This process is visualized in Figures \ref{fig:covarhst} and \ref{fig:almaautocorr}. This differs slightly from the technique used in \citet{2017ApJ...851...56W} in that the horizontal slice through the autocorrelation map was used to describe the 1D global covariance rather than an average. In that case, the ESO H$\alpha$ 569 disk was farther from edge-on and the peak in surface brightness from the central star was more pronounced in scattered light than for Oph163131. 

Structure in the autocorrelation map can point towards systematic deficiencies of the model to accurately describe the data, though this is difficult to interpret. Oph163131 is quite a flat disk, with a relatively narrow darklane. This produces the horizontal banding seen in this map as the dark lane and top/bottom nebulae constructively and destructively interfere with each other.
Additionally, the ALMA data show complex structure along the horizontal axis, possibly pointing towards a more complex radial structure than allowed for in our disk model.

Finally, for each pair of pixels $i,j$ the contribution from the global covariance $K_{i,j}^{G}$ is computed by interpolating the analytic autocorrelation function at the distance between the two pixels $r_{i,j} = \sqrt{(x_{i}-x_{j})^{2}+(y_{i}-y_{j})^{2}}$, with a cutoff outside of 8 pixels to make computations of $C_{i,j}$ tractable. The resulting covariance matrices are shown in Figures \ref{fig:covarhst} and \ref{fig:almaautocorr}. Note that, in the case of the ALMA image, the autocorrelation map and all ALMA data products were first downsampled to match the spatial resolution of the HST images. 

For the combined run, we use a parallel-tempered sampler with 2 temperatures, 50 walkers, and 2000 iterations. Here we required a longer burn in stage of 250 iterations. The priors were the same as in the individual fits (Table \ref{table:params}) with one exception. Both individual MCMC runs produced values for the grain size exponent (p) consistent with the canonical value of 3.5 for the ISM \citep{1977ApJ...217..425M}. For the combined MCMC run we keep this value fixed at 3.5 to further limit parameter degeneracies. 

In summary, we conduct three independent MCMC runs to optimize for different observables: the F814W scattered light image alone, the ALMA millimeter continuum alone, and a combined fit. 
For the individual dataset runs, we use a more familiar $\chi^{2}$ based log likelihood estimation and for the joint fit we employ the covariance framework. 
In all cases, we use a parallel-tempered sampler with 2 temperatures and 50 walkers. The number of iterations varied between runs with 2000 required for the F814W scattered light run, only 1000 needed for the ALMA continuum run and 2000 iterations for the joint fit. A burn in stage of 100 iterations was used for the ALMA and F814W individual fits, while a longer burn in of 250 iterations was needed for the joint run.
The results are presented in Section \ref{Sec:results}.

\section{Modeling Results}
\label{Sec:results}

\begin{figure*}
\centering
\includegraphics[width=6.5in]{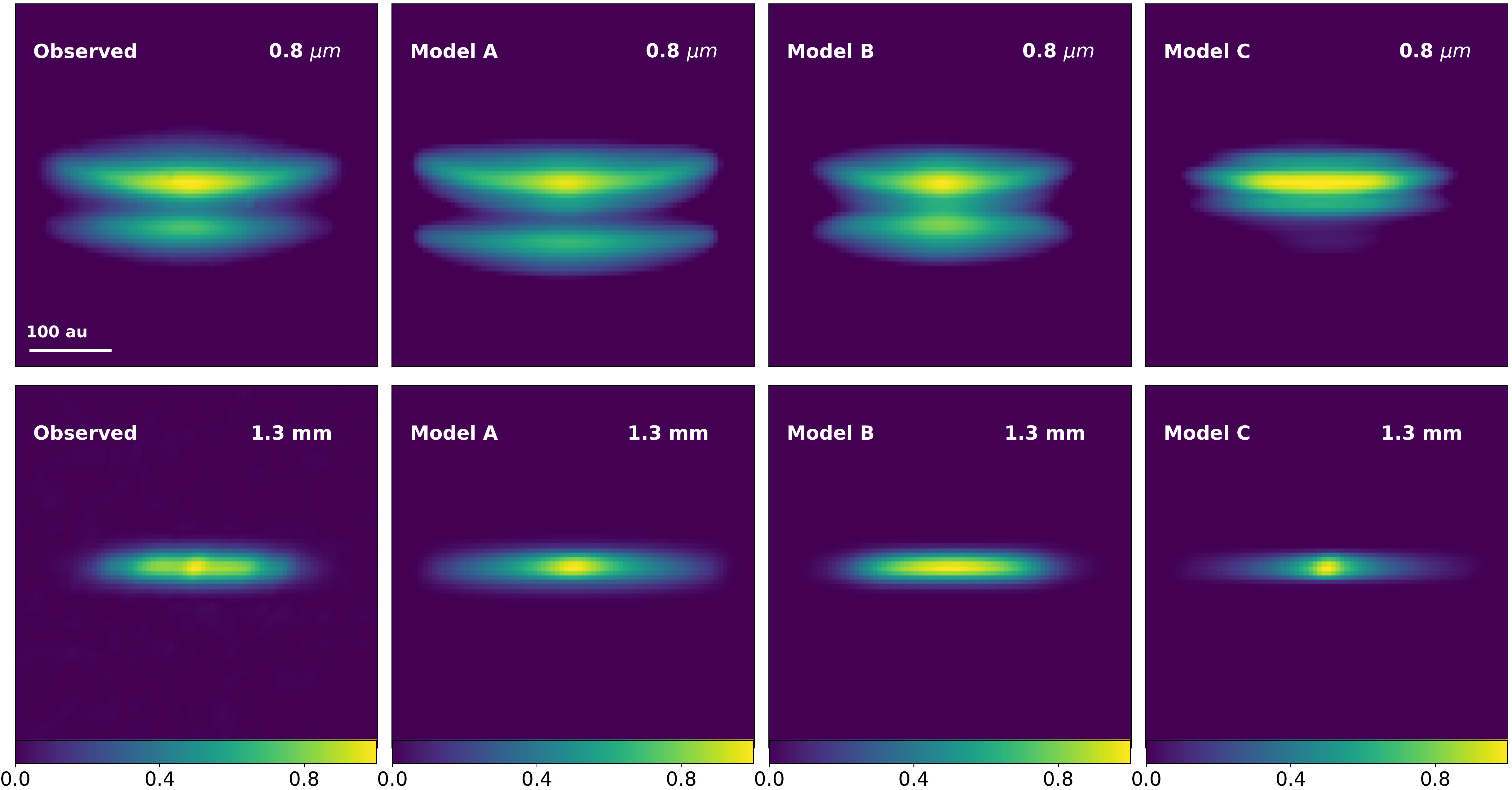}
\caption{Model images for Models A, B, and C as described in Table \ref{table:paramresults}. The top row shows images at 0.8 $\mu m$ (simulation of the F814W HST scattered light image) and the 1.3 mm models are shown on the bottom (simulation of the ALMA continuum image). The observations have been normalized to the peak intensity.  \label{fig:abcmods}} 
\end{figure*}

Here we describe the results from the three independent MCMC runs. The full posterior distributions are given in Appendix \ref{Appendix:triangle} and the parameter values are given in Table \ref{table:paramresults} with uncertainties given by the 16th and 84th percentiles. To facilitate comparisons between various observables, we draw a representative model from each MCMC run. 
Models A, B and, C represent characteristic `well-fit' models drawn from the posteriors generated from the F814W scattered light image, the ALMA continuum image, and the combination, respectively (Figure \ref{fig:abcmods}). These parameter values are also given in Table \ref{table:paramresults}.
Here we describe the results of each run individually and compare these models to the Keck 2.2 $\mu$m image and the literature compiled SED. 
Section \ref{Sec:discussion} gives a full discussion of the constraints this places on the individual parameters.

\begin{table*}[htb]
  \centering
  \caption{MCMC Disk Parameter Results} \label{table:paramresults}
  \begin{tabular}{l|c c|c c|c c}
    \toprule
    Parameter & \multicolumn{2}{c}{Scattered Light Only}  & \multicolumn{2}{c}{mm Continuum Only}   & \multicolumn{2}{c}{Combined MCMC} \\ 
     & Posterior & Model A & Posterior & Model B & Posterior & Model C   \\
     \midrule
    $i$ ($^{\circ}$) & $83.6^{+3.3}_{-5.3}$ & $84.5$ & $87.31^{+0.31}_{-0.30}$ & $87.2$ & $88.6^{+0.7}_{-2.8}$ & $86.7$ \\ 
$H_0$ (au)  & $9.7^{+3.5}_{-3.7}$ & 7.2 & $9.2^{+2.1}_{-1.2}$ & $9.5$ & $5.1^{+4.9}_{-1.1}$ & $4.0$ \\ 
$\log{M}$ ($M_{\odot}$)   &  $> -4.5$ & $-3.3$ & $-3.8^{+0.1}_{-0.1}$  & $-3.6$ & $-5.17^{+1.55}_{-0.32}$ & $-3.6$ \\
$\gamma$ & $1.01^{+0.68}_{-0.68}$ &$1.01$ & $< 0.64$  & $0.1$ & $0.78^{+0.74}_{-0.64}$ & $0.8$ \\
$\beta$ & $1.48^{+0.35}_{-0.31}$ & $1.50$ &  $1.66^{+0.21}_{-0.22}$  & $1.54$ & $1.29^{+0.29}_{-0.12}$ & $1.15$ \\
$p$ &  $3.43^{+0.33}_{-0.37}$ & $3.5$ &  $3.33^{+0.24}_{-0.26}$ & $3.3$ & $-$\footnote{This quantity was held at a fixed value of 3.5 during the combined MCMC run.}  & $3.5$ \\ 
 $\log{\alpha}$ & $> -2.5$ & $-2.3$  & $> -2.1$ & $-1.1$ & $-2.5^{+1.1}_{-0.8}$ &  $-2.9$ \\ 
$R_{C}$ (au) & $111.8^{+56.5}_{-56.2}$ & $89$ & $74.5^{+4.}_{-7.1}$  & $74.5$ & $120^{+36}_{-28}$ & $98.7$ \\
\midrule
$\chi^{2}_{0.8 \mu m}$ & - & $11$ & -  & $10$ & - & $37$\footnote{In the joint fit, the model likelihood is not the sum of the $\chi^{2}$ values (see Section \ref{Sec:model}), and the individual $\chi^{2}$ values for Model C are reported for comparison to Models A and B only. } \\%
$\chi^{2}_{1 mm}$ & - & $5.9$ & -  & $1.2$ & - & $16^{\mathrm{b}}$ \\%
    \bottomrule
  \end{tabular}
   \vspace{1ex}

     {\raggedright \textbf{Note:} For each MCMC run we provide the best fit results with uncertainties and a representative model drawn from the chain. The posterior results present the 50th percentiles of the samples in the posterior distributions with uncertainties given by the 16th and 84th percentiles. In cases where the posterior distributions peak at the edge of the allowed parameter range, upper/lower limits are used quoting the 50\% quantiles. For models A, B, and C, the reduced $\chi^{2}$ values for each observable are provided. \par}
     
\end{table*}

\subsection{Scattered Light Image}
\label{Sec:scatres}

\begin{figure*}
\begin{center}
\includegraphics[width=5.5in]{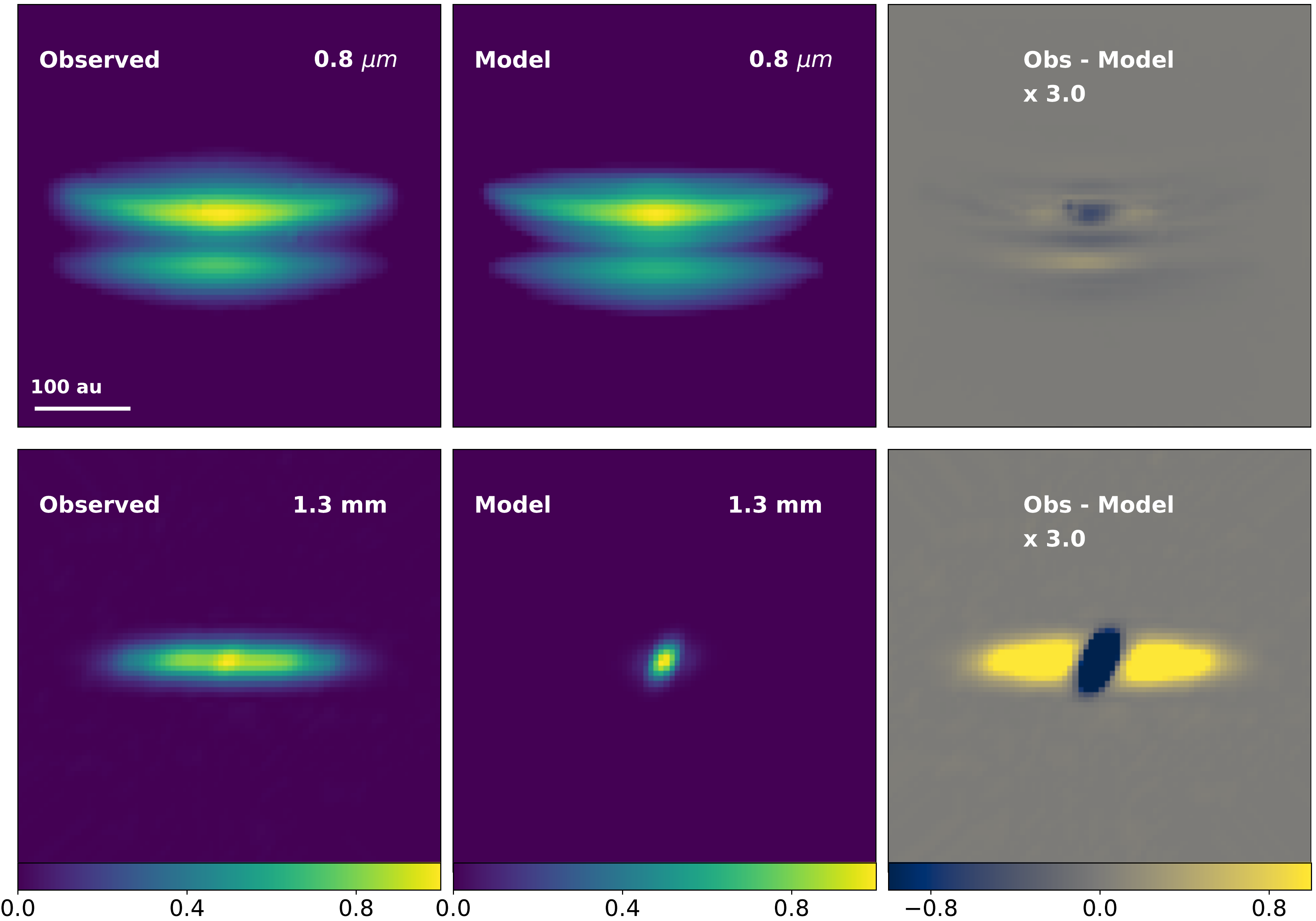}
\caption{Here we illustrate the results of the MCMC run optimized for the HST scattered light alone. The left panels show the observations, the middle panel represents the model images and corresponding residuals are shown on the right. 
To present the full complement of well-fitting models (see \ref{table:paramresults}) we show the mean of 1000 models drawn from the top 10\% of the MCMC chain rather than the single best fit model. The model images have been convolved with a PSF, recentered, and scaled to match the observations.
The 0.8 $\mu m$ imags are shown on the top and the 1.3 mm images are shown on the bottom. The observations have been normaized to the peak intensity. 
Residuals have been scaled to highlight features.
Note also that the $0.8 \, \mu m$ image is displayed on a linear scale while previously shown on a logarithmic scale in Figure \ref{fig:scatteredobs}. 
The simulated scattered light images provide a good fit to the observables, but the corresponding millimeter images are clearly inconsistent with the data.
\label{fig:f814models}} 
\end{center}
\end{figure*}

The results for the MCMC run of the F814W scattered light image are presented in Figure \ref{fig:f814models} where we show a median of the 1000 models with the lowest $\chi^{2}$ values. The posterior distributions and parameter correlations are shown in Figure \ref{fig:f814wcorner}. In this case, the posterior distributions are wide and it is difficult to place tight constraints on any of the free parameters. 
The inclination is the best constrained parameter with a best fit value of $83^{\circ}.6^{+3.3}_{-5.3}$. The scale height and grain size exponent are both marginally constrained with values of $9.7^{+3.5}_{-3.7}$\,au and $3.43^{+0.33}_{-0.37}$ respectively. Posterior distributions for the surface density exponent, flaring exponent and critical radius are nearly flat, while both the dust mass and the viscous settling parameter are pushed up against the upper limits imposed by the log uniform prior distributions. 

At first glance, the flat posterior for the $R_{C}$ parameter is puzzling, as the outer edge of the disk is evident in the HST images. Several factors contribute to this effect. First, recall that the critical radius and the outer radius describe different features of the disk. The $R_{C}$ value represents the radial scale of the exponential decay in the outer disk (See Eq. \ref{Eq:taper}), and is also related to the turning point in the radial slope of the surface density distribution. 
Previous efforts to fit the geometry of edge-on disks using radiative transfer modeling (and a sharp edged disk model) show that the flatter the disk, the less tight of a constraint can be placed on the disk geometry and, specifically, the surface density exponent. For an example of this, compare the derived geometries for a single wavelength fit to HH 30 \citep{2004ApJ...602..860W} and the more vertically compact HK Tau B \citep{2011ApJ...727...90M}. 
Finally, while it may seem straight forward to chose an outer radius for the disk, it is difficult to disentangle the exact edge of the flared disk from the surrounding material in the Ophiuchus SFR and the 'halo' observed in the Keck $2.2 \, \mu m$ and HST $0.8 \, \mu m$ images. Here we employ a mask including only those pixels containing $> 3\sigma$ disk flux in our goodness of fit metric. A less conservative mask would probably allow a tighter constraint on the  $R_{C}$ parameter but would also have unforeseen consequences on other parameter results.

We compute an effective sample size (ESS) for each parameter to test the convergence of the chain. This provides a measure of the effective number of independent samples in a correlated chain, using the relation $ESS = N_{samples}/(2 \tau_{x})$ where $\tau_{x}$ is the integrated autocorrelation time.
We compute the autocorrelation time via the method described in \citet{goodmanandweare} and find values ranging from 0.92 - 1.03 with corresponding ESS values of $\approx$ 50000 for all parameters.

The resulting scattered light model does a decent job of fitting the observations. The largest discrepancy can be seen in the width of the dark lane (Figure \ref{fig:f814models}, second panel) where the larger disk vertical extent preferred by the models misplaces the flux in the bottom nebula.
This could indicate that the true disk scale height is smaller than our model predicts. However, it is very difficult to decouple this from the effects of the disks' inclination and dust opacities. 

On average, the models preferred by the scattered light observations with the lower dust masses and farther from edge-on inclinations do not allow sufficient disk material along the line of sight at millimeter wavelengths to reproduce the ALMA observations at 1.3 mm. Note, however, that some well-fit models within the wide posterior distributions do provide a good fit to the ALMA observations. Model A (Figure \ref{fig:abcmods}) was chosen specifically to demonstrate that there is a subset of models within the scattered light results that are able to provide a good fit to both observables. 
To recreate the more extended ALMA structure and the observed plateau, we require the disk model to be more optically thick at mm wavelengths. This could be accomplished with smaller disk scale height, a more edge-on inclination, or a more settled disk. However, these parameters also have a marked effect on the amount of light from the central star coming through the upper disk surface which would negatively impact the scattered light model fit given that this is the highest SNR region. 

The flux scaling used in the $\chi^{2}$ calculation (the total intensity of each model is normalized to the 0.8 $\mu$m observed image) can also be used to diagnose these discrepancies. We find values around unity for the 0.8 $\mu$m image. However, if we compute the 1.3 mm images for these models and compare them to the ALMA observations, we find we are under-predicting the flux at millimeter wavelengths by several orders of magnitude.

\subsection{ALMA Continuum Map}
\label{Sec:almares}

\begin{figure*}
\centering
\includegraphics[width=5.5in]{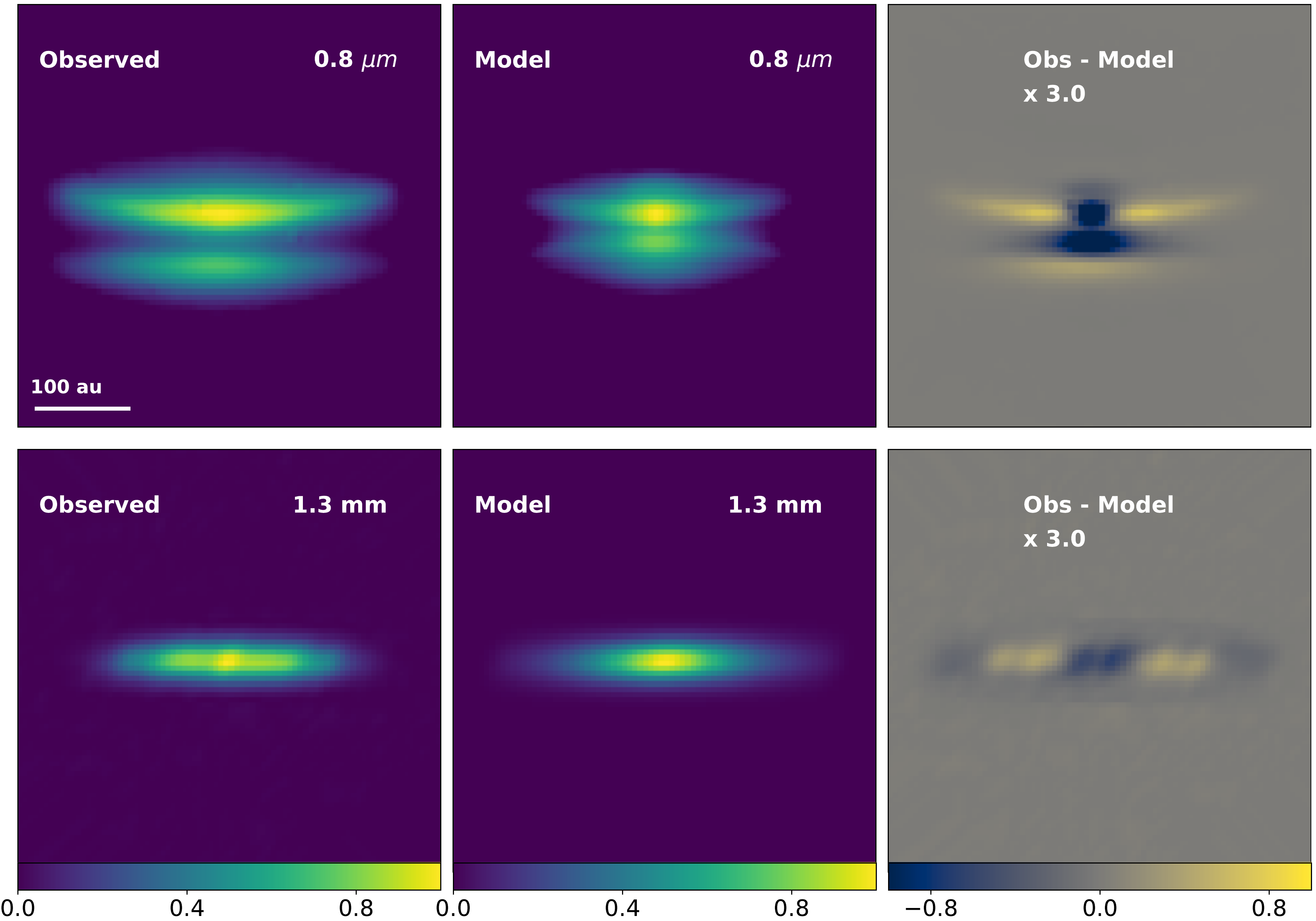}
\caption{The same as Figure \ref{fig:f814models} but with the fit optimized for the ALMA millimeter continuum image.
In this case, the 1.3 mm model is a very good fit to the observations with no remaining structure in the residual map. The scattered light image, however, is too compact with a more concentrated peak in intensity.
\label{fig:almamodels}} 
\end{figure*}

The results for the MCMC run for the ALMA millimeter continuum image are shown in Figure \ref{fig:almamodels} with the posteriors in Figure \ref{fig:almacorner}. 
In general, the fit to the ALMA data shows tighter posteriors than the scattered light fit. While still optically thick at 1.3 mm, the lower opacity of the larger grains leads to fewer parameter degeneracies. 
All parameters have clearly peaked posterior distributions except for the surface density exponent and the viscous settling parameter, which both prefer values at the edge of the parameter range allowed by the prior distributions. We provide 3$\sigma$ limits for these parameters of $< 0.64$ and $> -2.1$, respectively. 
All best-fit parameter values are consistent across the scattered light and continuum image fits within the (admittedly large) 1-2 $\sigma$ uncertainties. We take this as evidence that a joint fit to both observables should be possible within this disk model formalism. 
We confirm convergence using the same method described in \ref{Sec:scatres}. 
The integrated autocorrelation times ($\tau_{x}$) varied between 0.8 and 1.2 for all parameters with an associated ESS ranging from 20000-30000.

The best fit models for the ALMA MCMC run are presented in Figure \ref{fig:almamodels}. The ALMA data are well reproduced by the models.
The observation - model residuals for the 1.3 mm image (right panel of Figure \ref{fig:almamodels}) show that the models generally under-predict the flux at the location of the central point source (seen at an amplitude of $\sim$ 5-10\%), over-predict the flux in the plateau region (by $\lesssim$ 5\%), and are more radially extended than the observations. 
The residuals are all under 10\%, which is on par with the uncertainties in the observations. 

This substructure could point towards a global limitation in our model. A similar pattern is seen in the autocorrelation map first introduced in Section \ref{Sec:covar} and shown in Figure \ref{fig:almaautocorr}. 
A bright peak at the location of the central star is followed by symmetric dips at the location of the flux plateau and lastly by correlated limbs, moving radially outward.
If we had allowed for a more complex radial structure with higher density rings of material rather than a smooth distribution, we would likely obtain a better fit to the ALMA observations. This is discussed further in Section \ref{Sec:discussion} 

The 0.8 $\mu m$ average models and residuals corresponding to the ALMA MCMC results are also shown in the top panel of Figure \ref{fig:almamodels}.
Generally, the models preferred by the ALMA data tend to leave a free optical path to the central star, which the HST observations clearly rule out.
These models over-predict the flux in the top nebula and under-predict the flux in the bottom nebula which implies that the slightly lower scale height and, more importantly, the closer to edge-on inclination preferred by the ALMA image are too extreme to overcome the smaller residuals seen in Figure \ref{fig:f814models}. 
There is also a residual arced wedge on either side of the disk on the top nebula that indicates a poor fit to the disk flaring exponent and/or the surface density exponent.
It is worth noting that the compact disk preferred by ALMA produces roughly the same $\chi^{2}_{0.8 mu m}$ value as the HST optimized fits (Table \ref{table:paramresults}). In the ALMA case, the disagreement is concentrated to the central region where the photon noise is highest, whereas the faint over-subtraction seen with the more extended disk preferred by the scattered light observation encompasses more pixels in a lower noise region. 

The mean flux scaling factor for the 1.3 mm models was 1.1 while the corresponding 0.8 $\mu$m models had a mean flux scaling value of 0.4. The lower scaling factor for the 0.8 $\mu$m models is presumably due to the more compact and centrally peaked structure.
Ultimately, given the large degeneracies in the parameters when compared to the optically thick scattered light image as visualized in the MCMC densities shown in Figure \ref{fig:f814wcorner}, it is difficult to link any feature seen in the residual map to a single modeled parameter.

\subsection{Scattered Light and mm Continuum Combined}

Before we discuss the combined MCMC results below, we first examine if the results for the individual fits contradict each other to an extent that would prohibit a single joint fit to explain both datasets. Specifically, we are concerned about the critical radius ($R_{C}$).
We would expect strong radial drift to result in a markedly smaller critical radius ($R_{C}$) for the millimeter size dust grains probed by the ALMA observations than the micron sized grains seen in scattered light. No such discrepancy is observed, though the posterior distribution for the critical radius seen in the scattered light MCMC results is very flat (see discussion in Section \ref{Sec:scatres}). 
In the ALMA results, the lower preferred value of $\gamma$ results in a density profile which drops very fast outside of $R_{C}$, resulting in a more sharply peaked posterior distribution.
While we find that a single critical radius can accurately describe both datasets, we cannot rule out mechanisms such as radial drift, or more complex radial density structures.

Finally, in both the 0.8 $\mu$m and 1.3 mm individual MCMC runs, the dust particle size exponent was consistent with the canonical value of 3.5. Allowing this parameter to vary appears to broaden all parameter distributions while adding an extra degree of freedom to our covariance framework. In order to improve the computation time in the combined MCMC we choose a value of p = 3.5 and fix this parameter.

\begin{figure*}
\centering
\includegraphics[width=5.5in]{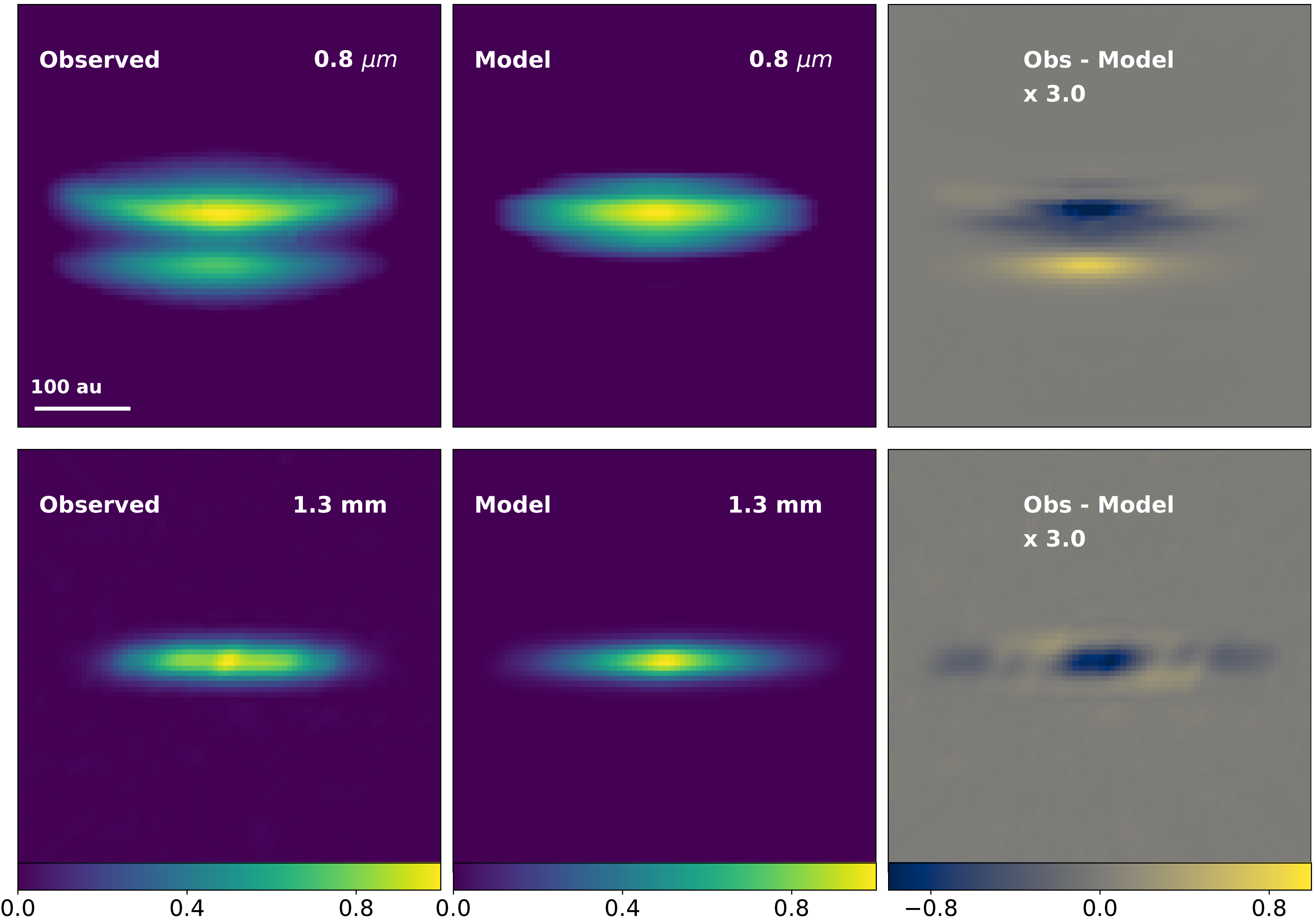}
\caption{The same as Figure \ref{fig:f814models} but with the fit optimized for a combination of the HST scattered light image and the ALMA millimeter continuum image using the MCMC covariance framework.
These models represent a compromise between the two datasets. At 0.8 $\mu$m the horizonatal slice to the top nebula is a good match, but the curvature is not reproduced and the small scale heights neglect the bottom nebula entirely. At 1.3 mm the general shape of the disk is replicated, though the central peak and plateau isn't exact.  \label{fig:combomodels}} 
\end{figure*}

The results for the MCMC run using the covariance framework with the combined HST scattered light and ALMA millimeter continuum datasets are shown in Figure \ref{fig:almacorner}. The best fit parameter values are shown in Table \ref{table:paramresults} and Figure \ref{fig:combomodels} shows the model and residuals for both datasets using the best fit values from the combined MCMC run. 
Clearly, this framework provides a markedly worse fit to both the scattered light and millimeter continuum observations than either of the individual fitting efforts described above. 
These results give us insight into the unresolved substructures of the disk and provide direction for future modeling efforts with higher resolution datasets. It is apparent that the simple disk model we have employed here lacks sufficient complexity to explain both datasets simultaneously. Nevertheless, we present the combined MCMC results here and discuss the implications further in Section \ref{Sec:discussion}.

The posterior distributions shown in Figure\ref{fig:combocorner} are complex and demonstrate a strong bimodality most notable in the dust mass, but also showing degeneracies with the inclination, surface density, and settling parameters. 
Using the convergence test described in Section \ref{Sec:scatres}, only the inclination and surface density exponent remain unconstrained with ESS values of 8 and 20, respectively. 
The surface density exponent is peaked at the lower edge of the allowed parameter space, but also exhibits a flat distribution above $\gamma = 0.5$. The posterior distribution for the inclination is narrow with values confined to  $85^{\circ} - 90^{\circ}$ in agreement with the other MCMC results. 
A more detailed discussion of the surface density results is presented in Section \ref{Sec:gamma}.

The scale height, flaring exponent, and critical radius all display sharply peaked posterior distributions with best fit values of $5.1^{+4.9}_{-1.1}$\,au, $1.3^{+0.3}_{-0.1}$, and $120^{+36}_{-28}$\,au. In the case of the gas scale height, this is inexplicably smaller than the value preferred by either of the previous fits, though consistent within 1$\sigma$ uncertainties. 

The dust mass posterior distribution is strongly bimodal with peaks at both ends of the allowed parameter range. 
The lower dust mass is preferred, but this does not agree with the best fit dust mass values from either of the individual dataset MCMC results, and one might expect the ALMA-preferred mass to be more reliable. 
This dichotomy is discussed in more detail in Section \ref{Sec:mass} and in Appendix \ref{Appendix:triangle}. 
To diagnose the source of this discord, we examine the flux scaling values used when computing the residuals. We find that models with a mass of $\log M (M_{\odot}) \sim -5$ require a scale factor of $\sim$1 for the 0.8 $\mu$m image and a factor of $\sim$9 for the 1.3 mm image. For $\log M (M_{\odot}) \sim -3.6$, this trend is reversed with a scale factor for the ALMA data near unity and the HST data being under-luminous by a factor of $\sim$9. 
Superficially, this bimodal structure appears to be a trade-off between the dust masses preferred by the two datasets with the ALMA data preferring a higher mass disk and the scattered light preferring a lower mass disk. However, recall that both datasets are probing optically thick disks, and thus a direct relationship between integrated flux and dust mass cannot be drawn. Indeed, in scattered light, the integrated brightness is most strongly affected by the flaring exponent. The set of lower dust mass models also tend to have a higher flaring exponent and a correspondingly brighter scattered light disk.

Lastly, the viscous settling parameter posterior distribution is fairly flat for $\log{\alpha} > -3.5$, but the lowest values of $\alpha$ are ruled out and some degree of settling is certainly required to explain the variation in the apparent dust scale height between the scattered light and mm continuum datasets. 
Upon closer inspection, there is a strong correlation with dust mass, where the more plausible higher mass models require less dust settling ($\log \alpha \simeq -3.1$) than the lower mass models (see Appendix \ref{Appendix:triangle}). This disagrees with the values for $\alpha$ preferred by the individual MCMC fits, though we don't expect a resolved image at a single wavelength to place a tight constraint on the dust settling parameter.
The implications for dust settling are discussed further in Section \ref{Sec:settling}.

As the modeling framework increases in complexity, it becomes more difficult to disentangle the correlations between parameters and how these relate to features in the datasets. 
Returning to Figure \ref{fig:combomodels}, we see that the scattered light image model misses the bottom nebula entirely while the ALMA model is too centrally concentrated.
Another way to express the bimodality 
in relation to the model subtracted residuals is as a compromise between fitting for the central peak in the ALMA dataset, requiring a low mass and high inclination, and the less turbulent, higher mass disk with a steeper density profile required to reconcile the ALMA and HST observations.
Ultimately, the results of an MCMC are limited by how well the model describes the data. In this case, our assumption of a tapered power law in surface density clearly lacks the degree of complexity needed to approach the true disk structure. 
The smooth formalism for dust settling across observations probing sub-micron to millimeter dust grain sizes may also be a culprit. 
We discuss this further in Section \ref{Sec:efficacy}.

\subsection{Comparison to the SED}
\label{sed:sedres}

\begin{figure}
\includegraphics[width=3.3in]{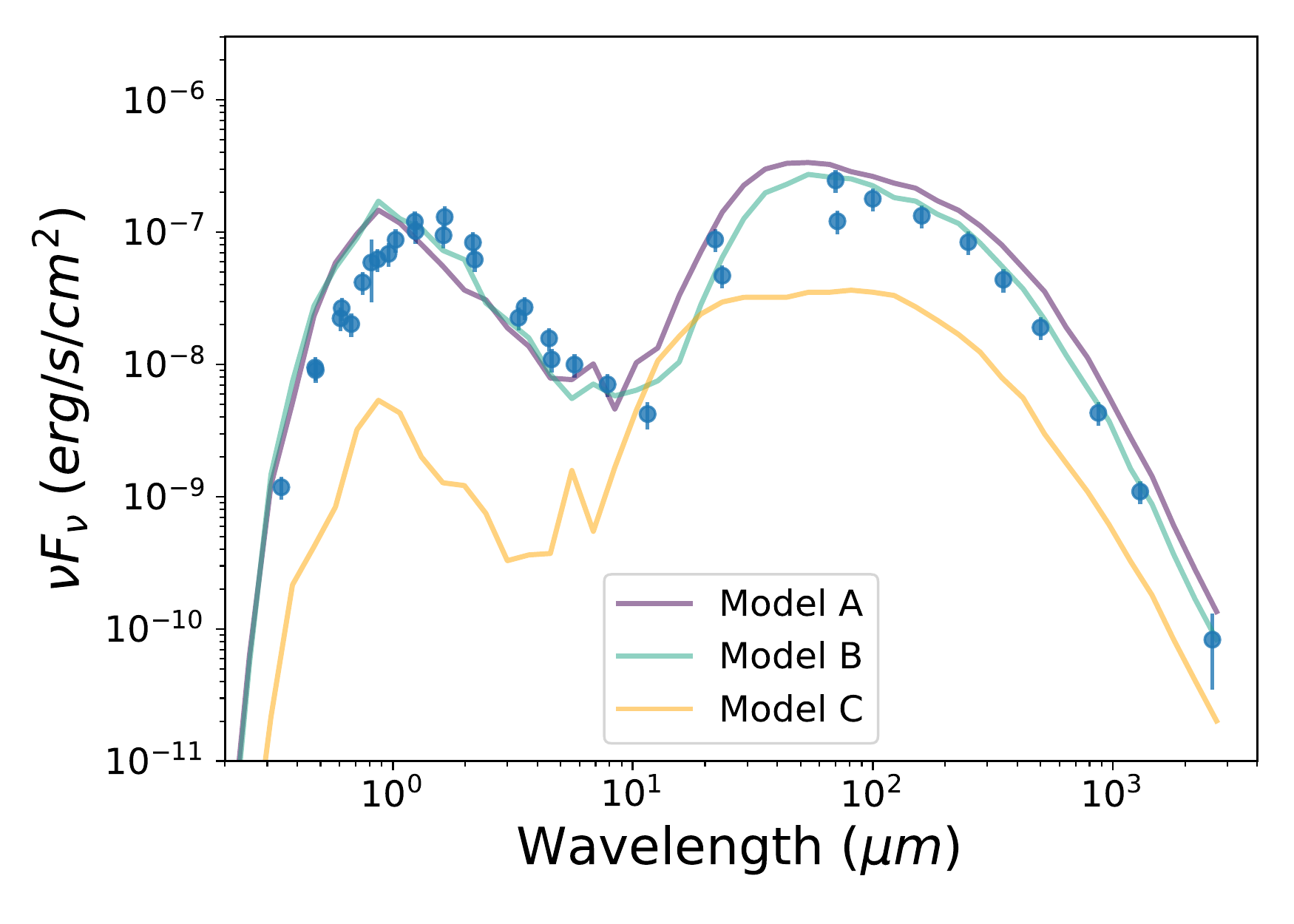}
\caption{The modeling results are compared to the literature compiled spectral energy distribution for Oph163131. This also includes HST, ALMA, and CARMA data presented in this work. See Table \ref{table:sed} for a complete description of the data and uncertainties. The  models from the three separate MCMC runs are shown assuming 2 magnitudes of extinction. 
\label{fig:sedmodels}}
\end{figure}

We compare the SEDs produced by the representative models for all three MCMC runs in Figure \ref{fig:sedmodels}. The model preferred by the ALMA data (Model B) provides the best fit to the SED with a $\chi^{2}$ value of 3.1. The best fit model produced by the 0.8 $\mu m$ MCMC run (Model A) has a comparable fit ($\chi^{2} = 3.9$) and over-predicts the flux slightly at longer wavelengths. 
The combined MCMC run SED (Model C) has a  $\chi^{2}$ value of 22 and under-predicts the disk flux at all wavelengths. 
Of the examples, Model C is the closest to edge-on, most vertically compact, and less flared. As the disk models become less flared, the grazing angles of the stellar photons decrease, lowering the number of low energy photons that penetrate down to the disk midplane. Consequently, less heat is absorbed into the midplane and the temperature decreases resulting in a lower millimeter continuum flux.

Despite not including the SED in the model optimization, the independently driven  HST 0.8 $\mu$m and ALMA MCMC results (represented by Models A and B) provide a decent fit to the spectral information. 
All three fits suffer at shorter wavelengths where both extinction and our assumption of an inner radius of 1 au have the biggest impact. 
However, despite allowing for a normalization factor in our fit to the millimeter continuum observations, Models A and B reproduce the longer wavelength fluxes seen in the SED, providing further confidence in our dust mass estimates.
Of the parameters we tested, the SED is most affected by the disk inclination and to a lesser extent the total dust mass. The SED appears to favor models with an inclination below $\sim 88^{\circ}$ and a dust mass of a few $\times 10^{-4} \, M_{\odot}$.

\subsection{Comparison to the 2.2 $\mu m$ Image}

\begin{figure*}
\begin{center}
\includegraphics[width=1.3in]{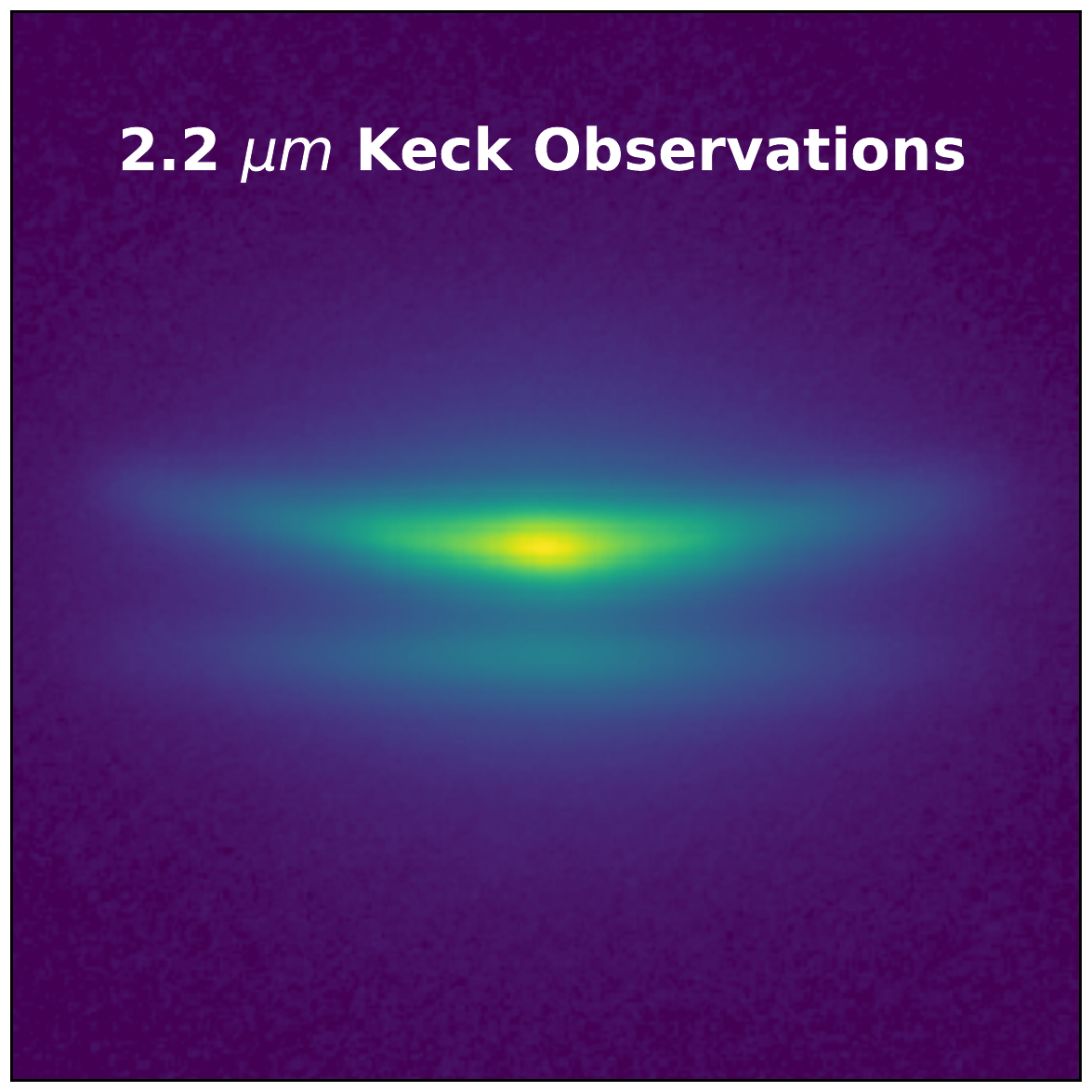}
\includegraphics[width=1.3in]{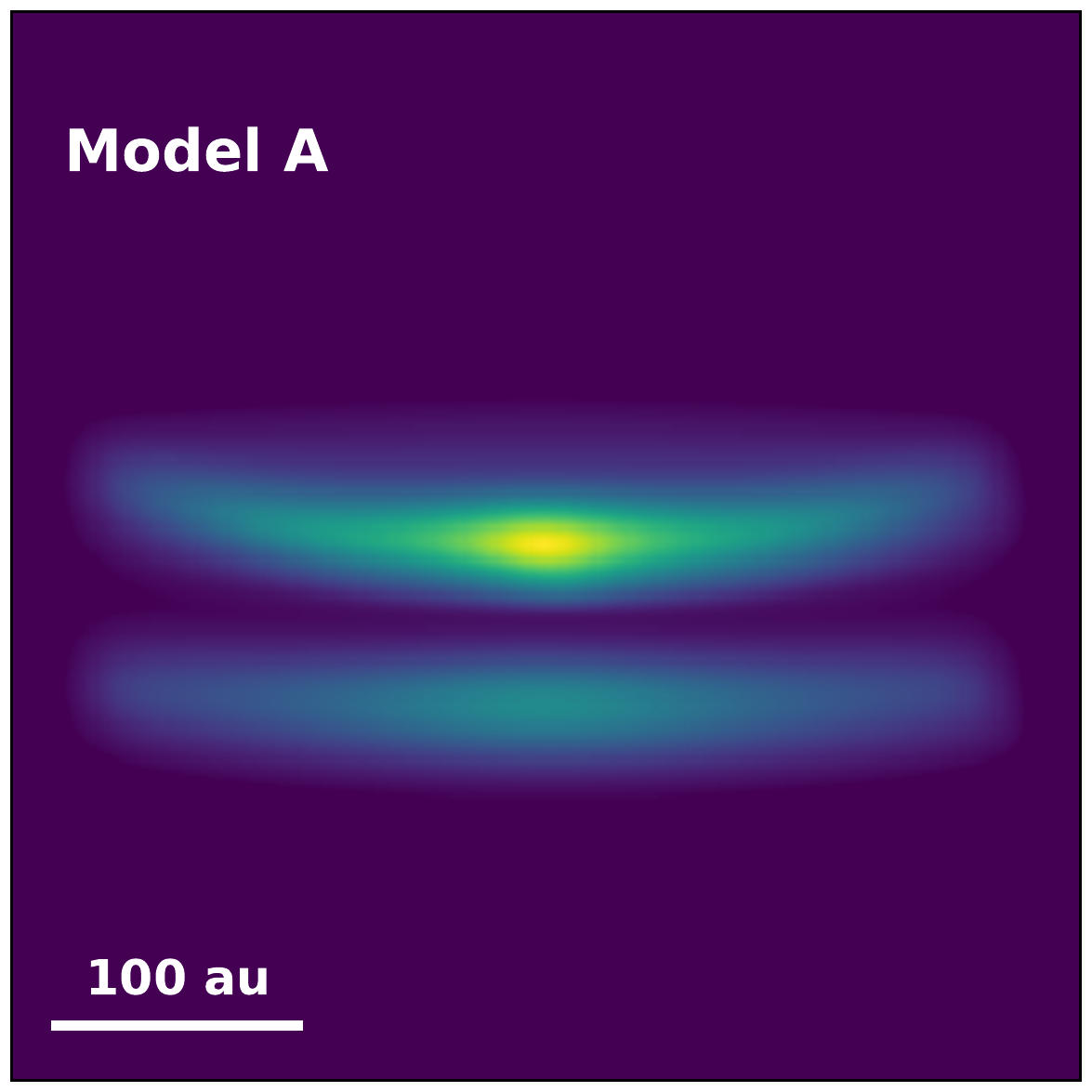}
\includegraphics[width=1.3in]{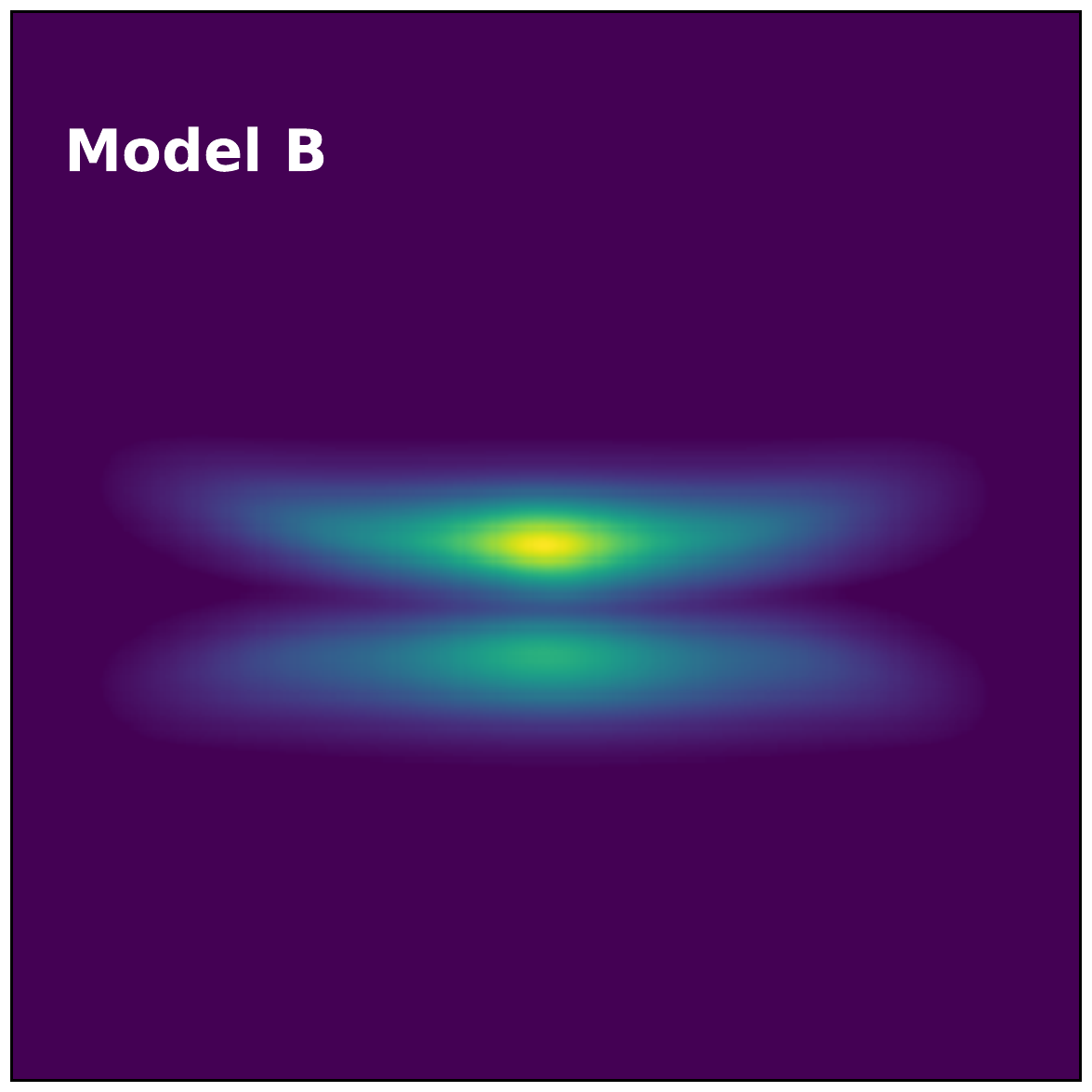}
\includegraphics[width=1.3in]{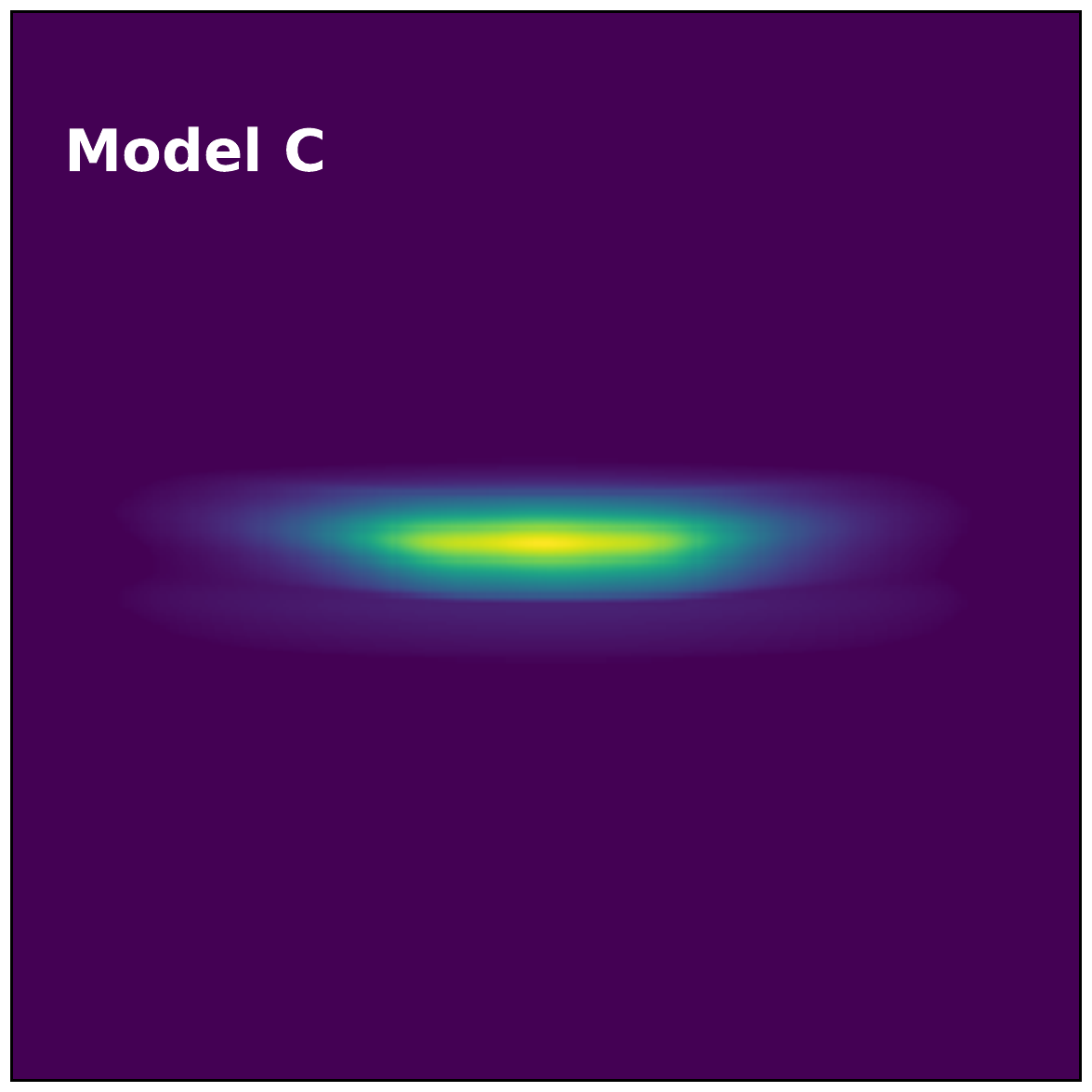}
\includegraphics[width=1.7in]{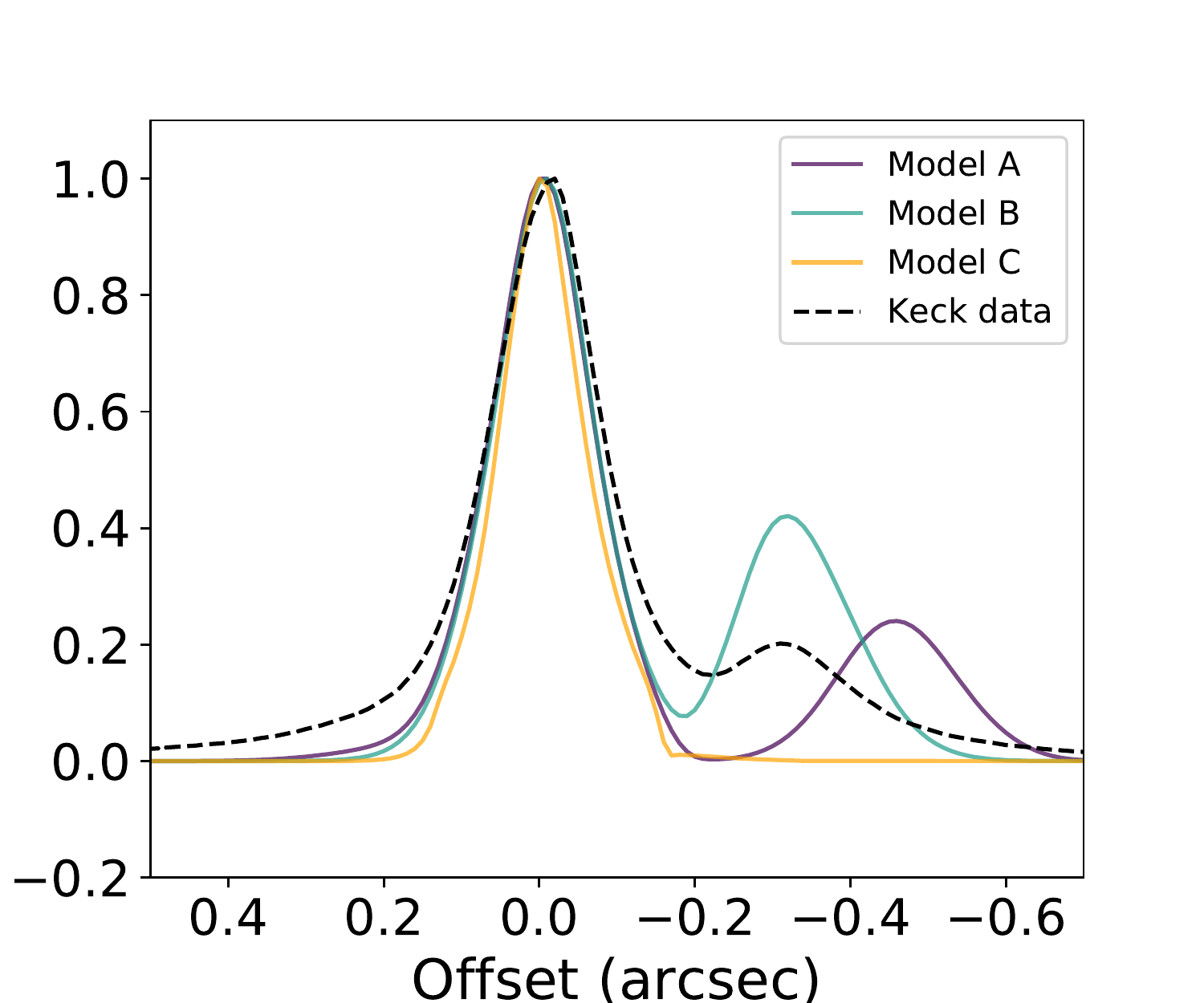}
\caption{
Here we compare the radiative transfer models for all three MCMC runs evaluated at $2.2 \, \mu m$ and compare the the Keck scattered light image. The three model images are show on the left, while the vertical slices through these models are compared the the vertical profile of the data on the right. None of the modeling efforts are optimized at this wavelength. \label{fig:keckmodels}}
\end{center}
\end{figure*}

While we do not model the Keck scattered light image directly, we relate this data to the best fit Models A, B and, C evaluated at a wavelength of 2.2 $\mu m$ in this section. Figure \ref{fig:keckmodels} shows the model image for each of the three MCMC runs at 2.2 $\mu m$ and compares vertical profiles to the Keck observations. The model images have been convolved with the asymmetric PSF (see Figure \ref{fig:keckobs}), recentered, and scaled to match the integrated flux of the observations. 

No single model reproduces both the width of the darklane and the top/bottom flux ratio.
The HST-optimized model (Model A) is able to closely reproduce the top/bottom nebulae flux ratio, but over-predicts the width of the dark lane. Conversely, the ALMA-optimized model (Model B) does a very good job in fitting the width of the darklane, but the ratio between the top and bottom disk nebulae is under-predicted by a factor of $\approx2$. The same effect is seen in Model B evaluated at 0.8 $\mu m$ when compared to our HST images. 
Generally, in scattered light the flux ratio is driven primarily by the phase function and the observed inclination. At the higher inclination (closer to edge-on) preferred by the ALMA dataset, one naturally gets a closer to 1-to-1 flux ratio.
Model C is far too vertically compact to reproduce the Keck results. When convolved with the $2.2 \mu m$ PSF, the top and bottom disk nebulae are indistinguishable.

Moving to longer wavelengths in scattered light, we would expect the apparent vertical height of the disk surface to decrease as we probe larger dust grains and the opacity decreases. This effect is seen in the observations (Fig. \ref{fig:darklane}). The magnitude of this change depends most significantly on the dust properties (these have the largest effect on opacity), the settling parameter and the observed inclination of the disk. Comparing the parameter values across the two most comparable models, we find that some combination of a $3.3-3.5$ grain size exponent, an inclination of $84-87^{\circ}$, and a viscous settling parameter of $5 \, \times 10^{-3} - 10^{-2}$ is required, but degeneracies limit further constraints. 

\subsection{Towards Comprehensive Multi-Wavelength Modeling}
\label{Sec:efficacy}

In recent years, ground based ExtremeAO instruments (GPI, SPHERE) sensitive to scattered optical and near-infrared starlight scattered off of disks' surfaces and mm interferometers (ALMA) probing thermal emission from larger dust grains in disk midplanes are producing remarkable disk images with unprecedented spatial resolution. 
Future instruments like JWST will provide mid-IR spectral information and probe grains of intermediate sizes, allowing a comprehensive understanding of disk physics and putting pressure on the simplistic assumptions about disk structures currently in use.
Each of these observables probe different disk regions and are sensitive to different aspects of the ongoing physical processes. 
The ability to reconcile different datasets into a single global model would provide the strongest constraint on disk properties, but this is non-trivial. 

For edge-on disks, this process is further complicated given that the disks are optically thick even at mm wavelengths. 
Combined radiative transfer modeling for scattered light and thermal images of edge-on disks have been conducted for several systems. 
\citet{2012A&A...543A..81M} combine Hubble Space Telescope (HST) images of the HH 30 disk with a continuum map from the  Plateau de Bure Interferometer (PdBI) at 1.3 mm, and an SED measured with IRS on the Spitzer Space Telescope in the mid-infrared. The authors find evidence for a partially cleared depletion zone and dust evolution using a simulated annealing modeling approach. 
CB 26 was characterized in \citet{2009A&A...505.1167S} via a HST images, a Sub-millimetre Array SMA map, and an SED. In this case, the SMA map is only marginally resolved and is represented by a peak flux and 1D radial profile. 
\citet{2003ApJ...588..373W} investigates an HST scattered light image and an OVRO Millimeter Array map at 1.3 and 2.7 mm for the ``Butterfly Star,'' IRAS 04302+2247. The authors used a grid modeling approach and found that the envelope and embedded disk must have different dust properties.

In all cases, the combination of scattered light and thermal continuum observations provided more stringent constraints on the disk properties than either alone.
It is worth noting here, however, that all the thermal continuum maps discussed above had a resolution of 3-5 beams along the disk major axis, and none were resolved in the vertical direction.
With the higher resolution data now available more complex modeling of radial structures in circumstellar disks must be employed at earlier evolutionary stages. As the model complexity increases, the computational problems also become more difficult. The covariance framework presented here provides a means of combining different observables and accounting for differing model sensitivities, but falls short of a complete description of the disk structure. While it is possible to include more complex surface density distributions within this framework, it would require a drastic increase in the already prohibitive computation times.\footnote{The modeling efforts presented in this paper represent $\sim 4$ months of CPU time parallelized on a 4 core Linux machine with a Intel Xeon(R) CPU E3-1240 v5 processor. 
}

\section{Discussion}
\label{Sec:discussion}

Here we compare the results across all MCMC runs and discuss the constraints we are able to place on the individual parameters. We relate the parameter values to the ongoing physical processes within the disk and compare the inferred temperature structure to results from ALMA CO observations as presented in the companion paper (Flores et al., in prep.).

First, we compare the best fit parameter estimations across all MCMC runs and describe the global constraints. The radial and vertical structures are discussed in Sections \ref{Sec:gamma} and \ref{Sec:settling}, respectively. \\

\noindent \textbf{Inclination:} The disk is nearly edge-on in inclination with a range of accepted values from $84 - 89$ degrees. The estimates for the inclination agree to within $1 \sigma$ across all MCMC runs, with the ALMA data preferring slightly larger values than the HST data. 
In scattered light, the flux ratio between the top and bottom nebula provides some information about the disk inclination but also depends on the scattering properties of the dust grains. 
The ALMA data is less optically thick and doesn't require assumptions about the grain scattering properties but the disk is not resolved in the vertical direction, so the inclination is difficult to decouple from the gas scale height, even without the complication of dust settling.
In the end, we find an inclination of $\sim 87 \pm 1^{\circ}$ is most probable. \\

\noindent \textbf{Disk Dust Mass:}  There are orders of magnitude disagreement between the dust masses inferred from the different observables, though the uncertainties are correspondingly large. The integrated flux in an ALMA continuum map is often used to directly determine the value in the optically thin case, but this disk remains optically thick even at 1.3 mm. The ALMA data prefers a dust mass of $\sim 2 \, \times 10^{-4} \, M_{\odot}$. 
The scattered light results, having an optical depth far greater than the millimeter continuum images are only able to place a lower limit on the disk dust mass and require $M > 3 \times 10^{-5} \, M_{\odot}$. 

\begin{figure}[h]
\includegraphics[width=3.3in]{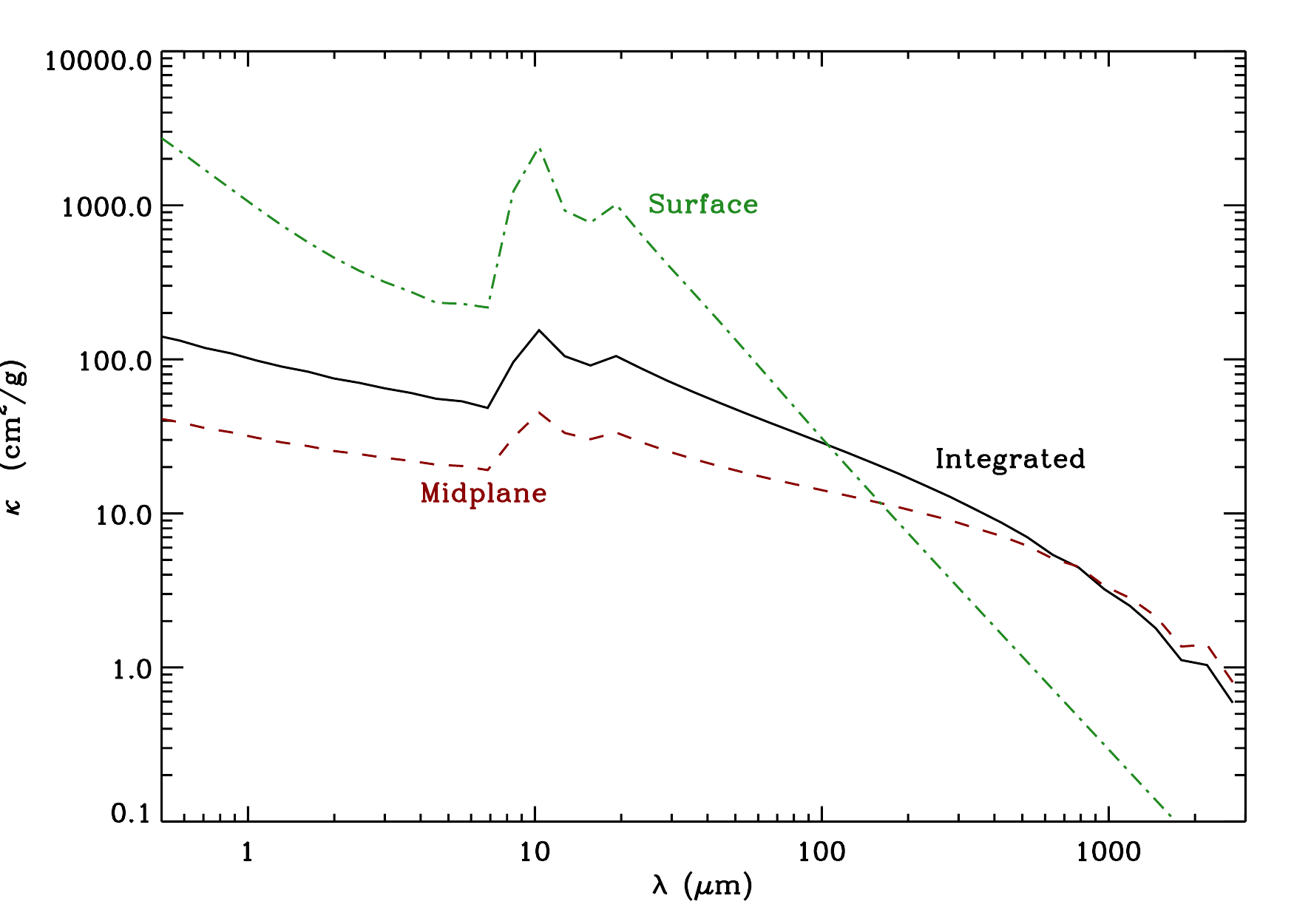}
\caption{Opacities are presented for Model A  (Table \ref{table:paramresults}) computed at a radius of 100 au. Dust settling allows for a change in opacity from the disk surface (22 au in this case) to the midplane (measured at 0 au).
\label{fig:opacity}} 
\end{figure}

Great caution is needed when considering the grain size distributions, as the HST and ALMA observations not only probe grain sizes that differ by roughly three orders of magnitude, but also very distinct locations in the disk. Our prescription for the vertical density profile, dust settling, and integrated size distribution allow us to interpolate between these populations, but this is likely done in too simplistic a fashion. To illustrate the spatial variations in dust properties implied by our model, we consider the opacity law in the midplane and at the disk surface.
If we examine Model A more closely, we find that the midplane opacity law (measured at an elevation of 0 au) is much shallower than the surface opacity (at an elevation of 22 au) as seen in Figure \ref{fig:opacity}. Furthermore, at millimeter wavelengths, the integrated opacity is roughly equivalent to that of the midplane, since this is where the large grains are located. The surface and mid-plane opacity slopes agree well with other literature results assuming a grain size power law slope of 3.5 \citep[e.g.][]{2001ApJ...553..321D}.

The dust mass posterior distribution for the scattered light plus mm continuum MCMC is bimodal with peaks at both $7 \times 10^{-6}$ and $3 \times 10^{-4} \, M_{\odot}$, the latter being more consistent with the ALMA continuum results. When compared to the spectral energy distribution 
(Section \ref{sed:sedres}) the lower mass models are not able to reproduce the $>20 \, \mu m$ data,
which requires a dust mass $> 10^{-4.5} \, M_{\odot}$. 
\citet{2020arXiv200806518V} provide a lower limit for the dust mass of $7.7 \, \times \, 10^{-5} \, M_{\odot}$ under the assumption of 
isothermal dust emission and a characteristic dust temperature of 20 K. 
Oph163131 is expected to be optically thick even at mm wavelengths. Indeed, all of our models produce an optically thick disk at 1.3 mm. 
Precluding an unusually cold disk (discussed further in Section \ref{Sec:temp}), this provides a robust lower limit for the dust mass.
Based on a visual comparison of model images, a dust mass of $> 10^{-4} \, M_{\odot}$ is required to block sufficient light from the central star at inclinations below $89^{\circ}$ to reproduce both the scattered light and mm continuum observations.
For a closer look at the bimodality in the disk dust mass, see Appendix \ref{Appendix:triangle}.

\label{Sec:mass}

\subsection{Radial Structure}
\label{Sec:gamma}

The radial structure of the disk as described in Equation \ref{Eq:taper} depends most critically on the surface density exponent ($\gamma$) and the critical radius ($R_{C}$). \\

\noindent \textbf{Surface Density Exponent:} The surface density exponent remains unconstrained. The scattered light observations place no constraint on this parameter, while the ALMA continuum observations and combined fit prefer the smallest allowed values. Low values of $\gamma$ correspond to a surface density distribution that is either flat or increasing with radius with a very steep exponential falloff of material at the critical radius. This may point to the existence of a very optically thick outer ring. In any case, it is likely that our assumption of a tapered power law is too simplistic. The profile could have significant substructures (i.e. gaps, ice lines) that the model does not encapsulate. A more complex prescription is left for future work (see  Sec. \ref{Sec:efficacy}). \\

\noindent \textbf{Critical Radius:} The MCMC preferred values for the critical radius range from 75 - 120 au and agree to within $2 \sigma$. Note that this is separate from the outer radius which is defined by a $3 \sigma$ flux threshold in the scattered light image to be 191 au. 
While the posterior distribution for the scattered light fit is very broad, the ALMA  posterior is tighter with a correspondingly flat disk. It is possible that a good fit for the ALMA observations requires a sharp outer edge which is forcing the fit towards small values of $\gamma$ (steeper exponential falloff of material). This could point towards a difference in the outer radius in the scattered light and mm continuum observations. Radial migration of the larger dust grains could be the culprit, but this is not conclusive. \\

Discrepancies between the model parameters preferred by the different observables point to a fundamental limitation in the parameterization of our model. Inconsistencies exist in the surface density exponent and the critical radius. 
It follows that the most likely culprit is our prescription for the radial density distribution in the disk. 
Traditionally, radiative transfer models of scattered light observations have used a single power law density profile with a sharp outer edge. For millimeter continuum observations, the tapered-edge model was required to reproduce more diffuse material in the outer regions of the disk \citep[e.g.][]{2011ARA&A..49...67W}. In \citet{2017ApJ...851...56W} we demonstrated that the tapered-edge model provided a superior fit for scattered light observations of edge-on disks and continue this practice here. However, the ever increasing diversity in observations of young disk morphologies may require the use of more complex density structures in order to avoid hidden biases in the MCMC results.
Here we discuss possible avenues for future inquiry, though a complete exploration of the surface density profiles is left for future work. \\

\noindent Dust Growth and Radial Drift: Some mechanism to decouple the radial profiles of the small and large dust grains would allow us to reproduce both the sharp edge seen at ALMA wavelength and the more radially extended scattered light observations. Work by \citet{2014ApJ...780..153B} suggests that the dust transport processes in a viscous accretion disk produce a radially dependent gas-to-dust ratio with significant gas depletion and radial drift in the outermost regions of the disk. In this case, our model assumption of a dust surface density that is linearly dependent on the gas surface density is incorrect. Gas observations for Oph163131 are presented in Paper II and the gas is more radially extended than the dust both at HST and ALMA wavelengths (see their Figure 8). While this strongly hints at a radially variable gas-to-dust ratio, estimating the latter is left for a future comprehensive modeling analysis. We note, however, that
there remains significant disagreement in the literature on how to parameterize dust growth and radial drift \citep[e.g.][]{2014prpl.conf..339T}. 
\\

\noindent Broken Power Law and/or Rings+Gaps: These types of discontinuous structures have recently been observed in many high resolution ALMA and scattered light images viewed face-on even at early evolutionary stages. The DSHARP sample of bright, nearby protoplanetary disks found that the most common form of substructure is concentric bright rings with dark gaps \citep{2018ApJ...869L..41A}.
Ring/gap substructures are most commonly observed in more extended disks and may be ubiquitous. A recent ALMA survey of Taurus found that all disks with effective radii larger than 55 au presented substructures with smooth inner cores that resembled their more compact counterparts \citep{2019ApJ...882...49L}.
The scattered light observations of Oph163131 are too optically thick to probe the radial structure of the inner regions.
However, the radial profile of the ALMA map with a central peak, plateau, and gradual decline may hint at more detailed structure. Higher resolution ALMA maps should help to further illuminate the radial structure of the disk. 
\\

\subsection{Vertical Structure}
\label{Sec:temp}
\label{Sec:settling}

In this section, we discuss how dust settling and mid-plane temperature affect the vertical structure of the disk. \\

\noindent \textbf{Scale Height:} The best fit scale height values range from $ 5 - 10$\,au at a reference radius of 100 au, and  $1 \sigma$ agreement for all three runs. From a visual inspection, the best models at the low end of this range require low inclinations and large dust masses that aren't generally preferred. Values of $ 7 - 10$\,au represent the data well. \\

\noindent \textbf{Flaring Exponent ($\beta$):} The disk is flared with results for the flaring exponent ranging from $\sim 1.3-1.8$. All MCMC runs agree to within $1-2 \sigma$. Interestingly, the individual fits to the HST scattered light and ALMA continuum observations on average prefer unphysically large values of $\beta$, though the parameter is not well constrained. It is possible that the models are attempting to compensate for a more diffuse halo surrounding the disk with enhanced flaring (see discussion in Section \ref{sec:keck}). The covariance matrix framework used in the joint fit tends to marginalize over more diffuse material in the outer layers of the disk. There is also a known degeneracy between $\beta$ and the surface density exponent ($\gamma$) due to viewing geometry \citep{1996ApJ...473..437B,2004ApJ...602..860W} that will also broaden these distributions.  \\

\noindent \textbf{Dust Settling ($\alpha$):}  Both of the MCMC results for the individual HST and ALMA datasets prefer higher values for the $\alpha$ settling parameter calling for a turbulent disk ($\log \alpha \gtrsim -2.5$) with moderate dust settling. 
It is not obvious why the individual datasets would have a preference for a high degree of dust settling. The scattered light images is not expected to change with dust settling as the smallest dust grains are well coupled with the gas. For the ALMA dataset, the apparent vertical extent of the disk could be changed by invoking dust settling or by changing the dust scale height, and it is unclear why a model with a little settling and a lower scale height is preferred.
The covariance-based results are able to combine observations probing dust particles of different sizes and place stronger leverage on this dust settling parameter. 
In this case, the combined MCMC run produces bifurcated results with the higher mass models preferring a value of $\log \alpha = -3.1$ and the lower mass models preferring very high values with $\log \alpha > -1.9$. 
Given that we only have two measurements of a very complex process that likely varies non-linearly as a function of grain sizes, and the disk radial and vertical locations, we are not able to put a tight constraint on the value for $\alpha$. However, it is clear that some degree of dust settling is required to reconcile the two datasets.
\\

Here we examine the vertical extent of our best fit disk models and compare this to the midplane temperature structures.
If the small dust particles in the disk are well coupled with the gas, their vertical distribution is dictated by gas pressure support and depends on the disk midplane temperature.
For a vertically isothermal disk, the vertical density distribution is described by Equation \ref{Eq:scaleheight} \citep{1996ApJ...473..437B}: 

\begin{equation}
  H(r) \, = \sqrt{ \frac{ k_{B}T(r)r^{3}}{ G M_{\mathrm{star}} \mu m_{p}}} 
  \label{Eq:scaleheight}
\end{equation}

\noindent where $k_{B}$ is the Boltzmann constant, we assume a reduced mass, $\mu$ of 2.3, and $m_{p}$ is the proton mass. The midplane gas temperature profile, $T(r)$, is determined via radiative transfer as part of the MCFOST output for each model. 

\begin{figure}[h]
\includegraphics[width=3.3in]{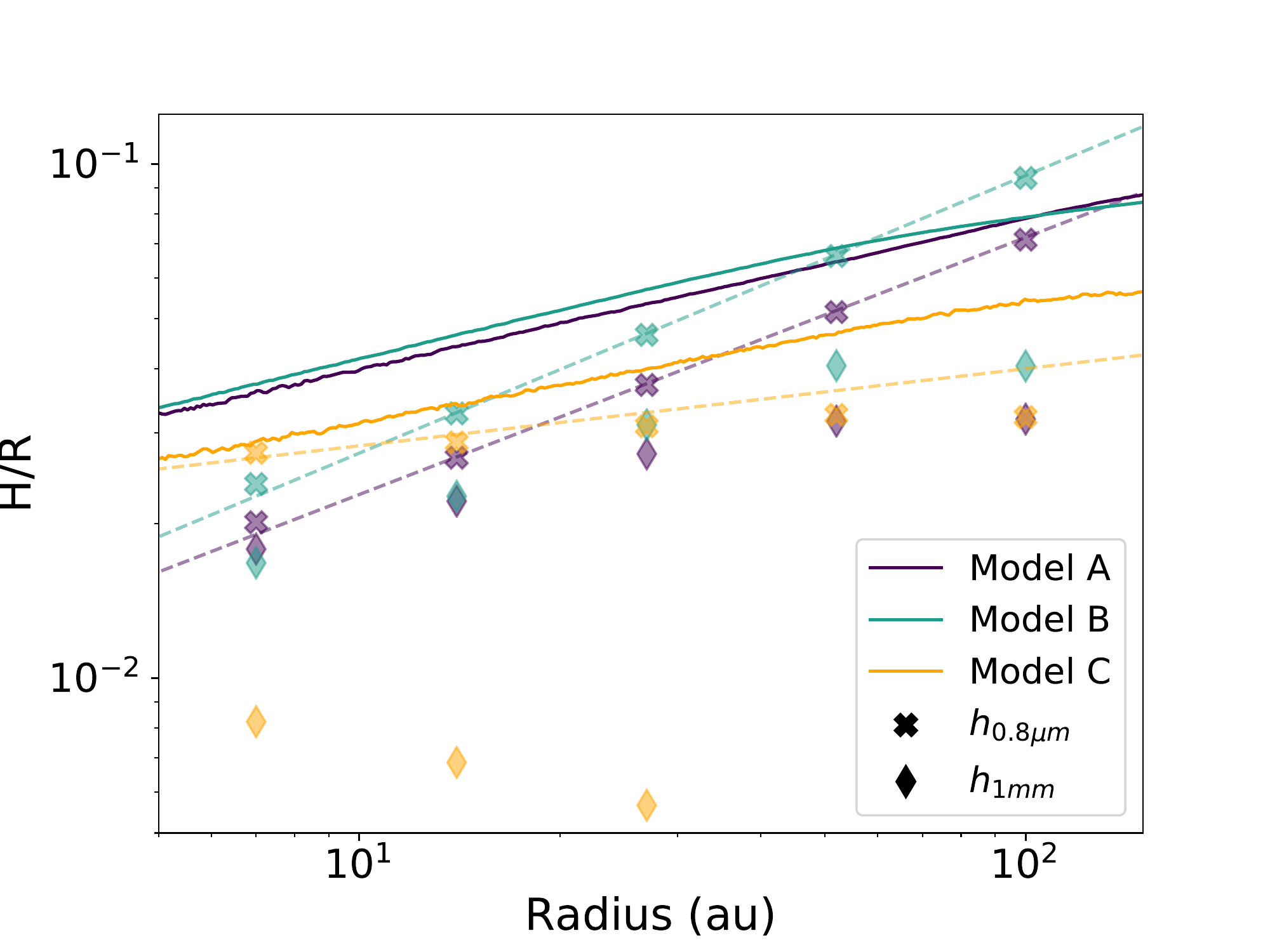}
\caption{The radial scale height dependence of the disk. Results are shown for the the three models presented in Table \ref{table:paramresults}. The solid lines indicate the scale height computed using the midplane temperature provided via MCFOST following equation \ref{Eq:scaleheight}, while the dashed lines show the scale height using the best fit values displayed in Table \ref{table:paramresults}. The radiative transfer temperature profile is shallower than the best fit flaring parameters would suggest for all models. 
For comparison, we show the dust scale heights for grain sizes probed by the scattered light observations $h_{0.8 \mu m}$ and the millimeter continuum data $h_{1mm}$ for these three models. 
\label{fig:scaleheight}} 
\end{figure}

Figure \ref{fig:scaleheight} compares the scale height derived using Eq. \ref{Eq:scaleheight} for Models A, B and C (see Table \ref{table:paramresults}) to the scale height defined in our disk model as $H(R) = H_{0} (R/R_{0})^{\beta}$ (where $R_{0}$ = 100 au) using the parameter values for $H_{0}$ and $\beta$ given in Table \ref{table:paramresults}. 
The temperature-derived scale height profile is shallower than the parameterized scale height would suggest for Models A and B (representing the MCMC results for the individual HST and ALMA datasets, respectively). 
For Model C (result of the combined observable MCMC run) the slope is a better match and agrees with the slope expected for a passive disk (a temperature radial dependency of $1/\sqrt{R}$ and corresponding scale height radial dependency of $R^{5/4}$ using Eq. \ref{Eq:scaleheight}).
However, the absolute values for the gas temperature inferred scale height are larger than predicted by the model geometry in all cases. 

To further illustrate this discrepancy, we verify the grain size dependent dust scale heights by fitting a Gaussian to the vertical density profiles in the MCFOST populated dust grids (Figure \ref{fig:scaleheight}). Unsurprisingly, the $0.8\, \mu$m dust scale heights match the parameterized gas scale height, while the 1\,mm dust scale heights demonstrate significant dust settling even for moderate turbulence values. 

There are two important caveats to consider when interpreting Figure \ref{fig:scaleheight}. 
First, our choice of the minimum grain size will affect the apparent scale height for a given midplane disk temperature. To test this effect, we recomputed Models A, B, and C with a minimum grain size of $1 \, \mu$m (compared to the modeled $0.03 \, \mu$m value) and found that this does result in a cooler disk. We saw a 3-10\% improvement in the percent difference between the temperature dependent and geometrically determined scale heights at a reference radius of 20 au. However, this decrease was insufficient to account for the observed discrepancy (50, 30 and, 29\% for Models A, B, and C). 
Secondly, the Gaussian vertical density profile defined in MCFOST assumes a vertically isothermal temperature distribution and is inconsistent with the expectation that the upper layers of a flared disk will be warmer than the midplane. For a disk in hydrostatic equilibrium, we would expect the warm upper layers to be puffed up resulting in an even higher scale height for a given disk midplane temperature, exacerbating the discrepancy.

This difference between the height of the disk inferred using the midplane gas temperature and the submicron-sized dust modeled geometry implies that either 1) the disk midplane is colder than would be expected for an object of this stellar type or 2) even the submicron-sized dust grains are settled relative to the gas. 

In Paper II, we present a resolved ALMA map of the gas in the system and use CO isotopologues to reconstruct the temperature as a function of both height and radius throughout the disk. We find that the vertical extent of the gas matches the HST scattered light observations (though the gas is more extended radially).
Based on these results, scenario (1) appears most likely. 

If indeed the midplane temperature is over-predicted, either the luminosity we have assumed for the central point source is incorrect, or there is some temperature shielding mechanism. For a detailed discussion of the spectral type and dynamical mass for the central star, see Paper II. 
It is not possible to account for the observed temperature deficiency within the uncertainties in our stellar properties. 
Here we discuss the possibility that some structure in the innermost regions of the disk could provide temperature shielding to the outer disk midplane. As we've already demonstrated, this disk likely has a more complex radial density structure (see Sec. \ref{Sec:gamma} especially in the outer regions past the ALMA plateau ($R \gtrsim$ 75 au). The central peak in the ALMA continuum data requires warm dust down to at least $\sim$ 10 au but isn't sensitive to the innermost regions of the disk. Likewise, the SED shows no evidence for an inner clearing or enhanced inner wall. 
There is, however, evidence for variability
operating on timescales much shorter than the orbital timescales in the visible regions of the disk. GAIA photometry for this source shows variability on the order of 5\% over 21 observations \citep{2016AA...595A...1G}. Additionally, the HST observations presented here show clear differences in the left/right flux ratios between the two F814W epochs spaced $\sim 2.5$ years apart (see Figure \ref{fig:difference}). 

Various disk temperature screening mechanisms have been proposed in the literature. \citet{2012A&A...539A..20S} demonstrate that a puffed up inner rim can cause apparent ring+gap structures and midplane temperature shielding. 
Self shadowing from an inclined inner disk both lowers the midplane temperature and causes variability in the outer disk. This mechanism could be the result of an embedded protoplanet and has been invoked to explain features seen in several young disks including TW Hydrae \citep{2018ApJ...860..115P}, HD 143006 \citep{2018A&A...619A.171B}, and HD 139614 \citep{2020A&A...635A.121M}. However, this mechanism is most commonly invoked to explain surface features and it is unclear what effect this would have on the azimuthally averaged disk midplane temperature. 
Far infrared emission lines can also serve as gas coolants with up to $10^{-1} \, L_{\odot}$ of cooling observed in Class 0 objects \citep{2001ApJ...555...40G}, though this should be mitigated by the large dust opacity in Oph163131.

The degree of dust settling will also impact the apparent vertical height of the disk. It is clear from our modeling efforts that some amount of settling is required to explain the discrepancy in the apparent vertical extent of the scattered light and mm continuum observations. The exact degree of settling is difficult to quantify since the vertical height of the disk in the ALMA data is not resolved. Our best fit values range from a few $\times 10^{-3} - 10^{-2}$ which agrees well with values presented in the literature for other members of the Ophiuchus SFR \citep[$\alpha = 0.0005 - 0.08$;][]{{2009ApJ...700.1502A}} and more generally \citep[e.g.][]{2007ApJ...659..705A, 2010A&A...513A..79B}. 
The corresponding millimeter scale heights (at a reference radius of 100 au) for Models A, B, and C are $h_{1mm}$ = 3.3, 4.2, 0.3 au, respectively as shown in Figure \ref{fig:scaleheight} (compare to $h_{0.8 \mu m}$ = 7.1, 9.4, 3.2 au for models A, B, and C at 100 au). 
The combined model calls for by far the strongest degree of settling, but without resolving the disk vertically at mm wavelengths, this is likely only an upper limit. 
If we compare this to similar work for the HL Tau disk with an initial scale height $H_{0}$ = 10 au, the millimeter scale height values were $h_{1mm}$ = 2.1 and 0.7 au with associated disk viscosity values of $\alpha$ = $3\times 10^{-3}$ and $3 \times 10^{-4}$, respectively \citep{2016ApJ...816...25P}. It is clear that even relatively turbulent disks can have markedly reduced millimeter scale heights. 

The degree of settling depends on both the viscous $\alpha$ parameter and the Stokes number for grains of a given size. As a consistency check, we compute the Stokes numbers for Models A, B and C following \citet{2016SSRv..205...41B} for a given grain size ($a$) where $St = a \, \frac{\pi}{2 \sqrt{2 \pi}} \, \frac{\rho_{s}}{\rho_{g,mid} h_{g}}$ for a material density $\rho_{s} = 1.8 \, g/cm^{3}$ from \citet{1991ApJ...377..526R} and a midplane gas density ($\rho_{g,mid}$) and gas scale height ($h_{g}$) taken from the models. 
We find $St_{0.8 \mu m, 10 au} = 3 \times 10^{-6}, 2 \times 10^{-5},$ and $8 \times 10^{-6}$ and $St_{0.8 \mu m, 100 au} = 8 \times 10^{-5}, 1 \times 10^{-4},$ and $1 \times 10^{-4}$ for Models A, B and C respectively. If we expect settling to become significant in models with $St \gtrsim 10^{-2}$ \citep[see for example][]{1995Icar..114..237D,2009A&A...496..597F, 2019MNRAS.485.5221L}, this confirms that the small dust grains are well coupled with the gas even in the outermost regions of the disk. Likewise, we find $St_{1 mm, 10 au} = 4 \times 10^{-3}, 2 \times 10^{-2},$ and $1 \times 10^{-2}$ and $St_{1 mm, 100 au} = 1 \times 10^{-1}, 2 \times 10^{-1},$ and $2 \times 10^{-1}$ for the larger grains. As expected, the millimeter sized dust grains experience strong settling most notable in the outer regions of the disk.

A measurement of the accretion rate may place additional constraints on the turbulent properties of the disk. Unfortunately, Oph163131 exhibits only marginal accretion signatures and a measurement of the accretion rate is not yet possible. For a more complete discussion of accretion in this system see Paper II. 
A non-accreting disk is also expected to be less turbulent (smaller $\alpha$) with stronger settling.

\section{Summary and Conclusions}
\label{Sec:summary}

In this work we perform a multi-wavelength modeling study of the protoplanetary disk Oph163131 viewed edge-on and resolved in scattered light for the first time. Here we summarize the results.

\begin{itemize}

\item[.] We present scattered light images from the Hubble Space Telescope at 0.8, 0.6, and 0.48 $\mu$m, Keck observations at 2.2 $\mu$m, an ALMA 1.3 $mm$ continuum map \citep{2020arXiv200806518V}, and a spectral energy distribution compiled from the literature.  

\item[.] We perform radiative transfer modeling using an MCMC to constrain the geometry and dust properties of the disk. Separate MCMC explorations were performed against the 0.8 $\mu$m and 1.3 mm and we find that a relatively flat disk with some degree of dust settling is required to explain the vertical structure in both datasets.

\item[.] A global fit to both observables using a covariance-based log-likelihood estimation with a parallel-tempered MCMC was marginally successful. Bimodal posterior distributions are characteristic of a highly degenerate parameter space and likely imply a more complex underlying disk structure. 

\item[.] When combined with temperature profiles extracted from ALMA CO gas maps in a companion paper (Paper II) inconsistencies in the radial density profile, a colder than expected disk surface, and evidence for dust settling all point to complex structures hidden in the disk midplane.

\end{itemize}

\acknowledgements
This work is based on observations made with the NASA/ESA Hubble Space Telescope, obtained at the Space Telescope Science Institute, which is operated by the Association of Universities for Research in Astronomy, Inc., under NASA contract NAS 5-26555. These observations are associated with GO Program \#12514, whose funding support K. Stapelfeldt and D. Padgett acknowledge.  C. Pinte acknowledges funding from the European Commission's 7$^\mathrm{th}$ Framework Program (contract PERG06-GA-2009-256513) and from Agence Nationale pour la Recherche (ANR) of France under contract ANR-2010-JCJC-0504-01. F. M\'enard acknowledges funding from ANR of France under contract number ANR-16-CE31-0013. GD acknowledges support from NASA grants NNX15AC89G and NNX15AD95G/NExSS as well as 80NSSC18K0442. CF acknowledges support from the NASA Infrared Telescope Facility, which is operated by the University of Hawaii under contract 80HQTR19D0030 with the National Aeronautics and Space Administration

This paper makes use of the following ALMA data: ADS/JAO.ALMA\#2016.1.00771.S. ALMA is a partnership of ESO (representing its member states), NSF (USA) and NINS (Japan), together with NRC (Canada), MOST and ASIAA (Taiwan), and KASI (Republic of Korea), in cooperation with the Republic of Chile. The Joint ALMA Observatory is operated by ESO, AUI/NRAO and NAOJ.
A portion of the data presented herein were obtained at the W. M. Keck Observatory, which is operated as a scientific partnership among the California Institute of Technology, the University of California and the National Aeronautics and Space Administration. The Observatory was made possible by the generous financial support of the W. M. Keck Foundation. 
The authors wish to recognize and acknowledge the very significant cultural role and reverence that the summit of Maunakea has always had within the indigenous Hawaiian community. We are most fortunate to have the opportunity to conduct observations from this mountain.

\software{MCFOST \citep{2006A&A...459..797P}, Tiny Tim \citep{1995ASPC...77..349K}, emcee \citep{2013PASP..125..306F} mcfost-python (https://github.com/swolff9/mcfost-python), CASA \citep{2007ASPC..376..127M}, DrizzlePac \citep{2012drzp.book.....G}, matplotlib \citep{Hunter:2007}, Astropy \citep{astropy:2013, astropy:2018}.}

\nocite{*}

\appendix

\section{Spectral Energy Distribution}
\label{appendix:sed}
Table \ref{table:sed} provides a complete description of these observations including references. 
In filters where multiple observations are available, the flux values from the work with the highest signal to noise was chosen. In cases with low uncertainties, we have included a systematic 5\% error to account for variability in this young source.


\begin{deluxetable*}{ccccc}[!t]  
\tablecolumns{5}
\tablecaption{Spectral energy distribution photometry and references. \label{table:sed}}
\tablehead{   
  \colhead{$\lambda (\mu m)$} &
  \colhead{Flux (mJy)} &
  \colhead{Source} &
  \colhead{Instrument:Filter} &
  \colhead{Bandwidth ($\mu m$)} 
}
\tablewidth{6.8in}
\startdata
0.343 & 0.014 $\pm$ 0.001 & \citet{2012MNRAS.426..903P} & XMM-OT:U & 0.086 \\
0.475 & 0.15 $\pm$ 0.008 & This Work & HST/WFC3:F475W & 0.13 \\
0.477 & 0.14 $\pm$ 0.007 & \citet{2016arXiv161205243F} & PAN-STARRS/PS1:g & 0.13 \\
0.606 & 0.45 $\pm$ 0.023 & This Work & HST/ACS:F606W & 0.22 \\
0.613 & 0.54 $\pm$ 0.028 & \citet{2016arXiv161205243F} & PAN-STARRS/PS1:r & 0.14 \\
0.671 & 0.45 $\pm$ 0.02 & \citet{2016AA...595A...1G} & Gaia:G & 0.44 \\
0.748 & 1.04 $\pm$ 0.05 & \citet{2016arXiv161205243F} & PAN-STARRS/PS1:i & 0.13 \\
0.814 & 1.61 $\pm$ 0.08 & This Work & HST/ACS:F814W & 0.15  \\
0.865 & 1.80 $\pm$ 0.09 & \citet{2016arXiv161205243F} & PAN-STARRS/PS1:z & 0.10 \\
0.960 & 2.19 $\pm$ 0.11 & \citet{2016arXiv161205243F} & PAN-STARRS/PS1:y & 0.06 \\
1.0 & 3.0 $\pm$ 0.2 & \citet{2007MNRAS.379.1599L} & UKIDSS:Y & 0.10 \\
1.2 & 5.0 $\pm$ 0.2 & \citet{2003tmc..book.....C} & 2MASS:J & 0.16 \\
1.2 & 4.2 $\pm$ 0.2 & \citet{2007MNRAS.379.1599L} & UKIDSS:J & 0.16 \\
1.6 & 7.1 $\pm$ 0.4 & \citet{2003tmc..book.....C} & 2MASS:H & 0.25 \\
1.6 & 5.1 $\pm$ 0.3 & \citet{2007MNRAS.379.1599L} & UKIDSS:H & 0.29 \\
2.2 & 6.0 $\pm$ 0.3 & \citet{2003tmc..book.....C} & 2MASS:Ks & 0.26 \\
2.2 & 4.5 $\pm$ 0.2 & \citet{2007MNRAS.379.1599L} & UKIDSS:K & 0.34 \\
3.3 & 2.5 $\pm$ 0.1 & \citet{2013yCat.2328....0C} & WISE:W1 & 0.66 \\
3.6 & 3.2 $\pm$ 0.2 & \citet{2009ApJS..181..321E} & Spitzer/IRAC:3.6 & 0.75 \\
4.5 & 2.4 $\pm$ 0.1 & \citet{2009ApJS..181..321E} & Spitzer/IRAC:4.5 & 1.02 \\
4.6 & 1.7 $\pm$ 0.1 & \citet{2013yCat.2328....0C} & WISE:W2 & 1.04 \\
5.8 & 1.9 $\pm$ 0.1 & \citet{2009ApJS..181..321E} & Spitzer/IRAC:5.8 & 1.43 \\
8.0 & 1.9 $\pm$ 0.1 & \citet{2009ApJS..181..321E} & Spitzer/IRAC:8.0 & 2.91 \\
12 & 1.1 $\pm$ 0.1 & \citet{2013yCat.2328....0C} & WISE:W3 & 5.51 \\
22 & 65.0 $\pm$ 3.2 & \citet{2013yCat.2328....0C} & WISE:W4 & 4.10 \\
24 & 37.0 $\pm$ 3.4 & \citet{2009ApJS..181..321E} & Spitzer/MIPS:24 & 5.3 \\
70 & 290 $\pm$ 34 & \citet{2009ApJS..181..321E} & Spitzer/MIPS:70 & 19 \\
70 & 574 $\pm$ 2.9 & \citet{pacs} & Herschel/PACS & 10.6 \\
100 & 595 $\pm$ 43 & \citet{pacs} & Herschel/PACS & 17.0 \\
160 & 709 $\pm$ 85 & \citet{pacs} & Herschel/PACS & 30.2 \\
250 & 700 $\pm$ 70 & \citet{spire} & Herschel/SPIRE & $\sim 85$ \\
350 & 510 $\pm$ 89 & \citet{spire} & Herschel/SPIRE & $\sim 117$ \\
500 & 317 $\pm$ 64 & \citet{spire} & Hershcel/SPIRE & $\sim 167$ \\
870 & 124.8 $\pm$ 2.4 & \citet{2017ApJ...851...83C} & ALMA:Band 7 & -- \\
1300 & 44.8 $\pm$ 4.5 & This Work & ALMA:Band 6 & -- \\
2600 & $<$ 13 & This Work & CARMA & -- \\
\enddata
\tablecomments{
Note that we have increased photometric uncertainties to 5\% where appropriate to account for the variability of young stellar systems. }
\end{deluxetable*}


\section{MCMC Results}
\label{Appendix:triangle}
Here we provide the resultant triangle plots for each of the three MCMC runs. The results represented here are given in Table \ref{table:paramresults}. 
Blue crosshairs show parameter estimates based on the 50th percentiles of the samples in the posterior distributions with uncertainties shown with dashed lines given by the 16th and 84th percentiles. The red crosshairs indicate the parameters used in the characteristic models A, B, and C to represent the HST, ALMA and combined MCMC runs, respectively.

\begin{figure*}
\includegraphics[width=7.1in]{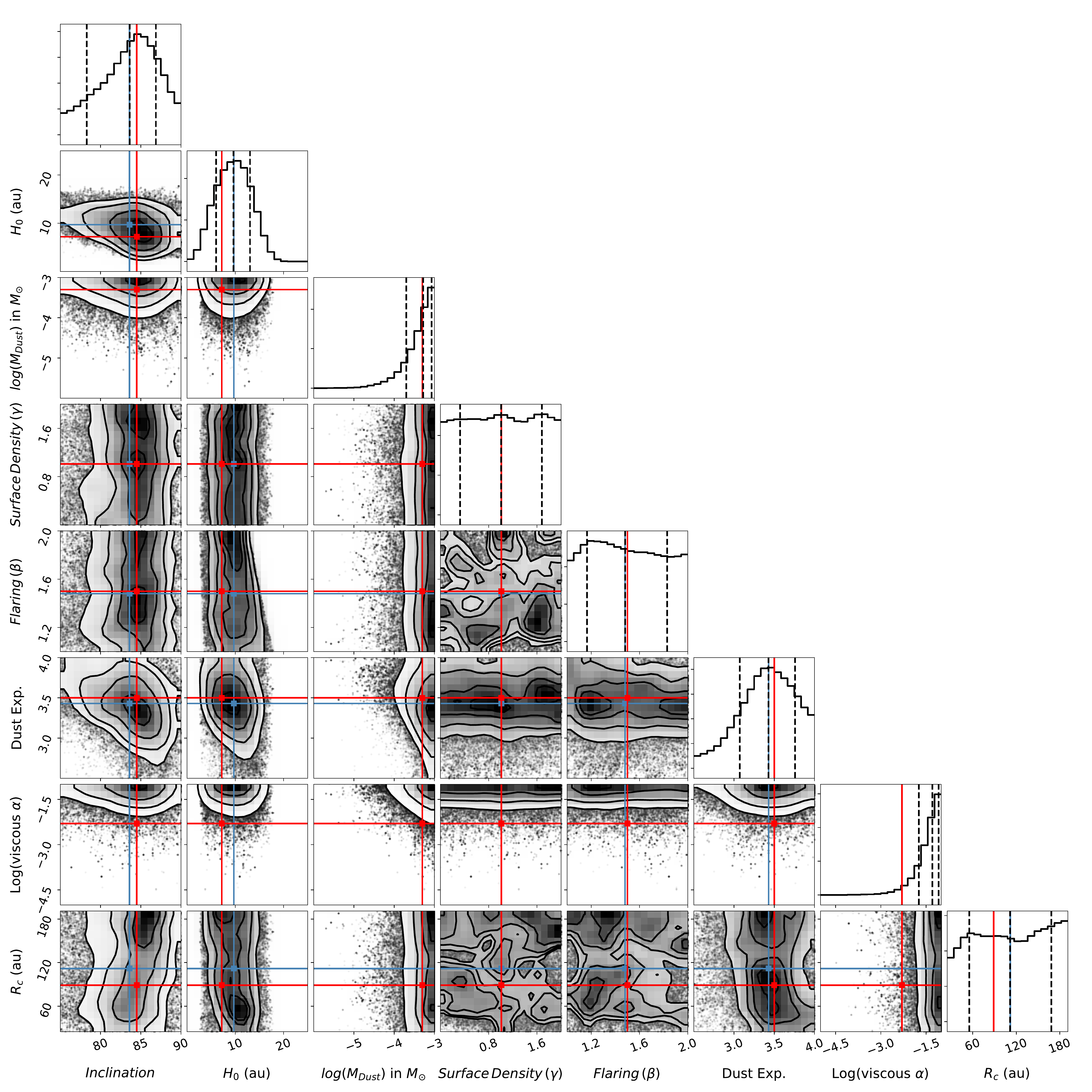}
\caption{The results of the F814W scattered light image MCMC run. The blue crosshairs indicate the best fit value for each parameter, taken as the median. The red crosshairs give the parameter values in Model A. Contours are drawn at the 1-4$\sigma$ levels. Inside of the contours, the density of the parameter space sampling is shown. Dashed vertical lines represent the 16th, 50th, and 84th percentiles of the samples in the posterior distributions. Only the inclination is well constrained, while the dust mass and viscous settling parameter both favor the highest allowed values. \label{fig:f814wcorner}} 
\end{figure*}

\begin{figure*}
\includegraphics[width=7.1in]{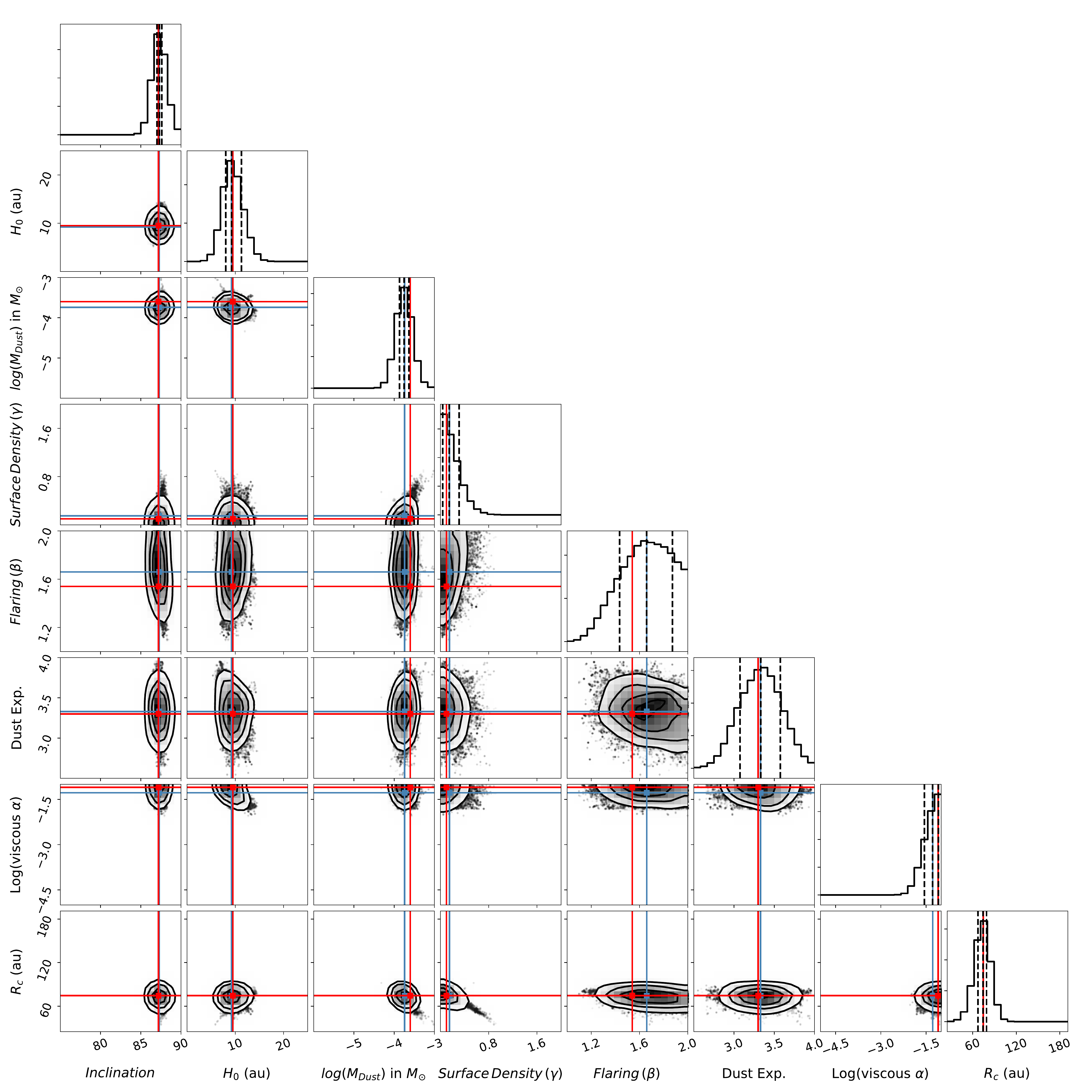}
\caption{The results of the ALMA millimeter continuum image MCMC run.  Markings are the same as shown in Figure \ref{fig:f814wcorner}. Here the red crosshairs give the parameter values in Model B. The ALMA data places a much tighter constraint on nearly all of the parameters when compared with the F814W image alone.  \label{fig:almacorner}}
\end{figure*}

\begin{figure*}
\includegraphics[width=7.1in]{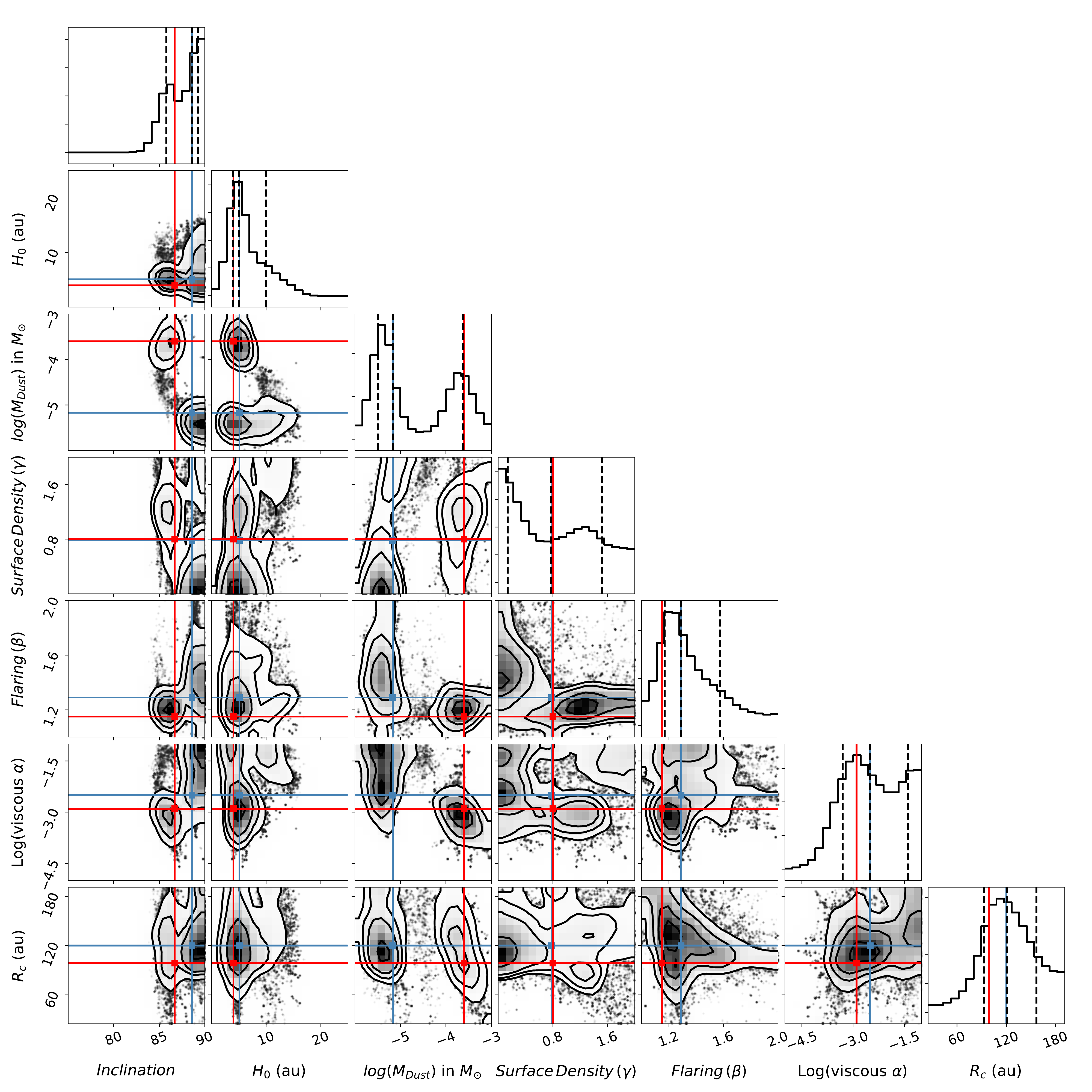}
\caption{The results of the MCMC run for the composite dataset (both the scattered light and mm continuum). Markings are the same as shown in Figure \ref{fig:f814wcorner}. Here the red crosshairs give the parameter values in Model C. The results represent a compromise between the two datasets. The inclination, scale height and alpha viscosity parameter remain unconstrained.   \label{fig:combocorner}}
\end{figure*}

\begin{deluxetable}{l|c|c}[htbp!]  
\tablecolumns{3}
\tablecaption{Covariance based MCMC Results show two families of models split by Dust Mass at $10^{-4} \, M_{\odot}$.  \label{table:combo}
}
\tablehead{   
  \colhead{Parameter} &
   \colhead{High Mass Results} &
    \colhead{Low Mass Results}
}
\tablewidth{3.7in}
\startdata
$i (^{\circ})$ & $86.0^{+0.4}_{-0.9}$ & $> 89.0$ \\  
$H_0$ (au)  & $4.8^{+0.9}_{-0.4}$ & $6.6^{+4.7}_{-3.2}$ \\
log M ($M_{\odot}$)   & $-3.7^{+0.3}_{-0.2}$ & $-5.4^{+0.2}_{-0.2}$ \\
$\gamma$ & $1.1^{+0.4}_{-0.6}$ & $< 0.4$ \\
$\beta$ & $1.2^{+0.1}_{-0.1}$ & $1.4^{+0.2}_{-0.2}$ \\
log $\alpha$ & $-3.1^{+0.4}_{-0.5}$ & $> -1.9$ \\
$R_{C}$ (au) & $119^{+34}_{-36}$ & $122^{+36}_{-22}$ \\
\enddata
\tablecomments{In cases where the posterior distributions peak at the edge of the allowed parameter range, upper/lower limits are used quoting the 50\% quantiles.}
\end{deluxetable}

\begin{figure*}
\includegraphics[width=7.1in]{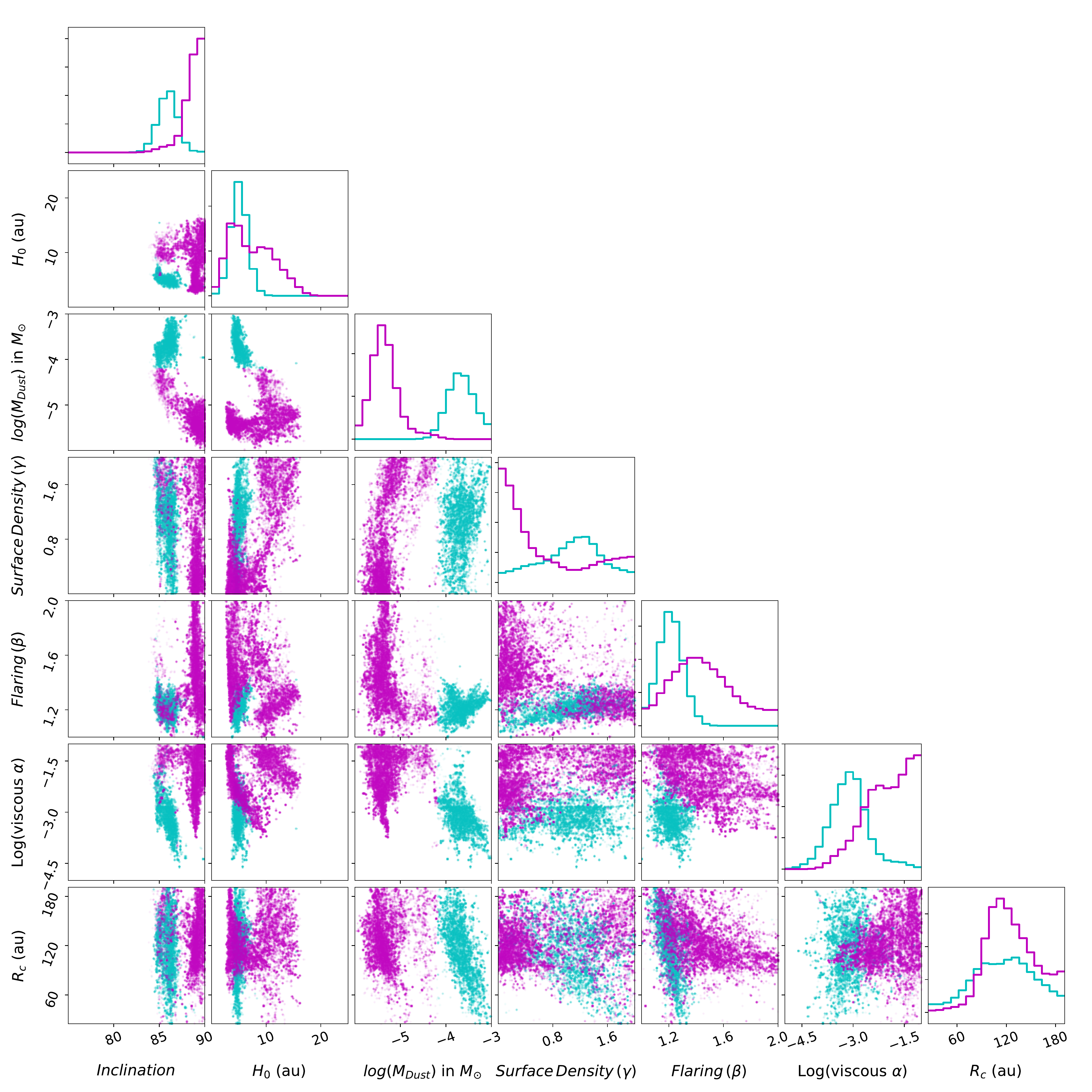}
\caption{The results of the MCMC run for the composite dataset (both the scattered light and mm continuum). Here we plot individual points for each model color-coded to distinguish between the low mass model with $M_{dust} < 10^{-4}$ shown in magenta and the high mass models with $M_{dust} \ge 10^{-4}$ shown in cyan. The histograms have been smoothed with a 1$\sigma$ Gaussian kernel (this results in the overlap in the mass histograms in column 3). \label{fig:colorcombo}}
\end{figure*}

The results for the covariance based MCMC run using the combined ALMA and F814W datasets show a bimodal parameter distribution as presented in Section \ref{Sec:covar} and further discussed in Section \ref{Sec:discussion}. This is most apparent in the disk dust mass parameter, but ultimately points to the existence of two discrete classes of models preferred by the two different datasets. 
To better demonstrate the locations of these two sub-populations within the other parameter distributions, we present a color coded triangle plot in Figure \ref{fig:colorcombo} (adapted from Figure \ref{fig:combocorner}). The best fit values to the posterior distributions are also separated in Table \ref{table:combo}.
We conclude that the higher mass population is most likely; a conclusion driven in part by the constraints on the dust mass provided by the 1.3 mm continuum observations. 
Furthermore, for the higher mass modeling subset we find that 1. the associated parameters are more in line with the individual fits (see in particular the inclination and disk mass) and 2. the flaring exponent associated with the low-mass models is unreasonably high. 

The subset of high mass models produce better constrained posterior distributions for nearly all parameters. The sole exception is in the critical radius where the difference is marginal and both the high and low mass models prefer a critical radius of $\sim$ 120 au. The low mass models hit the allowed parameter boundaries for the inclination, surface density exponent and the viscous settling parameter. It is unclear why the covariance modeling framework has a slight preference for the lower mass models when our expectation from the dataset is for a higher mass disk. This further illustrates the challenges of modeling the interiors of optically thick disks while invoking physical processes that generate wavelength-dependent morphologies.

\bibliography{oph163131}

\end{document}